\documentclass[letterpaper,11pt]{article}
\usepackage{jheppub} 
\usepackage{lineno}
\usepackage[utf8]{inputenc}

\pdfoutput=1 
\usepackage{jheppub}
\usepackage{graphicx,verbatim}
\usepackage{blindtext}
\usepackage[T1]{fontenc}
\usepackage{epsf}
\usepackage{mathtools}

\usepackage{lmodern}
\usepackage[mathscr]{euscript}

\usepackage{ytableau}
\usepackage{youngtab}

\usepackage{preambles_Saebyeok}

\usepackage{amsmath}
\usepackage{amsfonts}
\usepackage{amssymb}
\usepackage{mathrsfs}
\usepackage{slashed}
\usepackage{xcolor}
\usepackage{dsfont}
\usepackage{tikz, float}
\usetikzlibrary{patterns,shapes.misc}
\usepackage{circuitikz}

\def\qe{\mathfrak{q}}
\def\kq{\mathfrak{q}}

\def\fD{\mathfrak{D}}

\def\fR{\mathfrak{R}}

\def\rx{\mathrm{x}}

\def\ii{\mathrm{i}}

\def\bx{\mathbf{x}}
\def\ba{\mathbf{a}}
\def\bm{\mathbf{m}}
\def\bu{\mathbf{u}}
\def\bv{\mathbf{v}}
\def\bM{\mathbf{M}}

\def\bc{\mathbf{c}}
\def\bd{\mathbf{d}}

\def\bs{\mathbf{s}}

\def\bL{\mathbf{L}}
\def\bH{\mathbf{H}}
\def\bI{\mathbf{I}}
\def\bM{\mathbf{M}}
\def\bS{\mathbf{S}}
\def\bN{\mathbf{N}}
\def\bK{\mathbf{K}}
\def\bU{\mathbf{U}}
\def\bX{\mathbf{X}}
\def\bI{\mathbf{I}}

\def\bQ{\mathbf{Q}}

\def\bna{\boldsymbol\nabla}

\def\BC{\mathbb{C}}
\def\BR{\mathbb{R}}

\def\BZ{\mathbb{Z}}

\def\BP{\mathbb{P}}

\def\CalA{\mathcal{A}}

\def\CalR{\mathcal{R}}
\def\CalZ{\mathcal{Z}}
\def\CalH{\mathcal{H}}

\def\CalO{\mathcal{O}}
\def\CalM{\mathcal{M}}

\def\CalC{\mathcal{C}}

\def\qe{\mathfrak{q}}

\def\CalX{\mathcal{X}}

\def\ve{{\varepsilon}}

\def\fgl{\mathfrak{gl}}
\def\hfgl{\widehat{\mathfrak{gl}}}

\def\fD{\mathfrak{D}}

\def\fZ{\mathfrak{Z}}

\def\EQ{\EuScript{Q}}
\def\EN{\EuScript{N}}
\def\EY{\EuScript{Y}}

\def\EB{\EuScript{B}}
\def\ER{\EuScript{R}}
\def\ET{\EuScript{T}}
\def\EM{\EuScript{M}}

\def\EC{\EuScript{C}}
\def\ED{\EuScript{D}}
\def\EK{\EuScript{K}}

 \def\p{\partial}
 
 \def\a{\alpha}
 \def\b{\beta}
 \def\g{\gamma}
 \def\d{\delta}

 \def\th{\theta}
 
 \def\k{\kappa}
 \def\l{\lambda}
 \def\m{\mu}

 \def\r{\rho}

 \def\s{\sigma}
 
 \def\th{\theta}

 \def\G{\Gamma}
 \def\D{\Delta}

 \def\P{\Pi}
 \def\S{\Sigma}
 \def\L{\Lambda}
 \def\O{\Omega}
 \def\o{\omega }

\def\blm{\boldsymbol\lambda}
\def\bLm{\boldsymbol\Lambda}

\def\blm{\boldsymbol\lambda}
\def\bet{\boldsymbol\eta}
\def\bxi{\boldsymbol\xi}
\def\bka{\boldsymbol\kappa}
\def\bLm{\boldsymbol\Lambda}

\title{{$Q$-operators, $q$-opers, and R-matrices\\in 5d $\EN=1$ gauge theory}}
\author[a]{Saebyeok Jeong}
\author[b]{and Norton Lee}

\affiliation[a]{Department of Theoretical Physics, CERN, \\ 1211 Geneva 23, Switzerland}
\affiliation[b]{Center for Geometry and Physics, Institute for Basic Science (IBS),\\Pohang 37673, Republic of Korea}

\emailAdd{saebyeok.jeong@cern.ch}
\emailAdd{norton.lee@ibs.re.kr}

\preprint{CERN-TH-2025-138}

\vspace{2cm}
\abstract{We study the quantization of the moduli space of multiplicative Higgs bundles through the lens of five-dimensional $\EN=1$ supersymmetric gauge theories in $\Omega$-background. We extend the 4d $\EN=2$ gauge theoretical construction of key geometric and representation-theoretic structures, established in earlier works, to the five-dimensional uplift. We construct and analyze the $Q$-operators and $q$-opers associated with the canonical codimension-two defect: the $Q$-operators are defined via the insertion of the defect, while the $q$-opers arise as the $q$-difference chiral ring equations in its presence. The $q$-oper difference equations are further identified with the Baxter TQ equations for XXZ spin chains constructed from tensor products of bi-infinite evaluation modules over quantum affine algebras of type ${\mathfrak{gl}}(n)$. We define a $q$-difference module structure on the space of monodromy codimension-two defect partition functions and show that the eigenstates of the $Q$-operators, constructed from monodromy defects, simultaneously diagonalize the quantum Hamiltonians of the XXZ spin chain. A Fourier transformation exchanges the $Q$-operators associated with two XXZ spin chains bispectral dual to each other. Finally, we relate these constructions to the quantum cluster algebra arising from the BPS quiver of the 5d theory, and re-express the R-matrices in terms of the cluster variables.}

\begin{document}
\maketitle

\section{Introduction}

The rich interplay between supersymmetric gauge theories and geometry of moduli spaces of bundles and connections yields novel insights on both sides of the correspondence. A prominent example is 4d $\EN=2$ theories of class $\mathcal{S}$ \cite{gai1} which, under a proper compactification, give rise to a two-dimensional topological $A$-model with the target space given by the moduli space $\EM_{\text{Higgs}} (G,\EC;S)$ of parabolic $G$-Higgs bundles on an algebraic curve $\EC$ with ramifications at discrete points $S \subset \EC$ \cite{Gaiotto:2009hg,Nekrasov:2010ka}. This hyper-K\"{a}hler moduli space enjoys an algebraically integrable structure with respect to one of its holomorphic symplectic forms \cite{hitchin1987}. A quantization of this integrable structure is implemented by the $\ve_2 \to 0$ limit \cite{Nekrasov:2009rc,Nekrasov:2010ka,Jeong:2018qpc,Jeong:2023qdr} of the $\O_{\ve_1,\ve_2}$-background \cite{Nekrasov:2002qd}, which was shown to be compatible with the brane quantization in the sigma model perspective \cite{Gukov:2008ve}.

In the topological $A$-model, the quantization of the integrable structure is implemented in two stages. First, the commutative algebra of holomorphic functions on the moduli space $\EM_{\text{Higgs}} (G,\EC;S)$ is deformed into a sheaf of algebras $-$ the algebra of open strings stretched between space-filling canonical coisotropic branes $\EB_{\text{cc}}$ \cite{Kapustin:2001ij} $-$ whose product is given by the joining of strings. Then, any $A$-brane defines a sheaf of modules over this noncommutative algebra, with the module action also realized via string joining. The 4d $\EN=2$ gauge theory origin of the topological $A$-model suggests an alternative presentation of the sections of these sheaves of modules \cite{Nekrasov:2010ka,Jeong:2023qdr}. Namely, they are given by path integrals of the 4d $\EN=2$ gauge theory on $\BR^4$ in the presence of the $\O_{\ve_1,\ve_2}$-background, which reduce to finite-dimensional integrals of equivariant characteristic classes over the resolved moduli space of framed instantons on $\BR^4$. The technique of equivariant localization then allows these integrals to be exactly evaluated as summations over fixed points under the action of the global symmetry group \cite{Nekrasov:2002qd}. 

The 4d $\EN=2$ gauge theory has a natural uplift to 5d $\EN=1$ gauge theory compactified on a circle, which yields at low energy a 4d $\EN=2$ effective gauge theory incorporating the Kaluza-Klein modes. In this uplifted setting, the target space of the associated two-dimensional sigma model is replaced by the moduli space $\EM_{\text{mHiggs}} (G,\EC;S)$ of multiplicative Higgs bundles, in which the Higgs field becomes group-valued rather than Lie-algebra-valued \cite{Nekrasov:2012xe,Nekrasov:2013xda,Frassek:2018try}. The corresponding classical integrable models have been extensively studied in \cite{Nekrasov:1996cz,Cherkis:2000cj,Cherkis:2000ft}. A natural and compelling question is how the quantization of the integrable structure, implemented via the $\O$-background in four dimensions, generalizes in the five-dimensional uplift.

The path integral of the 5d $\EN=1$ gauge theory on $\BR^4 \times S^1$ in the presence of the $\O_{\ve_1,\ve_2}$-background yields the K-theoretic uplift of the 4d counterpart, i.e., the generating function of the equivariant indices of the Dirac operator on the resolved moduli spaces of framed instantons on $\BR^4$ \cite{Lawrence:1997jr,Nekrasov:2002qd}. Equivariant localization remains applicable in this five-dimensional setting, allowing one to compute the equivariant indices exactly as summations over the fixed points under the global symmetry action. The construction of the key ingredients in the topological $A$-model required for the quantization $-$ namely, the Hecke operators, the opers, the sections of the Hecke eigensheaf corresponding to a given oper, and the R-matrices of the affine Kac–Moody algebra $-$ was realized within the framework of 4d $\EN=2$ gauge theory in \cite{Jeong:2018qpc,Jeong:2023qdr}. This construction admits a natural lift to the 5d $\EN=1$ gauge theory. The principal aim of the present work is to carry out this uplift explicitly in the case where $G= GL(N)$, $\EC= \BC^\times = \BP^1 \setminus \{0,\infty\}$, in which the multiplicative Higgs bundles are assigned with framings at boundary, and $\vert S \vert =M$ with minimal regular ramifications (in particular, $M=2$ for most of the contents).

Namely, we will provide the 5d $\EN=1$ gauge theoretical constructions of $Q$-operators, $q$-opers, a $q$-difference module and its basis diagonalizing the actions of the $Q$-operators and the quantum Hamiltonians simultaneously, and finally R-matrices of quantum affine algebras intertwining evaluation modules as we now briefly explain.

\subsection{$Q$-operators and $q$-opers from canonical codimension-two defect}

In any complex structure different from $\pm I$, the moduli space $\EM_{\text{Higgs}}(GL(N),\EC;S)$ of Higgs bundles is viewed as the moduli space of $GL(N)$-local systems on $\EC$ with regular monodromies at $S$. A distinguished holomorphic Lagrangian submanifold is the spanned by opers, for which the local systems can be expressed as scalar differential operators \cite{BD1,BD2}.

In our multiplicatively uplifted setting with $\EM_{\text{mHiggs}}(GL(N),\EC;S)$, the local systems get replaced by the \textit{$q$-difference connections} with framings at $\{0,\infty\}\subset \BP^1$ and regular singularities at $S$ \cite{Elliott:2018yqm}. The moduli space of $q$-connections contains a distinguished holomorphic Lagrangian subspace spanned by \textit{$q$-opers}, for which the $q$-connections can be expressed as scalar $q$-difference operators \cite{Chanski1998,Semenov1998,Koroteev:2018jht,Feigin2024}. We will construct these $q$-opers (in fact, their $q_2 \neq 1$ refinement) as the quantized chiral ring relation for the canonical codimension-two defect, obtained by coupling a 3d $\EN=2$ $U(1)$ gauge theory whose flavor symmetry is gauged by the 5d gauge field at the codimension-two locus.

The 3d $\EN=2$ $U(1)$ gauge theory compactified on a circle admits a complexified Coulomb parameter $X$ and a complexified FI parameter $Z$, both of which are valued in $\BC^\times$ due to the large gauge transformation. We call the 3d theory to be in the Coulomb phase when $X$ can set to be arbitrary and the canonical codimension-two defect defines the $Q$-observable of the 5d gauge theory. Meanwhile, the 3d theory is said to be in the Higgs phase when $Z$ can set to be arbitrary and the canonical codimension-two defect defines the $H$-observable. We provide a gauge theoretical definition of the $Q$-operators as inserting the $Q$-observables (resp. the $H$-observables) into the correlation function.

In the limit $q_2 \to 1$, we have their vacuum expectation values factorized as
\begin{align}
  \lim_{q_2 \to 1}\left\langle \mathbf{H}(Z) \right\rangle_\ba = e^{\frac{\widetilde{\EW}(\ba)}{\log q_2}} \mathbf{H}(\ba;Z) ,\qquad   \lim_{q_2 \to 1}\left\langle \mathbf{Q}(X) \right\rangle_\ba = e^{\frac{\widetilde{\EW}(\ba)}{\log q_2}} \mathbf{Q}(\ba;X).
\end{align}
We show that the 5d Coulomb moduli $\ba$ parametrize the affine space of $GL(M)$ (resp. $GL(N)$) $q$-opers, which annihilate the vacuum expectation value $\mathbf{Q}(\ba;X)$ (resp. $\mathbf{H}(\ba;Z)$). We will see that the so-obtained $q$-oper difference equation for $\mathbf{Q}(\ba;X)$ (resp. for $\mathbf{H}(\ba;Z)$) is precisely the Baxter TQ equation for XXZ spin chain constructed on $N$ (resp. $M$) bi-infinite evaluation modules of the quantum affine algebra $U_q (\widehat{\fgl}(M))$ (resp. $U_q (\widehat{\fgl}(N)$).

The two observables obtained in the Coulomb and the Higgs phases are connected through a Fourier transformation. We establish the exact transformation between the two, which schematically reads
\begin{align}
    \mathbf{H} (Z) = \oint_{\CalC} \frac{\mathrm{d} X}{X} Z^{-\frac{\log X}{\log q_1}} \mathbf{Q}(X) ,
\end{align}
where $\CalC$ is a Barnes-type contour in the $\BC^\times _X$-plane. The associated $GL(N)$ and $GL(M)$ $q$-opers are precisely mapped to each other under this Fourier transformation. Since the coefficients of the $q$-opers determine the spectra of the corresponding quantum Hamiltonians of the XXZ spin chains, the Fourier transformation exactly exhibits the bispectral duality between the two associated quantum integrable models.

\subsection{Eigenstates of $Q$-operators from monodromy codimension-two defect}

For the ordinary moduli space of Higgs bundles, the $(\EB_{\text{cc}},\EB_{\text{cc}})$-strings give rise to the sheaf of differential operators on the moduli space $\text{Bun}_{GL(N)} (\EC;S)$ of parabolic $G$-bundles on $\EC$. For any $A$-brane $\EB'$ the $(\EB_{\text{cc}} , \EB')$-strings thus yield a twisted $\ED$-module on $\text{Bun}_{GL(N)} (\EC;S)$. In our multiplicative uplift, the twisted $\ED$-modules are replaced by the $q$-difference modules \cite{Mochizuki:2019xtr}. Instead of geometrically formulating it as a sheaf of modules on $\text{Bun}_{GL(N)} (\EC;S)$, we only note that this $q$-difference module gives rise to the quantization of the algebra of holomorphic functions on $\EM_{\text{mHiggs}}(GL(N),\EC;S)$, which is expected to provide a module over the quantum affine algebra $U_q (\widehat{\fgl}(N))$ \cite{Nekrasov:2013xda,Elliott:2018yqm}.

We propose to realize this module by the vector space spanned by the vacuum expectation values of the monodromy codimension-two defect in the 5d $\EN=1$ gauge theory, defined over a $\BZ$-lattice of Coulomb vacua. First, we form the vector space $\CalH = \BC[[ u_\o ^{\pm 1};\o\in [N] ]]$ of Laurent series in the monodromy parameters. Then, a distinguished basis is constructed by the vacuum expectation values of the monodromy codimension-two defect, enumerated by the Coulomb moduli $\ba$,
\begin{align}
    \psi(\ba) \in \CalH,\qquad U_q (\widehat{\fgl}(N))\to \text{End}(\CalH).
\end{align}
We show this by constructing the R-matrices of the quantum affine algebra $U_q (\widehat{\fgl}(N))$ from the analytic constraints imposed on the correlation functions of the monodromy codimension-two defect and the canonical codimension-two defect. As a result, we find the space $\CalH$ carries a $U_q (\widehat{\fgl}(N))$-module given by a tensor product of $M$ evaluation modules. The generators of the quantized enveloping algebra $U_q (\fgl(N))$ are represented by certain $q$-difference operators, with the quantum parameter given by the exponentiated $\O$-background parameter $q = q_1 ^{-\frac{1}{2}}$.

In the perspective of the bispectral duality, the $q$-difference module carried by $\CalH$ realizes a module over the quantum affine algebra $U_q (\widehat{\fgl}(M))$ at the same time. We show that it is indeed a tensor product of $N$ evaluation modules over $U_q (\widehat{\fgl}(M))$. This bi-module structure emerges from the isomorphism between $\EM_{\text{mHiggs}}(GL(N),\BC^\times;S)$ and $\EM_{\text{mHiggs}}(GL(M),\BC^\times;D)$ through the Fourier-Mukai transformation, from which the Hecke eigensheaf gives rise to modules over two quantum affine algebras. Note that the rank of the Lie group and the number of minimal regular ramifications are exchanged $M \leftrightarrow N$.

It turns out that the basis elements $\psi(\ba) \in \CalH$ diagonalize the action of the $Q$-operators. Recall that the action of $Q$-operators are defined by inserting the $H$-observable or the $Q$-observable, for the $U_q (\widehat{\fgl}(N))$-module and the $U_q (\widehat{\fgl}(M))$-module respectively, on top of the monodromy codimension-two defect. Then, the two codimension-two defects can be arbitrarily separated from each other in the limit of $q_2\to 1$, due to the cluster decomposition. Namely, we have their correlation function factorized into the product of the vacuum expectation values of each,
\begin{align}
    \left\langle \mathbf{H}(Z) \boldsymbol\Psi \right\rangle_\ba = e^{\frac{\widetilde{\EW}(\ba)}{\log q_2}} \mathbf{H}(\ba;Z) \psi(\ba),\qquad \left\langle \mathbf{Q}(X) \boldsymbol\Psi \right\rangle_\ba = e^{\frac{\widetilde{\EW}(\ba)}{\log q_2}} \mathbf{Q}(\ba;X) \psi(\ba).
\end{align}
In this sense, the distinguished basis provided by the vacuum expectation values of the monodromy codimension-two defect provides the eigenstates of the $Q$-operators, with the eigenvalues given by the vacuum expectation value of the $H$-observable or the $Q$-observable, respectively.

\subsection{Quantum Hamiltonians and their eigenstates}

Finally, we demonstrate that the $Q$-operators constructed within our 5d gauge theory satisfy the operator-valued $q$-oper difference equation for the module $\CalH$ of the quantum affine algebra $U_q (\widehat{\fgl}(N))$, from the chiral ring relation in the presence of the monodromy codimension-two defect and the canonical codimension-two defect inserted on top of each other. Combining with the scalar $q$-oper differential equation satisfied by the vacuum expectation value of the $Q$-observable, we verify
\begin{align}
    0 =\left( \hat{t}(X) - t(\ba;X) \right)\psi(\ba),
\end{align}
where $\hat{t}(X)$ is the generating function of the quantum Hamiltonians of the XXZ spin chain and $t(\ba;X)$ is the generating function of the vacuum expectation values of the codimension-four observables in the 5d $\EN=1$ gauge theory. In this way, we confirm that the eigenstates of the $Q$-operators are also eigenstates of the quantum Hamiltonians.\footnote{It was shown in \cite{Grassi:2014zfa} that a non-perturbative correction in the effective twisted superpotential provides the exact quantization condition for the $L^2$-normalizable wavefunction along a certain choice of real slice. In the present work, we construct eigenstates of the XXZ spin chain by imposing a different quantization condition, namely that the Coulomb moduli take values on a $q_1$-lattice, leading to a bi-infinite module over the quantum affine algebra. It would be interesting to investigate how these two quantization schemes are related within the framework of 5d $\EN=1$ gauge theory. Relatedly, it would be interesting to study the $q$-deformation of the analytic Langlands correspondence \cite{Etingof:2019pni,Etingof:2023drx} and its quantum variant \cite{Gaiotto:2024tpl}, in the 5d $\EN=1$ gauge theory framework. See \cite{Gaiotto:2021tsq,Gaiotto:2021tzd} for studies in the context of the GL-twisted 4d $\EN=4$ gauge theory and vertex algebras.}

\subsection{Quantum cluster algebra for BPS quiver}

Finally, we revisit the 5d $\EN=1$ gauge theoretical construction of the R-matrices of the quantum affine algebra from the IR perspective. Namely, we recover these R-matrices in the context of the quantum cluster algebra for the 5d BPS quiver.

The 5d BPS quiver is the quiver for the $\EN=4$ supersymmetric quantum mechanics encoding the BPS spectra of the 5d $\EN=1$ gauge theory in the IR \cite{Closset:2019juk}. In the perspective of the geometric engineering, when the 5d theory is compactified on a circle $S^1$, the 5d BPS quiver arises as the quiver dictating the contents of the 4d $\EN=1$ quiver gauge theory realized on the worldvolume of D3-branes probing a toric Calabi-Yau threefold \cite{Duan:2020qjy}. The bipartite graph dual to such a quiver is called the dimer model \cite{Kenyon:2003eyz} or the brane-tiling \cite{Hanany:2005ve,Franco:2005rj,Franco:2005sm}. 

As pointed out in \cite{goncharov2011dimers}, a dimer model realized as a bipartite graph on an oriented 2-torus give rise to a cluster integrable system, whose spectral curve $\boldsymbol\Sigma$ coincides with the Seiberg-Witten curve of the 5d $\EN=1$ gauge theory. The holomorphic curve is also important in the type IIB construction of the 4d $\EN=1$ quiver gauge theory. The D5-branes are suspended between NS5-branes wrapping $\boldsymbol\Sigma$, which is represented by the bipartite graph of the dimer model. The IIB fivebrane construction is T-dual to the D3-branes probing the toric Calabi-Yau threefold \cite{Douglas:1996sw,Feng:2000mi,Feng:2001xr}, and  
$\boldsymbol\Sigma$ is the mirror curve of the mirror Calabi-Yau threefold \cite{Hori:2000ck,Hori:2000kt,Feng:2005gw}.
The classical and quantum integrable models realized on the cluster algebras of quivers and the dimer models have been extensively studied in \cite{Kenyon:2003eyz,goncharov2011dimers,Eager:2011dp,Huang:2020neq,Kho:2025fmp}. 

In our case of the 5d $\EN=1$ $A_{M-1}$-quiver $U(N)$ gauge theory, we find this quantum integrable model coincides with the one arising from the quantization of the moduli space of multiplicative Higgs bundles, i.e., the XXZ spin chain built upon $M$ bi-infinite evaluation modules of the quantum affine algebra $U_q (\widehat{\fgl}(N))$ (and also $N$ bi-infinite evaluation modules of the quantum affine algebra $U_q (\widehat{\fgl}(M))$, by the bispectral duality). In particular, we recover the R-matrices of the quantum affine algebras, which were used to construct the XXZ spin chains, in terms of the quantum cluster variables.




\subsection{Outline}

We begin with reviewing the necessary ingredients of the 5d $\EN=1$ gauge theory compactified on a circle in section \ref{sec:5dg}. In particular, we introduce the canonical codimension-two defect and the monodromy codimension-two defect, obtaining their observable expressions. In section \ref{sec:tq},we derive the analytic constraints satisfied by the vacuum expectation values of the codimension-two defects and their correlation functions. We call these equations (fractional) quantum TQ equations. We show they are $q_2 \neq 1$ deformation of the $q$-oper difference equation. In section \ref{sec:rmat}, we rearrange the fractional TQ equation into the R-matrices of the quantum affine algebra. The construction leads to the operator-valued $q$-oper difference equation satisfied by the correlation function of the two codimension-two defects. By using the factorization property in the limit $q_2\to 1$, we conclude the eigenstates of the $Q$-operators also diagonalize the quantum Hamiltonians of the associated XXZ spin chain. In section \ref{sec:cluster}, we revisit the gauge theoretical construction of the R-matrices in the IR perspective. We establish them in the context of the quantum cluster algebra for the 5d BPS quiver. We conclude with discussions in section \ref{sec:dis}. The appendices contain computational detail of the dual $Q$-observable and a review of quantum affine algebra of $\fgl(n)$.

\paragraph{Acknowledgment}
The authors thank Sibasish Banerjee, Sebastian Franco, Alba Grassi, Nathan Haouzi, Nafiz Ishtiaque, Minsung Kho, Taro Kimura, Shota Komatsu, Kimyeong Lee, Nikita Nekrasov, Rak-Kyeong Seong, Dmytro Voloshyn, Philsang Yoo, and Yehao Zhou for discussions and collaboration on related subjects. SJ is grateful to Davide Gaiotto for helpful discussion and support during his visit to Perimeter Institute, where a part of the work was done. The work of SJ is supported by CERN and CKC fellowship. The work of NL is supported by IBS project IBS-R003-D1.

\section{Five-dimensional gauge theory and codimension-two defects} \label{sec:5dg}

We introduce our basic setup of the 5d $\EN=1$ gauge theory compactified on a circle. Throughout this work, we focus on the $A_1$-quiver $U(N)$ gauge theory; however, the generalization to the case of the $A_{M-1}$-quiver $U(N)$ gauge theory would be straightforward.

We introduce the canonical codimension-two defect by coupling a 3d $\EN=2$ sigma model. The observables defined by its insertion is called $Q$-observable or $H$-observable, depending on whether the 3d theory is in the Coulomb phase or the Higgs phase. Also, we introduce the codimension-two monodromy defect defined by assigning singular behavior of fields along the codimension-two locus. Finally, we study the configuration of parallel defects where those defects are inserted on top of each other.

\subsection{5d $\EN=1$ gauge theory compactified on $S^1$}

The partition function of the 5d $\EN=1$ $U(N)$ gauge theory with $N$ fundamental and $N$ anti-fundamental hypermultiplets on $\BR^2 _{\ve_1} \times \BR^2 _{\ve_2} \times S^1$ is computed by
\begin{align}
        \CalZ (\ba,\bm^\pm ;\qe) = \sum_{\blm} \kq^{|\blm|} \; \hat{a} \left[ \frac{-SS^*+M^+S^*+ S(M^-)^* }{P_{12}^*} \right].
\end{align}
Given an equivariant K-theory class $X = \sum_{i} n_i X_i$, we denote its dual by $X^* = \sum_{i} n_i X_i^{-1}$. Also, the $\hat{a}$-class of a virtual line bundle is computed by
\begin{align}
    \hat{a}[X] = \frac{1}{X^{\frac 1 2} - X^{-\frac 1 2}} = - X^{\frac{1}{2}} \text{PE}[X],
\end{align}
extended multiplicatively to any virtual bundle by $\hat{a}[X_1+X_2] = \hat{a}[X_1] \hat{a} [X_2]$. For later use, note that
\begin{align}
\begin{split}
    \hat{a}\left[ -\frac{X}{P_1 ^*} \right] &= \prod_{l=0} ^\infty \hat{a} \left[- X q_1 ^{-l} \right] = \prod_{l=0} ^\infty \left( X^{\frac{1}{2}} q_1 ^{-\frac{l}{2}} - X^{-\frac{1}{2}} q_1 ^{\frac{l}{2}} \right)\\
    &\;\dot{=} \; X^{\frac{1}{2} \zeta(0)} q_1 ^{-\frac{1}{2} \zeta (-1) } \left( X^{-1} ;q_1 \right)_\infty = X^{-\frac{1}{4}} q_1 ^{\frac{1}{24}} \left( X^{-1} ;q_1 \right)_\infty \\
    &=: \left( X^{-1} ;q_1 \right)'_\infty,
\end{split}
\end{align}
where $\dot{=}$ indicates the zeta-function regularization. We also defined the expression in the third line by the one in the second.

\subsection{Canonical codimension-two defect}
Local or non-local BPS observables of the 5d $\EN=1$ gauge theory can be inserted into the path integral. By supersymmetric localization, these observables pass to the equivariant K-theory classes on the moduli space of instantons. The equivariant localization then leads to the observables on the ensemble of fixed points of the moduli space of instantons, enumerated by $N$-tuples of Young diagrams $\bl = (\l^{(0)} ,\cdots, \l^{(N-1)})$. Here, we will introduce half-BPS codimension-two defects relevant to our study and write their observable expressions explicitly, uplifting the study of surface defects in 4d $\EN=2$ gauge theory \cite{Jeong:2018qpc,Jeong:2020uxz,Lee:2020hfu,Nikita:IV,Jeong:2023qdr,Jeong:2024hwf,Jeong:2024onv}.\\

We will introduce two codimension-two defect observables $-$ the $Q$-observable and the $H$-observable $-$ defined by coupling a 3d $\EN=2$ gauge theory supported on $\BR^2 _{\ve_1} \times S^1 _R$ to the 5d $\EN=1$ gauge theory. In particular, the contents of the 3d theory are the $U(1)$ vector multiplet, $N$ chiral multiplets of charge $+1$, and $N$ chiral multiplets of charge $-1$. 

Due to the compactification along the circle with radius $R$, the real scalar $\s^{\BR}$ in the vector multiplet gets augmented by the holonomy of the gauge field, yielding a complex scalar
\begin{align}
    \s^\BC = \s^\BR + \frac{\ii}{R} \oint_{S^1} A,
\end{align}
with the periodicity $\s^\BC \sim \s^\BC + \frac{2\pi \ii}{R}$ induced by the large gauge transformation. We define the $\BC^\times$-valued Coulomb parameter by the vacuum expectation value of $X = \exp R \s^\BC$. Similarly, we may turn on the background gauge field $A^{(T)}$ for the topological $U(1)$ symmetry, augmenting the real FI parameter by its holonomy,
\begin{align}
    \zeta^\BC = \zeta^\BR + \frac{\ii}{R} \oint_{S^1} A^{(T)},
\end{align}
with the periodicity $\zeta^\BC \sim \zeta^\BC + \frac{2\pi \ii}{R}$ due to the large gauge transformation. We denote the resulting $\BC^\times$-valued complexified FI parameter by $Z = \exp R \zeta^\BC$. The mass parameters for the chiral multiplets also get $\BC^\times$-valued by the holonomy of the background gauge field for the flavor symmetry.

\subsubsection{Coupling 3d $\EN=2$ theory in Coulomb phase: $Q$-observable} \label{subsubsec:Qobs}

Let us first consider the partition function of the 3d $\EN=2$ gauge theory on its own, without coupling it to our 5d $\EN=1$ gauge theory. To define the partition function, the boundary conditions of the vector multiplet and the chiral multiplets at the boundary $\p (\BR^2 \times S^1) = S^1 _\infty \times S^1$ must be specified. In this work, we do not analyze general choices of boundary conditions, but instead present the specific choice that arises from the gauge origami construction of the 3d $\EN=2$ gauge theory coupled to the 5d $\EN=1$ gauge theory. For details on the choices of boundary conditions, see \cite{Yoshida:2014ssa,Dimofte:2017tpi,Bullimore:2020jdq,Dedushenko:2023qjq,Crew:2023tky}. 

The vector multiplet is assigned with the Dirichlet boundary condition, where the scalar $X$ is set to a given constant and all the other fields are set to zero. Further, we impose the Neumann boundary condition for the $N$ chiral multiplets of charge $+1$ and the Dirichlet boundary condition for the $N$ chiral multiplets of charge $-1$. 

For the $N$ chiral multiplets of charge $-1$ subject to Dirichlet boundary conditions, the boundary values of their complex scalars must be specified. Different choices are compatible with different 3d BPS field configurations, leading to distinct localization schemes. For the codimension-two defect giving rise to the $Q$-observable, we set all boundary values to zero, a choice for which the Coulomb parameter $X \in \BC^\times$ remains unconstrained. This boundary condition is therefore compatible with the Coulomb branch localization scheme. \\

When the FI parameter is turned off, we may freely tune the vacuum expectation value $X \in \BC^\times $, probing the Coulomb phase of the 3d theory. In this case, there is no non-perturbative sector of the 3d field configuration, and its partition function is simply given by the 1-loop fluctuations of the chiral multiplets around the vacuum specified by $X$. 

\begin{align}
    \mathbf{Q}(X)[\bl = \varnothing] = \prod_{\a=0} ^{N-1} \frac{ \left( \frac{a_\a}{X}; q_1  \right)'_\infty}{\left(\frac{m^+ _\a }{X};q_1  \right)'_\infty}.
\end{align}
The 1-loop determinant of the vector multiplet does not depend on $X$ and it will not affect our study of the difference equations in $X$. Thus, we discarded this in the above expression, keeping only the contributions from the chiral multiplets. \\

Now, we consider coupling this 3d $\EN=2$ theory to the 5d $\EN=1$ gauge theory by requiring the $U(N)_-$ flavor symmetry (rotating the charge $-1$ chiral multiplets) to be gauged by the 5d gauge field restricted to the support of the 3d theory. This coupling is implemented in the path integral by converting the contribution from the flavor bundle $N$ to the one from the universal sheaf $S$ of the instantons. As a result, the $Q$-observable $\mathbf{Q}(X)$ at a given instanton configuration $\bl$ attains its observable expression as
\begin{align} \label{eq:qobser}
\begin{split}
    \mathbf{Q} (X)[\bl] & = \hat{a} \left[ \frac{X(M^+-S[\bl])^*}{P_1^*} \right] \\
        & = \prod_{\a=0} ^{N-1} \frac{ \left( \frac{a_\a q_1 ^{l(\l^{(\a)})}}{X}; q_1  \right)'_\infty}{\left(\frac{m^+ _\a }{X};q_1  \right)'_\infty} \prod_{i=1} ^{l(\l^{(\a)})} \sqrt{\frac{X}{a_\a} q_1^{-i+1} q_2 ^{-\l_i ^{(\a)}}  }\left( 1- \frac{a_\a}{X} q_1 ^{i-1} q_2 ^{\l_i ^{(\a)}} \right).
\end{split}
\end{align}
Note that the $\mathbf{Q} (X)$ has simple poles at $X \in m^+ _\a q_1 ^{\BZ_{\geq 0}}$ for each $\a \in \{0,1,\cdots, N-1\}$. This is precisely the loci where one of the KK mode of a chiral multiplet becomes massless. \\

Recall that the observables define equivariant K-theory classes on the moduli space of instantons due to supersymmetric localization. When the stability chamber for the moduli space of instantons is chosen differently, the observable is modified accordingly \cite{Jeong:2024onv}. 

In the case of the ${Q}$-observable \eqref{eq:qobser}, we call its dual observable in the other stability chamber the dual $Q$-observable. In fact, this $Q$-observable can be constructed from the gauge origami, as we review in appendix \ref{subsec:dualq}. Here, we only adopt its observable expression,
\begin{align} \label{eq:dualqobs}
    \tilde{\bQ}(X) [\bl] = \frac{\kq^{\frac{\log X}{\log q_1}}}{\prod_{\a=0}^{N-1} \left(\frac{m^+_\a}{X};q_1 \right)' _\infty} \sum_{d=0}^\infty \kq^d q_2^{-\frac{d}{2}} \frac{(q_1q_2;q_1)_d}{(q_1;q_1)_d} \tilde{Q}_d(X)[\bl],
\end{align}
where we defined
\begin{align}
    \tilde{Q}_{d}(X) [\bl] = \frac{Q(X)[\bl]M(Xq_1^{d})}{Q(Xq_1^{d})[\bl]Q(Xq_1^{d+1} q_2) [\bl]},
\end{align}
with $M(X)=M(Xq_1^{-1})P(X)$.

In a more general case of the $A_{M-1}$-quiver $U(N)$ gauge theory, there are $M$ inequivalent $Q$-observables \cite{Jeong:2024onv}. In this work, we will restrict our attention to the main example of $M=2$, with two inequivalent $Q$-observables \eqref{eq:qobser} and \eqref{eq:dualqobs}.

\subsubsection{Coupling 3d $\EN=2$ theory in Higgs phase: $H$-observable} \label{subsubsec:hobs}
For one ($\a$-th) of the $N$ chiral multiplets of charge $-1$ imposed with the Dirichlet boundary condition, we may turn on non-zero boundary value for its complex scalar. This is compatible with the BPS locus given by the vortex configuration with a fixed Coulomb parameter $X = a_\a$, leading to the Higgs branch localization. 

When $\zeta^\BR \gg  0$, the 3d $\EN=2$ gauge theory flows to the non-linear sigma model with the target space given by the Higgs branch. For the 3d $\EN=2$ theory at hand, the Higgs branch is given by the total space of the bundle
\begin{align}
    \mathcal{O}(-1)\otimes \BC^N \to \BP^{N-1}.
\end{align}
The localization yields the K-theoretic count of the based quasimaps from $\BP^1$ to the Higgs branch, with the image of $\infty \in \BP^1$ fixed to be the $\a$-th massive Higgs vacuum \cite{Bullimore:2020jdq,Crew:2020psc}. 

Thus, the localization yields the vortex partition function, or equivalently the generating function of the equivariant indices of the virtual structure sheaf of the moduli space of quasimaps, given by
\begin{align}\label{eq:vortex}
\begin{split}
      \mathbf{H}^{(\a)} (Z)[\bl=\varnothing] &=  \sum_{X \in a_\a q_1 ^\BZ} Z^{-\frac{\log X}{\log q_1}} \mathbf{Q}(X)[\bl=\varnothing] \\
    &= Z^{-\frac{\log a_\a}{\log q_1}} \sum_{k=1} ^\infty  Z^k \, \mathbf{Q}(a_\a q_1^{-k}) [\bl=\varnothing],
\end{split}
\end{align}
where the summation truncates to be semi-infinite due to the zeros of $\mathbf{Q}(X)[\bl=\varnothing]$, ensuring the convergence of the series in $\vert Z \vert <1$. Here, $\a \in \{0,1,\cdots, N-1\}$ enumerates the choice of the massive Higgs vacuum at infinity. \\

We define the codimension-two $H$-observable of the 5d $\EN=1$ gauge theory by coupling the 3d $\EN=2$ gauge theory in this Higgs phase. Due to the coupling, the summand of \eqref{eq:vortex} is simply replaced by the $Q$-observable $\mathbf{Q}(X)[\bl]$ at a given 5d BPS configuration labeled by $\bl$. The so-obtained codimension-two $H$-observable at $\bl$ is given by 
\begin{align} \label{eq:fourier}
\begin{split}
    \mathbf{H}^{(\a)} (Z)[\bl] &=  \sum_{X \in a_\a q_1 ^\BZ} Z^{-\frac{\log X}{\log q_1}} \mathbf{Q}(X)[\bl] \\
    &= Z^{-\frac{\log a_\a}{\log q_1}} \sum_{k=-l(\l^{(\a)} )+1} ^\infty  Z^k \, \mathbf{Q}(a_\a q_1^{-k}) [\bl].
\end{split}
\end{align}
Note that, even though there are negative powers of $Z$ appear due to the effect of the coupling, the series is still convergent in $\vert Z \vert <1$ since the summation is truncated from below. Also, note that the $H$-observable is labeled by $\a \in \{0,1,\cdots, N-1\}$, enumerating the choice of the massive  Higgs vacuum at infinity.

\subsection{Monodromy codimension-two defect}
Another way to define a codimension-two defect in the 5d $\EN=1$ gauge theory is to impose a prescribed singular behavior of the fields along the codimension-two locus. Such a singularity can be modeled by placing the 5d gauge theory on an orbifold, $ \BR^2 _{\ve_1} \times \BR^2 _{\ve_2} /  \BZ_K \times S^1$, where the generator $\zeta = \exp \frac{2\pi \ii}{K}$ of $\BZ_K$ acts by $z_2 \mapsto \zeta z_2$.

The map $z_2 \mapsto z_2 ^K$ sends the 5d $\EN=1$ gauge theory on the orbifold to the one supported on the ordinary $\BR^4 \times S^1$. The presence of the orbifold is compensated by the singularity of fields along the codimension-two locus $z_2=0$. In this sense, the orbifold gives rise to the monodromy codimension-two defect.

The monodromy codimension-two defect is characterized by the Levi subgroups of the gauge group and the flavor symmetry group it preserves. This choice is conveniently encoded in the \textit{coloring functions} $\mathbf{c} = (c,c_f,c_{af})$ assigning the $\BZ_K$-weights to the framing and the flavor bundles by
\begin{align}
    c: N \to \BZ_K,\quad c_f = M^+ \to \BZ_K ,\quad c_{af}= M^- \to \BZ_K,
\end{align}
for which the preserved Levi subgroup of the global gauge symmetry group is $\prod_{\o\in [K]} U (\vert c^{-1} (\o) \vert)$. We define $N_\o = c^{-1}(\o)$. Note that $N = \bigoplus_{\o\in [K]} N_\o$. A similar argument applies to the flavor symmetry group.

\subsubsection{Monodromy codimension-two observable}
The partition function of the 5d $\EN=1$ gauge theory placed on the $\BZ_K$-orbifold is computed as the equivariant indices of the bundle of the fermionic zero-modes over the moduli space of chain-saw quiver variety. For a given coloring function $c : N\to \BZ_K$, the defining data of the chain-saw quiver comprise of $K$ gauge nodes $\hat{K}_\o = \BC^{k_\o}$ with framing nodes $\hat{N}_\o = \BC^{\vert c^{-1} (\o) \vert} $, $\o \in [K]$. The arrows of the quiver are given by
\begin{align}
\begin{split}
    &B_{1,\o} : \hat{K}_\o \to \hat{K}_\o, \qquad B_{2,\o} : \hat{K}_{\o} \to \hat{K}_{\o+1} \\
    &I_\o : \hat{N}_\o \to \hat{K}_\o,\qquad J_\o : \hat{K}_{\o} \to \hat{N}_{\o+1},
\end{split} ,\qquad \o \in [K].
\end{align}
The chain-saw quiver variety $\mathcal{M}_{\mathbf{k}} ^{\text{chain-saw}}$ is defined by imposing the moment map equations
\begin{align} \label{eq:momentcs}
    0 = B_{1,\o} B_{2,\o-1} - B_{2,\o-1} B_{1,\o-1} + I_\o J_{\o-1} ,\qquad \o \in[K],
\end{align}
with the stability condition
\begin{align} \label{eq:stabcs}
    \bigoplus_{\o \in [K]} \hat{K}_\o = \BC[B_1,B_2] \bigoplus_{\o \in [K]} I_\o (\hat{N}_\o),
\end{align}
where $\BC[B_1,B_2]$ is the non-commutative ring generated by $B_{1,\o}$ and $B_{2,\o}$ with all possible $\o \in [K]$. Finally, we take the quotient by the gauge group $\prod_{\o \in [K]} GL(k_\o)$, acting by
\begin{align}
(g_\o)_{\o\in [K]} : (B_{1,\o}, B_{2,\o} ,I_\o,J_\o)_{\o\in [K]} \mapsto (g_\o B_{1,\o} g_{\o}^{-1} , g_{\o+1} B_{2,\o} g_\o^{-1} , g_\o I_\o, J_\o g_\o ^{-1})_{\o\in [K]}.
\end{align}

The fixed point set $\left(\mathcal{M}_{\mathbf{k}} ^{\text{chain-saw}} \right)^{\mathsf{T}}$ under the maximal torus of the symmetry group is classified by the \textit{colored} partitions $\left\{\boldsymbol\L = (\L^{(\a)})_{\a=0} ^{N-1} \, \vert \, \vert \boldsymbol\L_\o \vert = k_\o,\; \o\in [K] \right\}$, where each $\L^{(\a)}$ is a Young diagram in which the box $\Box_{(i,j)}$ at the $i$-th row and the $j$-th column carries the $\BZ_K$-weight $c(\Box_{(i,j)}) = c(\a) + j-1$. The total number of boxes with $\BZ_K$-weight $\o$ in $\boldsymbol\L$ is restricted to be $k_\o$. The $K$ vector spaces $\hat{K}_\o$, $\o\in [K]$, define the tautological bundles over the chain-saw quiver variety, whose equivariant Chern characters at fixed point $\boldsymbol\L$ are given by
\begin{align}
\hat{K}_\o [\boldsymbol\L]= \sum_{\a=0} ^{N-1} \hat{a}_\o \sum_{\substack{\Box_{(i,j)}\in \boldsymbol\L^{(\a)} \\ c(\a) +j-1 = \o}} q_1 ^{i-1} \hat{q}_2 ^{j-1}  \qquad \o \in [K].
\end{align}
The universal bundle $\hat{S} = \bigoplus_{\o \in [K]} \hat{S}_\o$ is decomposed into the direct sum of $K$ virtual bundles according to the $\BZ_K$-weights, whose equivariant Chern characters at fixed point $\boldsymbol\L$ are given by
\begin{align}
    \hat{S}_\o[\boldsymbol\L] = \hat{N}_\o - P_1 \hat{K}_\o [\boldsymbol\L] + \hat{q}_2 P_1 \hat{K}_{\o-1} [\boldsymbol\L] ,\qquad \o \in [K].
\end{align}

The partition function of the 5d $\EN=1$ gauge theory on the $\BZ_K$-orbifold is obtained as the generating function of the equivariant indices of the Dirac operator on the chain-saw quiver varieties, which can be conveniently expressed using the projection to the $\BZ_K$-invariants as
\begin{align}
\begin{split}
    \left\langle \Psi_\mathbf{c} (\mathbf{u}) \right\rangle_\ba &= \sum_{\{\boldsymbol\L\}} \prod_{\o\in [K]} \qe_\o ^{\vert \boldsymbol\L_\o \vert} \boldsymbol\m_{\boldsymbol\L} ^{\BZ_K} \\
    &= \sum_{\{\boldsymbol\L\}} \prod_{\o\in [K]} \qe_\o ^{\vert \boldsymbol\L_\o \vert} \hat{a} \left[ \frac{-\hat{S} [\boldsymbol\L] \hat{S}^* [\boldsymbol\L] +\hat{M}^+ \hat{S}^* [\boldsymbol\L] +\hat{S} [\boldsymbol\L] (\hat{M}^-)^* }{P_1^* \hat{P}_{2} ^*}  \right]^{\BZ_K},
\end{split}
\end{align}
where $[\cdots]^{\BZ_K}$ indicates the projection to the $\BZ_K$-weight zero part. Here, $\qe_\o$ is the fractionalized gauge coupling, counting the instantons carrying the $\BZ_K$-weight $\o \in [K]$. We label the codimension-two defect by their ratios,
\begin{align}
    \bu = \left( u_\o \right)_{\o=0} ^{K-1},\qquad \qe_\o = \frac{u_{\o+1}}{u_\o},\qquad \o \in [K].
\end{align}
Note that this is the K-theoretic uplift of the 4d $\EN=2$ gauge theory partition function placed on the $\BZ_K$-orbifold obtained in \cite{Kanno:2011fw}. See also \cite{Bourgine:2019phm} for a realization in the representation web of a quantum toroidal algebra.\\

The chain-saw quiver variety can be viewed as a bundle over the usual ADHM quiver variety. Explicitly, we may define the projection $\r_{\mathbf{k }} : \mathcal{M}^{\text{chain-saw}} _{\mathbf{k}}  \to \mathcal{M}^{\text{ADHM}} _{k_0}$ by
\begin{align}
\begin{split}
    &b_1 = {B}_{1,0} ,\qquad b_2 = B_{2,N-1} \cdots B_{2,1} B_{2,0} \\
    &i = \bigoplus_{\o\in[K]} B_{2,N-1} \cdots  B_{2,\o+1} B_{2,\o} I_\o \qquad j= \bigoplus_{\o\in [K]} J_\o B_{2,\o-1} \cdots B_{2,1} B_{2,0}
\end{split}
\end{align}
It is straightforward to see that \eqref{eq:momentcs} implies the ADHM moment map equation,
\begin{align}
    0 = [b_1,b_2] + ij.
\end{align}
The stability condition also follows from \eqref{eq:stabcs}as
\begin{align}
    K_0 = \BC[b_1,b_2]i(N).
\end{align}
Finally, the gauge group $GL(k_0) \subset \prod_{\o \in [K]} GL(k_\o)$ acts by
\begin{align}
    g_0: (b_1,b_2,i,j)\mapsto (g_0 b_1 g_0 ^{-1} , g_0 b_2 g_0^{-1} , g_0 i, jg_0 ^{-1}).
\end{align}
We may extend the projection to the discrete union by $\r : \coprod_{\mathbf{k}} \mathcal{M}^{\text{chain-saw}} _{\mathbf{k}} \to \coprod_{k_{0}=0}^{\infty} \mathcal{M}^{\text{ADHM}} _{k_0}$.

This projection precisely reflects the map $z_2 \to z_2 ^K$, from which we compensate the presence of the $\BZ_K$-orbifold in the spacetime by the singularity of fields along the locus $z_2 =0$. In the latter point of view, the emergent monodromy codimension-two defect defines an equivariant K-theory class over the usual ADHM moduli space, obtained by the pushforward along the projection $\r$.

The pushforward along the projection leads to the fiber-wise integral, which thus allows to rewrite the generating function of the equivariant indices as
\begin{align}
\begin{split}
     \left\langle \Psi_\mathbf{c} (\mathbf{u}) \right\rangle_\ba &= \sum_{\mathbf{k}} \prod_{\o=0} ^{K-1} \qe_\o ^{k_\o} \text{ind}_\mathsf{T} \left( {\mathcal{M}^{\text{chain-saw}} _{\mathbf{k}}} , \hat{\mathcal{E}} \right) = \sum_{\mathbf{k}} \prod_{\o=0} ^{N-1} \qe_\o ^{k_\o} \int_{\mathcal{M}^{\text{chain-saw}} _{\mathbf{k}}} \text{Ch}_{\mathsf{T}}\left( \hat{\mathcal{E}} \right) \hat{A}_{\mathsf{T}} \left( {\mathcal{M}^{\text{chain-saw}} _{\mathbf{k}}}\right) \\
     &= \sum_{k_{0}=0} ^\infty \qe^{k_{0}} \int_{\CalM_{k_0} ^{\text{ADHM}}} \hat{A}_{\mathsf{T}} \left(\CalM_{k_0} ^{\text{ADHM}}\right) \left( \sum_{k_1,\cdots, k_{K-1}=0} ^\infty \prod_{\o=1}^{K-1} \qe_\o ^{k_{\o}-k_0} \text{Ch}_{\mathsf{T}} (R\r_* \hat{\mathcal{E}}) \right),
\end{split}
\end{align}
where the gauge coupling after the projection is given by $\qe =\prod_{\o=0}^{K-1} \qe_\o$, $\hat{\mathcal{E}}$ is the bundle of zero-modes of the Dirac operator, and $R\r_*$ is the derived pushforward. 

Consequently, the monodromy codimension-two defect, labeled by the coloring functions $\mathbf{c}$ and the monodromy parameters $\mathbf{u}$, produces an observable whose expression at the fixed point $\bl \in \left(\mathcal{M}^{\text{ADHM}} _{k_{0}} \right)^{\mathsf{T}}$ is given by
\begin{align}
\Psi_\mathbf{c} (\mathbf{u}) [\bl] = \sum_{\boldsymbol\L \in \r^{-1} (\bl)} \prod_{\o \in [K]} \qe_\o ^{\vert \boldsymbol\L_\o\vert -\vert \boldsymbol\L_{K-1} \vert} \, \hat{a}\left[\sum_{0\leq\o<\o' \leq K-1}  \frac{S_\o [\boldsymbol\L] S_{\o'}  ^* [\boldsymbol\L] - M_\o ^+ S_{\o' } ^* [\boldsymbol\L] +q_1 ^{-1} S_\o [\boldsymbol\L] (M_{\o'}^-)^*}{P_1 ^*} \right]  .
\end{align}
By augmenting the perturbative part, we express the full monodromy codimension-two observable as
\begin{align}
        \boldsymbol{\Psi}_{\bc} (\bu)[\bl] = \prod_{\o\in [K]} u_\o^{\frac{\sum_{\a \in c^{-1} (\o)} \log \frac{m_\a^+}{ a_\a}}{\log q_1}} \Psi_{\mathbf{c}} (\bu)[\bl].
\end{align}

In the special case where $K=N$ and $\mathbf{c} = (\text{id},\text{id},\text{id} )$, we call the monodromy codimension-two defect to be \textit{regular}. In this work, we will restrict our attention to the regular monodromy codimension-two defect. In this case, we will simply denote its observable as $\boldsymbol{\Psi}(\mathbf{u})$.

\subsection{Parallel codimension-two defects}
We have introduced two kinds of codimension-two defects, by coupling a 3d $\EN=2$ gauge theory and by assigning a monodromy. We now consider the configuration of \textit{parallel} codimension-two defects where the two defects are placed on top of each other. On top of the codimension-two defects, we may or may not introduce codimension-four observables supported on their interface.

If there is no codimension-four observable at their interface, the correlation function can be simply computed by inserting the two observables into the localization formula. For instance, the correlation function of the $Q$-observable and the monodromy defect observable is written as
\begin{align}
    \left\langle \mathbf{Q}(X) \Psi (\mathbf{u}) \right\rangle = \sum_{\{\bl\}} \qe^{\vert \bl \vert} \mathbf{Q}(X)[\bl] \Psi (\mathbf{u}) [\bl] \boldsymbol\m_{\bl}.
\end{align}

Let us consider a set of distinguished codimension-four observables defined by the $\hat{a}$-classes of the following equivariant K-theory classes on the chain-saw quiver variety,
\begin{align}
    \EuScript{O}_\o (X) [\boldsymbol\L] = \hat{a} \left[X \sum_{ \o'=\o+1} ^{N-1}  S_\o ^* [\boldsymbol\L] \right] ,\qquad \o\in [N].
\end{align}
We denote their correlation functions by
\begin{align}
    \EQ_\o (X,\mathbf{u}) = 
\sum_{\{\boldsymbol\L\} }\prod_{\o\in [K] } \qe_\o ^{\vert \boldsymbol\L_\o \vert} Q(X)[\r(\boldsymbol\L)]   \EuScript{O}_\o (X) [\boldsymbol\L] \boldsymbol\m_{\boldsymbol\L} ^{\BZ_N},\qquad \o \in [N].
\end{align}
We will find the constraints relating these correlation functions among each other give rise to an image of the quantum affine algebra.

\section{TQ equations and $q$-opers for codimension-two defects} \label{sec:tq}
In this section, we will derive the difference equations satisfied by the vacuum expectation values of the codimension-two defects and their correlation functions with codimension-four defects.

We first give a geometric construction of $q$-connections and $q$-opers. The quantum TQ equation satisfied by the $Q$-observable or the $H$-observable can be viewed as a $q_2\neq 1$ deformation of the $q$-oper difference equation. Finally, we consider the correlation function of the monodromy codimension-two defect and the canonical codimension-two defect, showing that it satisfies the fractional TQ equation.

\subsection{Moduli space of $q$-connections and affine space of $q$-opers}
We introduce the moduli space of $q$-connections on $\BP^1$ with fixed framings at $\{0,\infty\} \subset \BP^1$ and regular singularities at $D \subset \BP^1 \setminus \{0,\infty\} = \BC^\times$, and the affine space of $q$-opers as its subspace. See \cite{Elliott:2018yqm,Koroteev:2018jht,Frenkel:2020iqq} for related works on $q$-connections and $q$-opers.

Let $q \in \BC^\times$ which is not a root of unity. Consider the shift automorphism $q^{D_z} : \BP^1 \to \BP^1$ which acts by $z \mapsto qz$, where $z \in \BC^\times$ is the holomorphic coordinate on $\BP^1 \setminus \{0,\infty\} = \BC^\times$. Given a rank $M$ holomorphic vector bundle $E \to \BP^1$, let $E^q$ be its pullback with respect to the map $q^{D_z}$. We always assume $E$ is trivializable.

Consider a map $A : E \to E^q$. Upon picking a trivialization of $E$, the map $A$ is determined by the matrix $A(z) \in GL(M, \BC(z))$ representing a linear map $E_z \to E_{qz}$ with respect to the given basis. A change in trivialization by $g(z)$ modifies the matrix $A(z)$ by
\begin{align}
    A(z) \mapsto g(qz) A(z) g(z)^{-1},
\end{align}
or equivalently
\begin{align}
    \mathds{1}_n  - A(z) q^{-D_z} \mapsto g(qz) \left( \mathds{1}_n - A(z) q^{-D_z} \right) g(qz)^{-1}.
\end{align}

Fix $D^\pm = \{m^\pm_0,m^\pm_1,\cdots, m^\pm_{N-1}\} \subset \BP^1 \setminus \{0,\infty\}$. We always assume $q^\BZ m^\pm _\a \cap q^\BZ m^\pm _\b = \varnothing$ for any choice of $\pm$ and $\a,\b\in \{0,1,\cdots, N-1\}$. The map $A: E\to E^q$ is called \textit{meromorphic q-connection} on $E$ with framings at $\{0,\infty\}\subset \BP^1$ and regular singularities at $D^\pm$, if $A(0)$ and $A(\infty)$ lie in fixed conjugacy classes in $GL(n)$ and $A(z)^{\pm 1}$ has simple poles at $D^\pm$.

Let us restrict to the case of $M=2$. In this case, we represent a $q$-connection by
\begin{align}
    A(z) = \begin{pmatrix}
    \a(z) & \b(z) \\ \g(z) & \d(z)
    \end{pmatrix},\quad A(0) \sim \begin{pmatrix}
        \qe \sqrt{\frac{m^-}{a}} & 0 \\ 0 & \sqrt{\frac{m^+}{a}}
    \end{pmatrix}, \quad A(\infty) \sim \begin{pmatrix}
        \qe \sqrt{\frac{a}{m^-}} & 0 \\ 0 & \sqrt{\frac{a}{m^+}}
    \end{pmatrix},
\end{align}
where $m^\pm = \prod_{\a=0}^{N-1} m^+_\a$ and $\qe,a \in \BC^\times$  parametrizes the choice of the framing. By making a diagonal gauge transformation, it is always possible to make the lower-left component to be $\frac{1}{P^+(z)}$, at the price of creating singularities at the zeros of $\g(z)$. Then, a \textit{$q$-oper} is defined to be the $q$-connection without any additional singularity in this gauge, so that we can set $\g(z) = \frac{1}{P^+(z)}$. By making a further triangular gauge transformation by
\begin{align}
    g(z) = \begin{pmatrix}
        1 & P^+(z) \d(z) \\ 0 & 1
    \end{pmatrix},
\end{align}
we get
\begin{align} \label{eq:gauge}
    A'(z) = \begin{pmatrix}
        \frac{t(z)}{P^+(z)} & -\qe  {P^- (z)}\\ \frac{1}{P^+(z)} & 0
    \end{pmatrix},
\end{align}
where $t(z) = P^+(z) \a(z) + P^+(qz) \d(qz) $ is, up to a factor of $z^{\frac{N}{2}}$, a degree $N$ polynomial,
\begin{align}
    z^{\frac{N}{2}} t(z) = \left( q_1 ^{\frac{N}{2}} a^{-\frac{1}{2}} + \qe m  ^{-\frac{1}{2}} a^{\frac 1 2} \right) z^N + \sum_{i=1}^{N-1} t_i z^i +  a^{\frac{1}{2}} q_1 ^{-\frac{N}{2}} + \qe m^{\frac{1}{2}} a^{-\frac 1 2}.
\end{align} 
The unconstrained $N-1$ coefficients $(t_i)_{i=1}^{N-1}$ span the affine space of $q$-opers.

In the gauge \eqref{eq:gauge}, we can write the $q$-difference equation for horizontal section as
\begin{align} \label{eq:matqoper}
    0 = \left[ \mathds{1}_2 - \begin{pmatrix}
        \frac{t(z)}{P^+(z)} & -\qe P^- (z) \\ \frac{1}{P^+(z)} & 0 
    \end{pmatrix} q^{-D_z} \right] \begin{pmatrix}
        P^+ (qz) \EQ (qz) \\ \EQ (z)
    \end{pmatrix},
\end{align}
It turns out that the lower component of the horizontal section satisfies a second-order scalar $q$-difference equation,
\begin{align} \label{eq:qoper2}
    0 = \left[ P^+(q z) - t(z) q^{-D_z} + \qe P^- (z) q^{-2D_z} \right]\EQ(qz).
\end{align}
We refer to this $q$-difference equation as the $q$-oper equation, and call $\EQ(z)$ the $q$-oper solution.

Note that the $q$-oper difference equation is precisely the Baxter TQ equation for the XXZ spin chain. 
\\

Let us consider two independent solutions $\EQ(z)$ and $\tilde{\EQ}(z)$ to the $q$-oper equation. The $q$-Wronskian of the solutions are defined by
\begin{align} \label{eq:qwrons} 
\begin{split}
    W_q (z) &= \begin{pmatrix}
       P^+ (qz)\EQ (qz) \\ \EQ (z)
    \end{pmatrix} \wedge \begin{pmatrix}
        P^+ (qz) \tilde{\EQ} (qz) \\ \tilde{\EQ} (z)
    \end{pmatrix} \\
    & =  P^+ (qz) \left( \EQ( qz) \tilde{\EQ}(z) -  \tilde{\EQ}( qz) {\EQ}(z)\right).
\end{split}
\end{align}
It follows from \eqref{eq:matqoper} that
\begin{align}
    \frac{W_q (z)}{W_q (q^{-1}z)} = \qe \frac{P^-(z)}{P^+ (z)},\qquad W_q (z) = \qe \frac{M^- (z)}{M^+ (z)},
\end{align}
where we recall $M^\pm (z) = \prod_{\a=0} ^{N-1} \left( \frac{m^\pm _\a}{z} ;q_1 \right)'_{\infty}$ so that $P^\pm(z) = \frac{M^\pm (z)}{M^\pm(z q^{-1})}$. 

Note that the $q$-Wronskian $W_q (z)$ of the $q$-oper has $N$ semi-infinite $q$-lattice of simple zeros at $z \in m^- _\a q^{\BZ_{\geq 0}}$ and $N$ semi-infinite $q$-lattice of simple poles at $z \in m^+_\a q^{\BZ_{\geq 0}}$, $\a \in \{0,1,\cdots , N-1\}$. This semi-infinite lattices of zeros and simple poles are necessary feature of the bi-infinite evaluation module of the quantum affine algebra $U_q (\widehat{\fgl}(2))$.

\subsection{TQ equations for codimension-two defects and $q$-opers}

We derive difference equations, called the quantum TQ equations, satisfied by the vacuum expectation values of the $Q$-observable and the $H$-observable. In the limit $q_2 \to 1$, the quantum TQ equations precisely reduce to the $q$-oper difference equations. 

\subsubsection{$Q$-observables}
To establish the constraints on the vev of the $Q$-observable, we consider the $qq$-characters \cite{Nikita:I} on top of the codimension-two defect. The quantum TQ equation follows from the regularity of the correlation function of the configuration \cite{Nikita:II}. These regularity constraints encode the chiral ring equations satisfied by the codimension-four observables wrapping the circle lying on the defect \cite{Jeong:2019fgx}. The $qq$-characters can be effectively engineered in the gauge origami, i.e., the configuration of intersecting stacks of D3-branes in IIB string theory. \cite{Nikita:III}.

To be specific, we consider the IIB string theory on the 10-dimensional worldvolume $X \times \BC^\times$, where $X $ is a local Calabi-Yau fourfold that we choose at our need. For the construction of the quantum TQ equation, we choose $X = \BC^4/ \G_{34}$, where $\G_{34} = \BZ_3$ acts by the weight $(0,0,+1,-1)$ on the holomorphic coordinates. We can implement the $\O$-background, with the associated exponentiated $\O$-background parameters $q_i$, $i=1,2,3,4$, associated with the rotations. They are constrained by $q_1 q_2 q_3 q_4 =1$ to keep a supersymmetry.

We consider three stacks of D3-branes: $n_{12,0}=N$ regular D3-branes of wrapping on $\BC^2 _{12}$, one fractional D3-brane of $\BZ_3$-weight 1 wrapped on $ \BC^2_{13}$, and one fractional D3-brane of $\BZ_3$-weight $0$ wrapped on $ \BC^2_{34}$. The moduli space of D$(-1)$-branes on these stacks gives the moduli space of spiked instantons, on which the framing bundles for the D3-branes are defined. The equivariant Chern characters of these bundles are written as
\begin{align}\label{eq:gosetup}
\begin{split}
    & {\bN}_{12} = \sum_{\alpha=0}^{N-1} {a_\alpha} \CalR_0 + {m_\alpha^-q_1q_2q_3} \CalR_1 + {m^+_\alpha q_3^{-1}} \CalR_2, \\
    & {\bN}_{13} = X' q_1 q_3 \CalR_1, \\
    & {\bN}_{34} = X \CalR_0,
\end{split}
\end{align}
where $\CalR_i$, $i=0,1,2$, is the one-dimensional representation of $\BZ_3$ with weight $i$.

There are three gauge couplings $\kq_{0,1,2}$ counting the spiked instantons carrying the $\BZ_3$-weight $0$, $1$, and $2$. We take the decoupling limit $\kq_{1} = \kq_2 = 0$ and denote $\kq = \kq_0$, to focus on the $A_1$-quiver $U(N)$ gauge theory. As pointed out in \cite{Jeong:2023qdr}, the decoupling limit and the choice of the branes in \eqref{eq:gosetup} give us reduced 13-34 cross terms in the partition function, yielding
\begin{align}
\begin{split}
    \CalZ_\text{GO} & = \sum_{\boldsymbol{\lambda}} \kq^{|\boldsymbol{\lambda}|} \boldsymbol\m_{\boldsymbol\lambda} \left[ \left( \sqrt{\frac{X}{X'}} - \sqrt{\frac{X'}{X}} \right) {\EY}(Xq_{12}) + \kq \left( \sqrt{\frac{Xq_2}{X'}} - \sqrt{\frac{X'}{Xq_2}} \right) \frac{P(X)}{{\EY}(X)} \right] Q(X') \\
    & = \left\langle T(X,X')Q(X') \right\rangle_\ba.
\end{split}
\end{align}
Here, the $Q(X)$ and ${\EY}(X)$ are observables defined by
\begin{align}
\begin{split}
    & {\EY}(X) = \hat{a}[-X \bS_{12}^*]~,~  Q(X) = \hat{a} \left[ -X \frac{\bS_{12}^*}{P_1^*} \right]~,~\frac{{Q}(X)}{{Q}(Xq_1^{-1})} = {\EY}(X)~,~ \\
    & P^\pm(X) = \hat{a}[-X(M^\pm)^*]~,~ P(X) = P^+(X)P^-(X)~.~
\end{split}
\end{align}

By the compactness of the moduli space of spiked instantons \cite{Nikita:II}, the expectation value of the left hand side
$\left \langle T(X,X') {Q}(X') \right \rangle$
is singular only at $X = 0,\infty$; namely, it is a Laurent polynomial in $X$. Now, we specialize to two loci,
\begin{subequations}
\begin{align}
    & T(X=X',X') {Q}(X') = \kq  (q_2^{\frac12}-q_2^{-\frac12}) P(X') \frac{{Q}(X')}{{Y}(X')} = \kq (q_2^{\frac12}-q_2^{-\frac12}) P(X'){Q}(X'q_1^{-1}) ,\\
    & T(X=X'q_2^{-1},X') {Q}(X') = - (q_1^{\frac12}-q_1^{-\frac12}) {Y}(X'q_1) {Q}(X') = -(q_2^{\frac12}-q_2^{-\frac12}) {Q}(X'q_1).
\end{align}
\end{subequations}
By subtracting the two equations from each other, we arrive at
\begin{align}\label{eq:TQ-bulk}
   0 = \left\langle P^{+}(q_1 X)  \mathbf{Q}(q_1 X) - t( X) \mathbf{Q}(X)    + \kq P^{-}(X) \mathbf{Q}(q_1 ^{-1} X) \right\rangle,
\end{align}
where we used 
\begin{align}
    \mathbf{Q}(X) = \frac{Q(X)}{M^+(X)}, \qquad \frac{M^+(X)}{M^+(Xq_1^{-1})} = P^+(X).
\end{align}
Here, we defined
\begin{align}
    t( X') = \frac{T(X=X',X') - T(X=X'q_2^{-1},X') }{q_2^{\frac12}-q_2^{-\frac12}},
\end{align}
which is a Laurent polynomial in $X$ by construction. It is immediate from its definition that $X^{\frac{N}{2}}  t(X)$ is a degree-$N$ polynomial in $X$,
\begin{align}
    X^{\frac{N}{2} }  t(X) =  \left( a^{-\frac12}q_1^{\frac{N}{2}} + \kq m^{-\frac12} a^{\frac12}\right) X^{N} + \sum_{i=1}^{N-1} t_i X^{N-1} +  a^{\frac12}q_1^{-\frac{N}{2}} + \kq m^{\frac12} a^{-\frac12}.  
\end{align}
Note that the first and last coefficients are given by simple combinations of the Coulomb parameters and the mass parameters. The rest of $N-1$ coefficients $(t_i)_{i=1}^{N-1}$ generate the space of independent codimension-four observables in the 5d $\EN=1$ gauge theory.

Thus, we arrive at the $q_1$-difference equation \eqref{eq:TQ-bulk} as a constraint of the vevs of $Q$-observable and its correlation function with the generating function $t(X)$. We call this equation the quantum TQ equation. \\

Importantly, the quantum TQ equation involves a non-trivial contact term in the correlation function of the two observables $t(X)$ and $\bQ(X)$. When we take the limit $q_2 \to 1$ of the $\O$-background, the topological symmetry along the $\BR^2_{\ve_2}$-plane is restored. Thus, the $Q$-observable $\mathbf{Q}(X)$ and the generating function $t(X)$ of the codimension-four observables, both of which are local on this plane, can be arbitrarily separated from each other. Such a cluster decomposition yields the factorization of the correlation function of the two,
\begin{align} \label{eq:factori}
    \lim_{q_2\to 1}  \langle t( X) \mathbf{Q}(X) \rangle_\ba = e^{\frac{\widetilde{\EuScript{W}}(\ba)}{\log q_2}} t(\ba;X) \mathbf{Q}(\ba;X),
\end{align}
where $\mathbf{Q}(\ba;X)$ and $t(\ba;X)$ denote the vevs of the two observables at the vacuum specified by the Coulomb moduli $\ba$, respectively. Therefore, the quantum TQ equation \eqref{eq:TQ-bulk} reduces to the second-order $q_1$-difference equation satisfied by the vev $\mathbf{Q}(\ba;X)$ of the $Q$-observable, 
\begin{align}\label{eq:gaugeqoper}
    0 = \left[ P^+ ( q_1 X) - t(\ba;X) q_1 ^{- D_X} +\qe P^- (X ) q_1 ^{-2 D_X} \right] \mathbf{Q}(\ba; q_1 X).
\end{align}
Note that this is precisely the $q$-oper difference equation \eqref{eq:qoper2}. In this sense, the quantum TQ equation \eqref{eq:TQ-bulk} can be regarded as a $q_2\neq 1$ deformation of the $q$-oper difference equation. \\

As a second-order $q$-difference equation, the $q$-oper equation \eqref{eq:gaugeqoper} admits two independent solutions. Another solution is provided by nothing but the $q_2\to 1$ limit of the vev of the dual $Q$-observable $\tilde{\bQ}(X)$, as we show in appendix \ref{subsec:dualtq}. In more general case of the $A_{M-1}$-quiver $U(N)$ gauge theory, there are $M$ inequivalent $Q$-observables ($M=2$ here). They provide the solutions to the $M$-th order $q$-oper difference equation (see \cite{Jeong:2023qdr} for the 4d $\EN=2$ gauge theoretical construction of the $\hbar$-oper solutions. See \cite{Jeong:2017pai,Nikita:IV,Poghosyan:2016mkh} for related earlier works).

\subsubsection{$H$-observables}\label{subsubsec:hobs}
The correlation function of the codimension-to $H$-observable and the codimension-four observables is constrained in a similar way. This constraint can be derived easily by the Fourier transformation \eqref{eq:fourier} relating the $Q$-observable and the $H$-observable. Namely, applying this Fourier transformation to the quantum TQ equation \eqref{eq:TQ-bulk}, we obtain
\begin{align} 
   0 = \left[ Z P^+ \left( q_1 ^{-D_Z}  \right)     + \frac{\qe}{Z} P^- \left(q_1 ^{-D_Z +1} \right) - t\left( q_1 ^{-D_Z}  \right) \right] \bH^{(\a)}  (Z) , 
\end{align}
for each $\a \in \{0,1,\cdots, N-1\}$.

Note that there is a non-trivial contact term between the coefficients of $ t\left( q_1 ^{-D_Z}  \right) $ and $\bH^{(\a)}  (Z)$. In the limit $q_2 \to 0$, the correlation function factorizes into the product of the vevs of each, just as in \eqref{eq:factori} by the cluster decomposition. Thus, we obtain the $N$-th order $q_1$-difference equation satisfied by the vev of the $H$-observable. This precisely gives a gauge theoretical construction of the $GL(N)$ $q$-opers with framing at $\{0,\infty\} \subset \BC^\times _Z$ and regular singularities at $\{\qe,1\} \subset \BC^\times _Z$.\footnote{This $q$-oper difference equation is precisely the Baxter TQ equation for the $\fgl(N)$ XXZ spin chain. Thus, the $H$-observable also provides a $Q$-operator in this context, and the Fourier transformation \eqref{eq:fourier} relates the $Q$-operators of the two XXZ spin chains which are bispectral dual to each other. However, we call it $H$-observable to distinguish it from the $Q$-observable, which are realized by the Higgs and the Coulomb phase of the canonical codimension-two defect respectively.}

In this way, we establish a connection between the $GL(2)$ $q$-opers and the $GL(N)$ $q$-opers, realized through the Fourier transformation between the $Q$-observable and the $H$-observable whose vevs provide the $q$-oper solutions. This is precisely the $q$-deformation of the transition between the $GL(2)$ $\hbar$-opers on and the $GL(N)$ opers established in \cite{Jeong:2023qdr,Jeong:2024onv}. In our gauge theoretical construction, such a $q$-deformation is realized by uplifting to the 5d $\EN=1$ gauge theory compactified on a circle.

\subsection{Fractional TQ equation for parallel codimension-two defects}
Finally, we derive the constraints, called the fractional TQ equations, for the correlation function of monodromy codimension-two defect and the canonical codimension-two defect parallel to each other. These constraints follow from the regularity of the $qq$-characters implemented on the parallel defect configuration.

We replace the worldvolume of the IIB string theory by $X = \BC^4 /(\G_{24}\times \G_{34}) $, where the generator of $\Gamma_{24} = \BZ_N$ acts by the weight $(0,+1,0,-1)$ on top of the action of $\G_{34}=\BZ_3$ that we introduced earlier. For general coloring function $c: [N] \to \{0,\dots,N-1\}$, we denote
\begin{align}
    \hat{a}_{\o=c(\alpha)}.
\end{align}
We also define the shifted parameters $a_{\o} = \hat{a}_\o \hat{q}_2^{-\o}$ such that $a_\o$ carries weight $0$ under the action of $\BZ_N$. 

The $\Omega$-background parameters works similarly by $\hat{q}_1=q_1$, $\hat{q}_3=q_3$, $\hat{q}_4=q_1^{-1}\hat{q}_2^{-1}q_3^{-1}$. The gauge origami data can be expressed in terms of the shifted parameters as
\begin{align}
\begin{split}
    \bN_{12} & = \sum_{\o'=0}^{N-1} {a}_{\o'} \hat{q}_2^{\o'} \CalR_0 \otimes \fR_{\o'} + {m}^-_{[\o'-1]}q_1q_3 \hat{q}_2^{\o'} \CalR_1 \otimes \fR_{\o'} + {m}^+_{\o'}q_1^{-1} \hat{q}_2^{\o'} \CalR_2 \otimes \fR_{\o'}, \\
    {\bN}_{13} & = \sum_{\o'=0}^{N-1} X'_{\o'}q_1q_3 \hat{q}_2^{\o'} \CalR_1 \otimes \fR_{\o'}, \\
    {\bN}_{34} & = X \hat{q}_2^{\o} \CalR_0 \otimes \fR_{\o},
\end{split}
\end{align}
where $\fR_\o$ is the one-dimensional representation of $\BZ_N$ carrying weight $\o \in [N]$.

The gauge origami partition function is obtained by picking up the invariant contribution under the action of $\Gamma=\BZ_3\times \BZ_N$. With straightforward but tedious computation, we find the gauge origami partition function can be organized in to the following form 
\begin{align}
\begin{split}
    \hat\CalZ_\text{GO} & = \sum_{\hat{\boldsymbol\lambda}} \prod_{\o=0}^{N-1} \hat{\kq}_\o^{k_\o} \hat\CalZ_\text{defect}[\hat{\boldsymbol\lambda}] T_\o (X,\bX')[\hat{\boldsymbol\lambda}] {\EQ}(\bX')[\hat{\boldsymbol\lambda}] \\
    & = \left \langle T_\o(X,\bX') {\EQ}(\bX') \right \rangle_{\BZ_N} \hat\CalZ_{\hat\BC^2_{12}}
\end{split}
\end{align}
with $\bX'=\{X'_0,\dots,X_{N-1}'\}$. 
The fractional coupling $\hat\kq_\o$ obeys
\begin{align}
    \kq = \prod_{\o=0}^{N-1} \hat\kq_\o, \quad \hat\kq_{\o} = \frac{u_{\o+1}}{u_\o}, \quad u_{\o+N} = \kq u_\o.
\end{align}
The fractional $qq$-character, which is a correlation function of the gauge theory observables, consists both the fractional $Q$- and ${\EY}$-observables: 
\begin{align}
\begin{split}
    T_\o (X,\bX') {\EQ}_\o(\bX') = & \ \left( \sqrt{\frac{X}{X'_\o}} - \sqrt{\frac{X'_\o}{X}} \right) {\EY}_{[\o+1]}(Xq_1q_2^{\d_{\o,N-1}}) {\EQ}(\bX') \\
    & + \hat\kq_\o P_\o(X) \left( \sqrt{\frac{Xq_2^{\d_{\o,N-1}}}{X_{[\o+1]}'}} - \sqrt{\frac{X'_{[\o+1]}}{Xq_2^{\d_{\o,N-1}}}} \right) \frac{{\EQ}(\bX')}{{\EY}_\o(X)},
\end{split}
\end{align}
where we defined
\begin{align}
    {\EQ}(\bX') = \prod_{\o=0}^{N-1} Q_{\o}(X'_\o). 
\end{align}

We consider two specializations of the parameters:
\begin{subequations}
\begin{align}
    T_\o(X=X'_\o,\bX'){\EQ}(\bX') & = \hat\kq_\o P_\o(x'_\o) \left( \sqrt{\frac{X'_\o q_2^{\d_{\o,N-1}}}{X'_{[\o+1]}}} - \sqrt{\frac{X'_{[\o+1]}}{X'_\o q_2^{\d_{\o,N-1}}}} \right) \frac{{\EQ}(\rx')}{{Y}_{\o}(x'_\o)} \\
    & = \hat\kq_\o P_\o(X'_\o) \left( \sqrt{\frac{X'_\o q_2^{\d_{\o,N-1}}}{X'_{[\o+1]}}} - \sqrt{\frac{X'_{[\o+1]}}{X'_\o q_2^{\d_{\o,N-1}}}} \right) {\EQ}(\bX'q_1^{-e_\o}) \nonumber\\
    T_\o(X=X'_{[\o+1]}q_2^{-\d_{\o,N-1}},\bX') {\EQ}(\bX')
    & = - \left( \sqrt{\frac{X'_\o q_2^{\d_{\o,N-1}}}{X'_{[\o+1]}}} - \sqrt{\frac{X'_{[\o+1]}}{X'_\o q_2^{\d_{\o,N-1}}}} \right) {Y}_{[\o+1]}(X'_{[\o+1]}) {\EQ} (\bX'), \\
    & = - \left( \sqrt{\frac{X'_\o q_2^{\d_{\o,N-1}}}{X'_{[\o+1]}}} - \sqrt{\frac{X'_{[\o+1]}}{X'_\o q_2^{\d_{\o,N-1}}}} \right) {\EQ} (\bX'q_1^{e_{[\o+1]}}) .\nonumber
\end{align}
\end{subequations}
By subtracting the two equations from each other, we arrive at 
\begin{align}\label{eq:frac-T-Q}
\begin{split}
    &  \ET_{\o}(\bX) \langle \EQ(\bX) \rangle_{\BZ_N} \boldsymbol\Psi = \langle \EQ(\bX q_1^{\d_{[\o+1]}}) \rangle_{\BZ_N} \boldsymbol\Psi
    + \kq_\o P_\o(\bX_\o) \langle \EQ(\bX q_1^{-\d_\o}) \rangle_{\BZ_N} \boldsymbol\Psi,
\end{split}
\end{align}
where we defined $\d_\o = (\d_{\o,\o'})_{\o'=0}^{N-1}$ and
\begin{align} \label{def:ct}
    \langle \ET_\o(\bX') {\EQ}(\bX') \rangle_{\BZ_N} = \left \langle \frac{T_\o(X=X'_\o,\bX'){\EQ}(\bX') - T_\o(X=X'_{[\o+1]}q_2^{-\d_{\o,N-1}},\bX'){\EQ}(\bX')}{(X'_\o)^{\frac12}(X'_{[\o+1]})^{-\frac12}q_2^{\d_{\o,N-1}}-(X'_\o)^{-\frac12}(X'_{[\o+1]})^{\frac12}q_2^{-\d_{\o,N-1}}} \right \rangle_{\BZ_N}.
\end{align}
We call this equation the fractional TQ equation.

To calculate $\ET_\o(X)$, we write down the explicit form of $T_\o(X,\bX')$ in ${\EY}_\o(X)$:
\begin{align}\label{eq:T-Q-component}
\begin{split}
    & T_\o (X,\bX') \EQ(\bX') \\
    & = \left\{ \left( X (X'_\o)^{-\frac12} - (X'_\o)^{\frac12} \right) \times   (a_{[\o+1]}q_1^{-1}q_2^{-\d_{\o,N-1}})^{\frac12} \left( 1 - \frac{a_{[\o+1]}q_1^{-1}q_2^{-\d_{\o,N-1}}}{X} \right) \right.  \\
    & \qquad \times \prod_{\Box \in \EK_{\o+1}} q_1^{-\frac{1}{2}} \frac{1-\frac{{c_\Box}}{X}}{ 1-\frac{{c_\Box q_1^{-1}}}{X} } \prod_{\Box \in \EK_{\o}} q_1^{\frac{1}{2}} \frac{1- \frac{c_\Box q_1^{-1}}{X} q_2^{\delta_{N-1,\o}} } {1-\frac{{c_\Box}}{X}q_2^{\delta_{N-1,\o}} }   \\
    & \quad - \kq_\o \left( X(X'_{[\o+1]}q_2^{-\d_{\o,N-1}})^{-\frac12}  - (X'_{[\o+1]}q_2^{-\d_{\o,N-1}})^{\frac12} \right) \times \left( \frac{X}{\sqrt{m_\o^+m_\o^-}}  - \sqrt{\frac{m_\o^+}{m_\o^-}} - \sqrt{\frac{m_\o^-}{m_\o^+}} + \frac{\sqrt{m_\o^+m_\o^-}}{X} \right) \\
    & \qquad \left. \times \left[  \frac{ a_\o^{-\frac{1}{2}} }{1 - \frac{X}{{a_\o}}} \prod_{\Box\in \EK_{\o}} q_1^{\frac{1}{2}} \frac{{\frac{{c_\Box}}{X}} - 1}{ \frac{{c_\Box}q_1 }{X} - 1 } \prod_{\Box \in \EK_{\o-1}} q_1^{-\frac{1}{2}} \frac{\frac{c_\Box q_1}{X}q_2^{\delta_{0,\o}}-1}{\frac{{c_\Box}}{X}q_2^{\delta_{0,\o}}-1} \right] \right\} \EQ(\bX').
\end{split}
\end{align}
Since $\langle T_\o(X,\bX') \EQ(\bX') \rangle$ is holomorphic in $X \in \BC^\times$, it can be singular only at $X=0$ and $X =\infty$. We can simply determine its singular behavior by taking the limit $X \to 0$ and $X \to \infty $ of \eqref{eq:T-Q-component}. As a result, we obtain the Laurent polynomial, 
\begin{align}\label{eq:T-w-vev}
    \langle T_\o(X,\bX') \EQ(\bX') \rangle = \left\langle \left[ \spadesuit_\o X + \diamondsuit_\o + \frac{\clubsuit_\o}{X} \right] \EQ_\o(\bX') \right \rangle. 
\end{align}
The coefficients $\spadesuit_\o$, $\diamondsuit_\o$, and $\clubsuit_\o$ depend on $\bX'$. In particular, the coefficients of $X$ and $X^{-1}$ can be easily obtained as
\begin{align}
\begin{split}
    & \spadesuit_\o = (X'_\o q_1^{-1})^{-\frac12} a_{[\o+1]}^{-\frac12} q_1^{\frac{k_\o-k_{\o+1}}{2}} q_2^{\d_{\o,N-1}} + \hat\kq_\o (X'_{[\o+1]})^{-\frac12} a_\o^{-\frac12} (m_\o^+m_\o^-)^{-\frac12} q_2^{\d_{\o,N-1}} q_1^{-\frac{k_{\o-1}-k_\o}{2}}, \\
    & \clubsuit_\o =  (X'_\o q_1^{-1})^{\frac12} a_{[\o+1]}^{\frac12} q_1^{- \frac{k_\o-k_{\o+1}}{2}} q_2^{-\d_{\o,N-1}} + \hat\kq_\o (X'_{[\o+1]})^{\frac12} a_\o^{\frac12} (m_\o^+m_\o^-)^{\frac12} q_2^{-\d_{\o,N-1}} q_1^{\frac{k_{\o-1}-k_\o}{2}} .
\end{split}
\end{align}
The coefficient of $X^0$ can be computed by expanding around either large or small $X$. We will not need this in our work and do not write its explicit expression here.

Substituting this into \eqref{def:ct}, we obtain
\begin{align}
\begin{split}
    \langle \ET_\o(\bX')\EQ(\bX') \rangle
    & = \left\langle \frac{  \spadesuit_\o X'_\o + {\clubsuit_\o} (X'_\o)^{-1} -  \spadesuit_\o X'_{[\o+1]}q_2^{-\d_{\o,N-1}} + {\clubsuit_\o} (X'_{[\o+1]})^{-1} q_2^{\d_{\o,N-1}}    } {(X'_\o)^{\frac12}(X'_{[\o+1]})^{-\frac12}q_2^{\d_{\o,N-1}}-(X'_\o)^{-\frac12}(X'_{[\o+1]})^{\frac12}q_2^{-\d_{\o,N-1}}} \EQ(\bX')\right\rangle \\
    & = \left\langle \left[ (X'_{[\o+1]} a_{[\o+1]}^{-1} )^{\frac12} q_1^{\frac12(D_{u_{[\o+1]}}+1)} - (X'_{[\o+1]} a_{[\o+1]}^{-1} )^{-\frac12} q_1^{-\frac12(D_{u_{[\o+1]}}+1)} \right. \right. \\
    & \left. \left. \qquad + \kq_\o \left( (X'_\o a_\o (m_\o^+m_\o^-)^{-1})^{\frac12} q_1^{-\frac12 D_{u_\o}}  - (X'_\o a_\o (m_\o^+m_\o^-)^{-1})^{-\frac12} q_1^{\frac12 D_{u_\o}} \right) \right] \EQ(\bX') \right\rangle
\end{split}
\end{align}
The final result for the $\ET_\o(\bX')$ function as an operator is 
\begin{align}\label{eq:frac-TQ}
\begin{split}
     \ET_\o(\bX') & = (X'_{[\o+1]} (m_{[\o+1]}^+)^{-1} )^{\frac12} q_1^{\frac12(D_{u_{[\o+1]}}+1)} - (X'_{[\o+1]} (m_{[\o+1]}^+)^{-1} )^{-\frac12} q_1^{-\frac12(D_{u_{[\o+1]}}+1)}  \\
    & \qquad + \kq_\o \left[ (X'_\o (m_\o^-)^{-1})^{\frac12} q_1^{-\frac12 D_{u_\o}} - (X'_\o (m_\o^-)^{-1})^{-\frac12} q_1^{\frac12 D_{u_\o}} \right] 
\end{split}
\end{align}
acting on $\langle \EQ(\bX') \rangle \Psi$.

\section{R-matrices from parallel codimension-two defects} \label{sec:rmat}
We recast the constraints on the correlation functions of the codimension-two and codimension-four defects into the R-matrices of the quantum affine algebra intertwining evaluation modules.

In particular, the fractional TQ equations yield the R-matrices of both the quantum affine algebra of $\fgl(2)$ and the quantum affine algebra of $\fgl(N)$, reflecting the bispectral duality between the associated XXZ spin chains.\footnote{See \cite{Gaiotto:2013bwa} for the bispectral duality between XXZ spin chains involving finite-dimensional or highest-weight Verma modules. Our 5d $\EN=1$ gauge theoretical construction extends the bispectral duality to XXZ spin chains involving bi-infinite modules. See also \cite{Mironov:2013xva} for a related work.} We will see that the relevant evaluation modules are bi-infinite, i.e., neither highest-weight nor lowest-weight. Note that this is the multiplicative uplift of the bispectral duality between Gaudin model and XXX spin chain involving bi-infinite evaluation modules over the affine Kac-Moody algebra $U(\widehat{\fgl}(N))$ and the Yangian $Y(\fgl(M))$, established in \cite{Jeong:2024onv}. Upon establishing the gauge theoretical construction of the R-matrices, we find the monodromy codimension-two defect provides the common eigenstates of such bi-infinite XXZ spin chain system by its vevs.

\subsection{R-matrices of quantum affine algebra of $\fgl(2)$} \label{subsec:rgl2}
We will show how the fractional TQ equation \eqref{eq:frac-T-Q} leads to the R-matrices of the quantum affine algebra of $\fgl(2)$. By concatenating them we find the operator-valued $q$-oper equation satisfied by the correlation function of the $Q$-observable and the monodromy codimension-two defect. It follows that the eigenstates of the $Q$-operator, given by the vevs of the monodromy codimension-two defect, are also the common eigenstates of the quantum Hamiltonians of the associated $\fgl(2)$ XXZ spin chain.

\subsubsection{R-matrices from fractional TQ equations}
For the fractional TQ equation \eqref{eq:frac-T-Q}, let us take $\bX'=(X'_0,\dots,X_{N-1}') \in \left(\BC^\times\right)^N$ as follows:
\begin{align}\label{def:x'}
    {X}'_{\o'} = \begin{cases} X' & \text{if} \quad 0 \leq \o' \leq \o, \\ X' q_1^{-1} &\text{if} \quad \o < \o' \leq N-1 \end{cases}. 
\end{align}
Then, we define
\begin{align}
    \EQ_\o(X) = \prod_{\o'=0}^{\o} {Q_\o(X)} \prod_{\o'=\o+1}^{N-1} {Q_\o(Xq_1^{-1})}, \qquad \o \in [N].
\end{align}
Note that these $Q$-observables are defined with the periodicity $\EQ_{\o+N} (X ) = \EQ_\o (X q_1)$. The fractional TQ equations become
\begin{align}\label{eq:frac-TQ-x'}
    \ET_\o(X) \langle \EQ_\o(X) \rangle = \langle \EQ_{\o+1}(X) \rangle + \kq_{\o} P_\o(X) \langle \EQ_{\o-1}(X) \rangle 
\end{align}

To make the R-matrices of the quantum affine algebra $U_q (\widehat{\fgl}(2))$ appear explicitly in the standard form, let us take the $2 \times 1$ vector,
\begin{align}\label{eq:theta}
    \Theta_\o(X) = \G_\o (X) \begin{pmatrix}
        \langle \EQ_\o(X) \rangle \\ \langle \ER_\o(X) \rangle
    \end{pmatrix} =: \begin{pmatrix}
         \langle \boldsymbol{\EQ}_\o(X) \rangle \\ \langle \boldsymbol{\ER}_\o(X) \rangle
    \end{pmatrix}, \qquad \o \in [N],
\end{align}
where we defined 
\begin{align}
    \Gamma_\o(X) = \prod_{\o'=0}^\o \frac{1}{M_{\o'}^+(X)} \prod_{\o'=\o+1}^{N-1} \frac{1}{M^+_{\o'}(Xq_1^{-1})},
\end{align}
so that $ P^+_\o(X) = \frac{\G_{\o-1}(X)}{\G_\o(X)}$ and $\G_{\o+N} (X) = \G _\o (X q_1)$. Also, we defined
\begin{align}\label{def:ER}
    \langle \ER_{\o}(X) \rangle = \frac{1}{u_{[\o+1]}}\left( \frac{X}{{m_{\o+1}^+}}q_1^{\frac{1}{2}D_{u_{[\o+1]}}} - q_1^{-\frac12D_{u_{[\o+1]}}} \right) \langle \EQ_{\o}(x) \rangle -\sqrt{\frac{X}{m^+ _{\o+1}}} \frac{1}{u_{[\o+1]}} \langle \EQ_{\o+1}(X) \rangle.
\end{align}
Note that $\ER_{\o+N} (X) = \ER_{\o} (X q_1)$ so that $\Theta_{\o+N} (X) = \Theta_\o (Xq_1)$. Now, the fractional TQ equations \eqref{eq:frac-TQ-x'} imply the following $2 \times 1$ matrix-valued equations:
\begin{align} \label{eq:lax}
 \left(1- \frac{m^+_\o}{X} \right)  \Theta_\o (X) = \begin{pmatrix}
    1 & 0 \\ 0 & \qe^{\d_{\o,N-1}} \sqrt{\frac{m^+_\o}{m^+_{\o+1}}}
 \end{pmatrix} L_\o (X)  \Theta_{\o-1} (X), \qquad \o \in [N],
\end{align}
where 
\begin{align} \label{eq:r2gauge}
    L_\o (X) = L^+ _\o - \frac{m^+ _\o}{X} L^- _\o
\end{align} 
is a $2 \times 2$ matrix with the entries given by $q_1$-difference operators in the monodromy parameters $(u_\o)_{\o=0}^{N-1}$,
\begin{align}
\begin{split}
    &L^+ _\o = \begin{pmatrix}
        q_1 ^{\frac{1}{2} D_{u_\o}} & 0 \\ u_\o ^{-1} \left( q_1 ^{\frac{1}{2} D_{u_\o} } - q_1 ^{-\frac{1}{2} D_{u_\o}}  \right) \left(q_1 ^{\frac{1}{2} D_{u_\o}  } \sqrt{\frac{m^- _\o}{m^+ _\o}} - q_1 ^{- \frac{1}{2} D_{u_\o}}\sqrt{\frac{m^+ _\o}{m^- _\o}}  \right) & \sqrt{\frac{m^+ _\o}{m^- _\o}} q_1 ^{-\frac{1}{2} D_{u_\o} -\frac{1}{2}}
    \end{pmatrix}, \\
    &L^- _\o = \begin{pmatrix}
        q_1 ^{-\frac{1}{2} D_{u_\o}} & u_\o \\ 0 & \sqrt{\frac{m^- _\o}{m^+ _\o}} q_1 ^{\frac{1}{2} D_{u_\o} +\frac{1}{2}}
    \end{pmatrix}.
\end{split}
\end{align}

Now, we show that $L_\o (X) $ is precisely an R-matrix of the quantum affine algebra $U_q( \widehat{\fgl}(2))$. Let us consider the spaces $\CalH_\o$ of Laurent polynomials,
\begin{align}
    \CalH_\o = u_\o ^{\frac{\log \frac{m^+ _\o}{a_\o}}{\log q_1}} \BC(\!(u_\o)\!),\qquad \o \in [N].
\end{align}
The $q_1$-difference operators in the variable $u_\o$ naturally act on the space $\CalH_\o$, endowing it with a $q_1$-difference module structure.

Then, we define a map $\r_{\o} : U_q (\fgl(2)) \to \text{End}(\CalH_\o)$ by (see appendix \ref{subsubsec:qea} for the presentation of the quantized enveloping algebra $U_q (\fgl(2))$)
\begin{align}
    T^\pm \mapsto L^\pm _\o.
\end{align}
In terms of the Chevalley generators, the map reads
\begin{align}
\begin{split}
    &t_1 \mapsto q_1 ^{\frac{1}{2}D_{u_\o}},\qquad t_2 \mapsto \sqrt{\frac{m^+ _\o}{m^- _\o}} q_1 ^{-\frac{1}{2} D_{u_\o} -\frac{1}{2}},\\
    &e \mapsto -\frac{u_\o q_1 ^{\frac{1}{2} D_{u_\o}}}{q_1 ^{\frac 1 2} - q_1 ^{-\frac 1 2}},\qquad f\mapsto \frac{q_1 ^{-\frac{1}{2} D_{u_\o} } u_\o ^{-1}}{q_1 ^{\frac 1 2} - q_1 ^{-\frac 1 2}} \left( q_1 ^{\frac 1 2 D_{u_\o}} - q_1 ^{-\frac 1 2 D_{u_\o}} \right)\left( \sqrt{\frac{m^- _\o}{m^+ _\o}} q_1 ^{\frac 1 2 D_{u_\o}} - \sqrt{\frac{m^+ _\o}{m^- _\o}} q_1 ^{-\frac 1 2 D_{u_\o}} \right).
\end{split}
\end{align}
It is straightforward to check that this map is an algebra homomorphism, provided that the quantum parameter $q$ is identified with the $\O$-background parameter as
\begin{align} \label{eq:identhbar}
    q = q_1 ^{\frac{1}{2}}.
\end{align}

Thus, each space $\CalH_\o$ is endowed with a $U_q (\fgl(2))$-module structure, which is irreducible (for generic values of the mass parameters) and bi-infinite (i.e, neither highest-weight nor lowest-weight). Note that the quantum Casimirs are mapped to the mass parameters as
\begin{align}
    C^{(1)} _q = t_1t_2 \mapsto \sqrt{\frac{m^+ _\o}{m^- _\o q_1}} ,\qquad C^{(2)} _q = e f + \frac{q^{-1} t_1 t_2 ^{-1} + q t_1 ^{-1} t_2}{(q-q^{-1})^2} \mapsto \frac{\sqrt{\frac{m^+ _\o}{m^- _\o}} + \sqrt{\frac{m^- _\o}{m^+ _\o}}}{\left( q_1 ^{\frac{1}{2}} - q_1 ^{-\frac{1}{2}} \right)^2}.
\end{align}
Note that both quantum Casimirs are determined by a single combination of mass parameters, $m^- _\o/m^+_\o$, for each $\CalH_\o$.

Finally, we compose the map $\r_\o$ with the evaluation homomorphism $\text{ev}_{m^+_\o} : U_q (\widehat{\fgl}(2)) \to U_q (\fgl(2))$ with the evaluation parameter given by the mass $m^+ _\o \in \BC^\times$, to view $\CalH_\o$ as an evaluation module over the quantum affine algebra $U_q (\widehat{\fgl}(2))$. In this view, $L_\o (X) $, which is valued in $\text{End}(\BC^2) \, {\otimes} \, \text{End}(\CalH_\o)$, is identified with the R-matrix of the quantum affine algebra $U_q (\widehat{\fgl}(2))$ intertwining the representations assigned to the two spaces.\footnote{In \cite{Haouzi:2024qyo}, it was shown that the Miura operators of $q$-deformed $W$- and $Y$-algebras are R-matrices of quantum toroidal algebra of $\fgl(1)$ intertwining a vector representation and a Fock representation. The proposal for the Miura operators for the $W$- and $Y$-algebras was also explicitly verified at the perturbation theory of 5d Chern-Simons theory in \cite{Ishtiaque:2024orn}. The R-matrices were computed there by the transition amplitudes of the intersecting defect configuration, as in the Yangian R-matrix realization in 4d Chern-Simons theory \cite{Costello:2017dso}. The R-matrices of the quantum affine algebra we established in this work are expected to be related through the evaluation homomorphism of the quantum toroidal algebra.}

\subsubsection{Monodromy matrix and operator-valued $q$-opers}
Note that the column vectors $\Theta_\o$ \eqref{eq:theta} are valued in $\BC^2 \otimes \mathcal{H}$, where $\CalH$ is the space of Laurent polynomials in monodromy defect parameters $(u_\o)_{\o=0}^{N-1}$,
\begin{align}
    \mathcal{H} := \mathcal{H}_0 \, \widehat{\otimes} \, \CalH_1 \, \widehat{\otimes} \cdots \widehat{\otimes} \, \mathcal{H}_{N-1},
\end{align}
which provides a module over the quantum affine algebra $U_q (\widehat{\fgl}(2))$ by the map $\r : U_q (\widehat{\fgl}(2)) \to \text{End}(\CalH)$, defined by
\begin{align} \label{eq:wholerep}
    \r := (\r_{0} \otimes \cdots \otimes \r_{N-1} ) \circ (\text{ev}_{m^+ _0} \otimes \cdots \otimes \text{ev}_{m^+ _{N-1}} ) \circ \Delta^{N-1},
\end{align}
where $\Delta$ is the coproduct of $U_q (\widehat{\fgl}(2))$ (see appendix \ref{subsubsec:cop}). In the language of the quantum integrable model, the space $\CalH$ will be the space of states for our 
$\fgl(2)$ XXZ spin chain with $N$ sites.

The monodromy matrix $T(X)$, whose entries are operators on this space, is obtained by concatenating \eqref{eq:lax}, 
\begin{align}
  \frac{\left(m^+  \right)^{\frac 1 2}}{X^{\frac{N}{2}}} P^+ (X)  \Theta_{N-1} (X) = T(X) \Theta_{-1} (X) = T(X) q_1 ^{-D_X} \Theta_{N-1} (X).
\end{align}
We defined the monodromy matrix $T(X) \in \text{End}(\BC^2) \otimes \text{End}(\mathcal{H})$ as the ordered product of the R-matrices and twist matrices $K_\o = \text{diag}\left(1,\qe^{\d_{\o,N-1}} \sqrt{\frac{m^+ _\o}{m^+ _{\o+1}}}\right)$ as
\begin{align}
    T(X) := K_{N-1} L_{N-1}(X) \cdots K_2 L_2 (X) K_1 L_1 (X) \in \text{End}(\BC^2) \,{\otimes } \, \text{End}(\CalH),
\end{align}
where the product is taken as elements in $\text{End}(\BC^2)$. In the view of the representation \eqref{eq:wholerep} assigned to $\CalH$, the monodromy matrix $T(X)$ is precisely the R-matrix of the quantum affine algebra $U_q (\widehat{\fgl}(2))$ intertwining the two representations carried by $\BC^2$ and $\CalH$. \\

We may put the above relation into a matrix-valued first-order $q_1$-difference equation as
\begin{align} \label{eq:matqoper}
    0 = \left[  \frac{\left(m^+  \right)^{\frac 1 2}}{X^{\frac{N}{2}}} P^+ (X) - T(X) q_1 ^{-D_X} \right] \Theta_{N-1} (X)  .
\end{align}
We can convert this into a second-order $q_1$-difference equation for the vev of the $Q$-observable as follows. Let us explicitly write the components of the monodromy matrix as 
\begin{align}\label{eq:monodromy}
    T(X) =  \begin{pmatrix}
        A(X) & B(X) \\ C(X) & D(X)
    \end{pmatrix},
\end{align}
where $A(X),B(X),C(X),D(X) \in \text{End}(\CalH)$. Explicitly writing out the two components of the equation \eqref{eq:matqoper}, we get
\begin{align}
\begin{split}
    & \left(  \frac{\left(m^+  \right)^{\frac 1 2}}{X^{\frac{N}{2}}} P^+ (X)  - A(X)q_1^{-D_X} \right) \langle \boldsymbol{\EQ}_{N-1} (X) \rangle - B(X) q_1^{-D_X} \langle \boldsymbol{\ER}_{N-1}(X) \rangle = 0, \\
    & \left( \frac{\left(m^+  \right)^{\frac 1 2}}{X^{\frac{N}{2}}} P^+ (X) - D(X) q_1^{-D_X} \right) \langle \boldsymbol{\ER}_{N-1}(X) \rangle  - C(X) q_1^{-D_X} \langle  \boldsymbol{\EQ}_{N-1}(X)\rangle  = 0.
\end{split}
\end{align}
To completely cancel the terms involving $\boldsymbol{\ER}_{N-1}(X)$, we proceed with the following steps. First, act $\frac{\left( m^+ \right)^{\frac 1 2}}{X^\frac{N}{2} q_1 ^{-\frac N 2}} P^+ (Xq_1 ^{-1}) - q_1 ^{-\frac 1 2} D(X) q_1 ^{-D_X}$ to the first equation from the left. Second, shift $X \to X q_1 ^{-1}$ in the second equation and act $B(X)$ from the left. By adding the two equations, we obtain
\begin{align}
\begin{split}
    0 = &  \frac{\left( m^+ \right)^{\frac 1 2}}{X^{\frac N 2} q_1 ^{- \frac N 2}} P^+ (X q_1 ^{-1}) \left(  \frac{\left( m^+ \right)^\frac{1}{2}}{X^{\frac N 2}} P^+ (X)  - \left( A(X)+ q_1 ^{-\frac 1 2} D(X) \right) q_1 ^{-D_X} \right) \langle \boldsymbol{\EQ}_{N-1} (X) \rangle \\
    & + \left( q_1 ^{-\frac 1 2} D(X) A(X q_1^{-1}) -B(X) C(X q_1 ^{-1}) \right) q_1 ^{-2 D_X}  \langle \boldsymbol{\EQ}_{N-1} (X) \rangle \\
    &+ \left(q_1 ^{-\frac 1 2} D(X) B(X q_1 ^{-1}) - B(X) D(X q_1 ^{-1}) \right) q_1 ^{-D_X} \langle \boldsymbol{\ER}_{N-1} (X q_1 ^{-1}) \rangle.
\end{split}
\end{align}
The last line identically vanishes since the defining relation \eqref{eq:RTT} of the quantum affine algebra $U_q (\widehat{\fgl}(2))$ implies
\begin{align}
    q_1 ^{-\frac 1 2} D(X) B(X q_1 ^{-1}) = B(X) D(X q_1 ^{-1}) .
\end{align}
Also, the term in the parenthesis in the second line is given by the quantum determinant (see \eqref{eq:qdet}) of the monodromy matrix, 
\begin{align}
\begin{split}
    \text{qdet}\,T(X) &= D(X) A(X q_1 ^{-1}) - q_1 ^{\frac 1 2 } B(X) C(X q_1 ^{-1}) \\
    & = \prod_{\o=0} ^{N-1} \det K_\o \prod_{\o=0} ^{N-1} \text{qdet}\, L_\o (X) \\
    & = \qe X^{-N} q_1 ^{\frac {N+1} {2}} m^+ P^+ (X q_1 ^{-1}) P^- (X q_1 ^{-1}),
\end{split}
\end{align}
where we used the factorization property \eqref{eq:factor} of the quantum determinant under the coproduct in the second line. Recalling that $\boldsymbol\EQ_{N-1}(X) = \bQ(X)$, we arrive at the second-order operator equation,
\begin{align} \label{eq:opqop}
    0 = \left[P^+ (X) + \hat{t}(X) q_1 ^{-D_X} + \qe  P^- (Xq_1 ^{-1}) q_1 ^{-2D_X} \right]  \bQ (X).
\end{align}
We call this equation the operator-valued $q$-oper difference equation,
since it is exactly the same with the $q$-oper difference equation \eqref{eq:qoper2} except that the coefficients are operators in
$\text{End}(\CalH)$. In particular, we defined the transfer matrix $\hat{t}(X) \in \text{End}(\CalH)$ by
\begin{align}
    \hat{t}(X) := X^{\frac N 2} \left(m^+ \right)^{-\frac 1 2} \left( A(X) + q_1 ^{-\frac 1 2} D(X) \right).
\end{align}
It is straightforward to see that the coefficients of the transfer matrices, as operators in $\text{End}(\CalH)$, commute with each other due to the defining relation \eqref{eq:RTT} of $U_q (\widehat{\fgl}(2))$. They are precisely the quantum Hamiltonians of the $\fgl(2)$ XXZ spin chain system constructed on $\CalH$ as the space of states.

\subsubsection{Eigenstates of $Q$-operators and quantum Hamiltonians}
We show that the eigenstates for the $Q$-operators are also eigenstates for the quantum Hamiltonians of our $\fgl(2)$ XXZ spin chain, using our 5d $\EN=1$ gauge theoretical constructions.

Note that the vev of the monodromy codimension-two defect $\psi(\ba;\bu)$ in the limit $q_2 \to 1$ provides a distinguished basis of the space $\CalH$. It is straightforward that it diagonalizes the action of the $Q$-operators, since
\begin{align}
    \lim_{q_2\to 1} \Big\langle \bQ(X) \boldsymbol{\Psi}(\bu) \Big\rangle _\ba = e^{\frac{\widetilde{\EuScript{W}}(\ba)}{\log q_2}} \bQ(\ba;X) \boldsymbol\Psi(\ba;\bu)
\end{align}
due to the cluster decomposition in the limit $q_2 \to 1$.

On the other hand, by taking the $q_2 \to 1$ limit to the operator-valued $q$-oper equation \eqref{eq:opqop}, we get
\begin{align}
        0 = \left[P^+ (X) + \hat{t}(X) q_1 ^{-D_X} + \qe  P^- (Xq_1 ^{-1}) q_1 ^{-2D_X} \right]  \bQ (\ba;X) \boldsymbol\Psi(\ba;\bu).
\end{align}
Recall that the vev $\bQ (\ba;X)$ of the $Q$-observable is the solution to the $q$-oper equation,
\begin{align}
        0 = \left[P^+ (X) + {t}(\ba;X) q_1 ^{-D_X} + \qe  P^- (Xq_1 ^{-1}) q_1 ^{-2D_X} \right]  \bQ (\ba;X).
\end{align}
Also, note that exactly the same relations hold after replacing $\bQ(\ba;X)$ by $\tilde{\bQ}(\ba;X)$ (see appendix \ref{sec:dualQ}). Thus, we conclude
\begin{align}
    0 = \begin{pmatrix}
         \left(\hat{t}(X) - t(\ba;X)\right) \psi(\ba;\bu) & 0 \\ 0 & 0
    \end{pmatrix} \begin{pmatrix}
        \bQ(\ba;X) & \tilde{\bQ}(\ba;X) \\ \frac{1}{P^+(X)} \bQ(\ba; q_1 ^{-1} X) & \frac{1}{P^+(X)} \tilde{\bQ}(\ba; q_1 ^{-1} X)
    \end{pmatrix}
\end{align}
at generic $X \in \BC^\times$. The determinant of the second matrix is exactly the $q$-Wronskian \eqref{eq:qwrons} of the $q$-oper, with simple zeros only at discrete loci. Thus, at generic $X$ we can invert this matrix to yield
\begin{align}
    0 =  \left(\hat{t}(X) - t(\ba;X)\right) \psi(\ba;\bu).
\end{align}
Since $X$ is generic, the equation holds at each coefficient of the Laurent polynomial. They are nothing but the spectral equations for the mutually commuting quantum Hamiltonians of the $\fgl(2)$ XXZ spin chain, with the eigenvalues given by the vevs of the codimension-four observables.

We stress that the derivation of the $q$-difference spectral equations is exact. See \cite{Bullimore:2014awa} for other works on the gauge theoretical account of the spectral equations with different contents.


\subsection{R-matrices of quantum affine algebra of $\fgl(N)$} \label{subsec:rmatN}

We will demonstrate that the fractional TQ equations, which constrain the correlation functions of codimension-two defects and codimension-four observables, can be reorganized in terms of the representation theory of the quantum affine algebra $U_q (\widehat{\fgl}(N))$. 

Notably, in contrast to the previous realization of the R-matrices in section \ref{subsec:rgl2}, the rank of the base Lie algebra and the number of modules are exchanged: whereas the earlier construction involved $N$ R-matrices of $U_q (\widehat{\fgl}(2))$, the present framework features 2 R-matrices of $U_q(\widehat{\fgl}(N))$. This exchange is a hallmark of the bispectral duality.

\subsubsection{R-matrices from fractional TQ equations}
We show that the analytic constraint imposed by the fractional TQ equations can be cast into the form of R-matrices of the quantum affine algebra $U_q (\widehat{\fgl}(N))$.

Let us rearrange the fractional TQ equations into a recursive form, where the multiplication by $X$ appears in only one of the two equations:
\begin{subequations} \label{eq:frac-RQ-relation}
\begin{align}
    &\qe^{-\d_{\o,N-1}} \sqrt{\frac{m^+_{\o+1}}{m^- _\o}} u_\o \langle \boldsymbol{\ER}_\o (X)\rangle - q_1 ^{-\frac{1}{2} D_{u_\o} } u_\o \frac{m^+ _\o}{m^- _\o} \langle \boldsymbol{\ER}_{\o-1} (X) \rangle \nonumber \\ &\quad = \left( q_1 ^{\frac{1}{2} D_{u_\o} } - \frac{m^+ _\o}{m^- _\o} q_1 ^{-\frac{1}{2} D_{u_\o}}  \right) \langle \boldsymbol{\EQ}_\o (X) \rangle - \left( 1- \frac{m^+ _\o}{m^- _\o} q_1 ^{-D_{u_\o}} \right) \langle \boldsymbol{\EQ}_{\o-1} (X) \rangle, \\
    &\qe^{-\d_{\o,N-1}} \sqrt{\frac{m^+_{\o+1}}{m^- _\o}} u_\o \langle \boldsymbol{\ER}_\o (X) \rangle - q_1 ^{\frac{1}{2} D_{u_\o} } u_\o \langle \boldsymbol{\ER}_{\o-1} (X)\rangle \nonumber \\
    &\quad = X\left[ \left( (m^+ _\o)^{-1} q_1 ^{\frac{1}{2} D_{u_\o}} - (m^- _\o)^{-1} q_1 ^{-\frac{1}{2} D_{u_\o}} \right) \langle \boldsymbol{\EQ}_\o (X) \rangle +\left( (m^- _\o)^{-1} - (m^+_\o)^{-1} q_1 ^{D_{u_\o}} \right) \langle \boldsymbol{\EQ}_{\o-1} (X) \rangle \right]
\end{align}
\end{subequations}
At this point, we perform the Fourier transformation by considering
\begin{align}
\begin{split}
    & H_\o^+(Z) =  \oint_{\CalC} \frac{\mathrm{d}X}{X} e^{-\frac{\log X \log Z}{\log q_1}} \langle \boldsymbol{\EQ}_\o(X) \rangle,  \quad H_\o^-(Z) = \oint_{\CalC} \frac{\mathrm{d} X}{X} e^{-\frac{\log X \log Z}{\log q_1}} \langle  \boldsymbol{\ER}_\o(X) \rangle,
\end{split}
\end{align}
where the contour $\CalC$ is a Barnes contour enclosing an semi-infinite $q_1$-lattice of simple poles of the integrand. There are $N$ choices of the contour, enclosing the simple poles at $X = m^+ _\a q_1 ^{\BZ_{\geq 0}}$. They exactly correspond to $N$ discrete choices of the vacuum at infinity for the canonical codimension-two defect in the Higgs phase (see section \ref{subsubsec:hobs}). The construction of R-matrices follow from any choice, and we will omit the label for this choice below.

Note that $H^\pm _{\o+N} (Z) = Z H^\pm _\o (Z)$. Then, let us form vectors having them as components,
\begin{align}
    H^\pm (Z) := \begin{pmatrix}
        H_{0}^\pm(Z) \\ H_{1}^\pm(Z) \\ \vdots \\ H_{N-1}^\pm(Z)
    \end{pmatrix}.
\end{align}
With these vectors, \eqref{eq:frac-RQ-relation} can be written as the following $N \times N$ matrix equations:
\begin{subequations} \label{eq:laxNint}
\begin{align}
    &\left(\boldsymbol\k ^{-1} \mathbf{M^-} \mathbf{U} \bM^+ \bU^{-1} \mathbf{u} - (\bM^+)^2 \boldsymbol{\nabla} ^{-1} \bu \bU^{-1} \right) H^- (Z)\nonumber \\
    &\qquad\qquad = \left((\bM^-)^2 - (\bM^+)^2 \boldsymbol{\nabla}^{-2} \right) \left( \boldsymbol{\nabla}- \bU^{-1} \right)  H^+ (Z), \\
    & \left( \boldsymbol\k ^{-1} \bM^- \bU \bM^+ \bU^{-1} \bu - (\bM^-)^2 \boldsymbol{\nabla} \bu \bU^{-1}  \right) H^- (Z) \nonumber \\
    &\qquad \qquad = q_1 ^{-D_Z} \left( (\bM^-)^2 (\bM^+)^{-2 } \boldsymbol{\nabla} - \boldsymbol{\nabla} ^{-1} \right) \left( \mathds{1}_N- \boldsymbol{\nabla} \bU^{-1} \right)H^+ (Z),
\end{align}
\end{subequations}
where we used the following notations:
\begin{align}
\begin{split}
    & \bU  =  \begin{pmatrix}
    0 & 1 & 0 & \cdots & 0 & 0 \\
    0 & 0 & 1 & \cdots & 0 & 0 \\
    0 & 0 & 0 & \cdots & 0 & 0 \\
    & \vdots & & \ddots & & \vdots \\
    0 & 0 & 0 & \cdots & 0 & 1 \\
    Z & 0 & 0 & \cdots &0 & 0
    \end{pmatrix}, \qquad  \bM^\pm = \text{diag} (\sqrt{m^\pm_{0}},\dots,\sqrt{m_{N-1}^\pm}), \\ 
    & \bu = \text{diag} (u_{0},\dots,u_{N-1}), \ \bna = \text{diag}\left(q_1^{\frac12D_{u_{0}}},\dots,q_1^{\frac{1}{2}D_{u_{N-1}}}\right), \ \bka = \text{diag} \left(1,\dots,1, \qe q_1 ^{\frac{1}{2}} \right).  
\end{split}
\end{align}

To obtain the R-matrices of the quantum affine algebra $U_q (\widehat{\fgl}(N))$, we need to invert either the left or right hand side of the relations \eqref{eq:laxNint}. However, it is not straightforward to invert the right hand side, since there is a diagonal matrix whose entries are $q_1$-difference operators. To remedy this, we formally perform the $q$-Fourier transformation by considering
\begin{align}
\begin{split}
    & \tilde{H}^+ (\bv,Z) = \int   \prod_{\o=0}^{N-1} \mathrm{d} u_\o\, e_{q_1} (-u_\o v_\o) u_\o^{\frac{\log m_\o^--\log m_\o^+}{\log q_1}} H^+(\bu,Z) , \\
    & \tilde{H}^-(\bv,Z) = \frac{1}{q_1^{\frac12}-q_1^{-\frac12}} \int \prod_{\o=0}^{N-1} \mathrm{d} u_\o \, e_{q_1} (-u_\o v_\o) u_\o^{\frac{\log m_\o^--\log m_\o^+}{\log q_1}} H^-(\bu,Z) ,
\end{split}
\end{align}
where we defined the $q$-Fourier dual variables by $\bv=(v_\o)_{\o=0}^{N-1}$.

Passing to the $q$-Fourier dual space, the relations \eqref{eq:laxNint} are converted into
\begin{subequations} \label{eq:laxNint2}
\begin{align}
    &\left(\bI - \tilde\bna \bU^{-1}  \bka \right) \bka^{-1} \bU\bM^+\bU^{-1}\tilde{H}^- (Z) = \bM^+ \bv \left( \bI- \tilde\bna\bs^{-1} \bU^{-1} \right)  \tilde{H}^+ (Z), \\
    & \left( \bI - \tilde\bna^{-1} \bU^{-1} \bka  \right) \bka^{-1} \bU\bM^+\bU^{-1} \tilde{H}^- (Z) = q_1 ^{-D_Z} (\bM^+)^{-1} \bv \left( \bI-  \bs\tilde\bna^{-1}\bU^{-1} \right) \tilde{H}^+ (Z),
\end{align}
\end{subequations}
where we used the following notations:
\begin{align}
\begin{split}
    & \tilde\bna = \text{diag} \left(q_1^{\frac12D_{v_0}},\dots,q_1^{\frac12D_{v_{N-1}}} \right)~,~ \bv = (v_0,\dots,v_{N-1})~,~ \bs = \bM^+ (\bM^-)^{-1}q_1^{-\frac12}.
\end{split}
\end{align}
Let us define
\begin{align}
    \tilde\Pi^+(Z) := \left( \bI - \bs \tilde\bna^{-1} \bU^{-1} \right) \tilde{H}^+(Z)~,~ \tilde\Pi^-(Z) := \bv^{-1} (\tilde\bna^{-1}-\bU^{-1}\bka ) \bka^{-1} \bU \bM^+\bU \tilde{H}^-(Z).
\end{align}
Then, the relations \eqref{eq:laxNint2} turn into
\begin{subequations}
\begin{align}
    & \left( 1- \frac{\sqrt{m^+/a}}{Z} \right) \tilde\Pi^-(Z) = \bM^- \tilde{\mathbf{L}}_1 (Z) \tilde{\Pi}^+(Z), \\
    & \left( 1 - \frac{\kq q_1 ^{\frac 1 2}  \sqrt{\frac{a}{m^-q_1 ^N}}}{Z} \right)  q_1^{-D_Z} \tilde\Pi^+(Z) = \bM^+ \tilde{\mathbf{L}}_2(Z)  \tilde\Pi^-(Z),
\end{align}
\end{subequations}
where $\tilde{\mathbf{L}}_i$, $i=1,2$, are $N\times N$ matrices with entries given by $q_1$-difference operators in $(v_\o)_{\o=0}^{N-1}$. Explicitly, we get
\begin{align}
\begin{split}
    &\left(\tilde{\mathbf{L}}_1 \right)_{ij} = \d_{i,j} \left( 1 - \frac{\sqrt{m^+/a}}{Z} \right) q_1^{\frac12D_{v_{i-1}}} \bs_{i-1}^{-1}+  \tilde{B}_{1,i} \tilde{A}_{1,j} \left(\frac{\sqrt{m^+/a}}{Z} \right)^{\theta_{i<j}},  \\    &\left(\tilde{\mathbf{L}}_2\right)_{ij} = \d_{i,j} \left( 1 - \frac{\kq q_1^{\frac{1-N}{2}}\sqrt{a/m^-}}{Z} \right) q_1^{-\frac12D_{v_{i-1}}}+ \tilde{B}_{2,i} \tilde{A}_{2,j}\left( \frac{\kq q_1^{\frac{1-N}{2}}\sqrt{a/m^-}}{Z}\right)^{\theta_{i<j}} ,
\end{split}
\end{align}
for $i,j =1,2,\cdots,N$, where we defined
\begin{align}
\begin{split}
    \tilde{B}_{1,i} & = \left( \sqrt{\frac{m^+ _{i-1}}{m^-_{i-1} q_1}} q_1^{-\frac{1}{2}D_{v_{i-1}}} - \sqrt{\frac{m^-_{i-1} q_1}{m^+ _{i-1}}}q_1^{\frac12D_{v_{i-1}}}  \right) \prod_{k=1}^{i} \sqrt{\frac{m^+ _{k-1}}{m^-_{k-1} q_1}}  q_1^{-\frac12 D_{v_{k-1}} } , \\
    \tilde{A}_{1,j} & = \prod_{k=1}^{j} \sqrt{\frac{m^-_{k-1} q_1}{m^+ _{k-1}}} q_1^{\frac12 D_{v_{k-1}}},
\end{split}
\end{align}
and
\begin{align}
\begin{split}
     \tilde{B}_{2,i} & = \frac{1}{v_{i-1}} \left( q_1^{\frac12D_{v_{i-1}}} - q_1^{-\frac12D_{v_{i-1}}} \right) \prod_{k=1}^{i-1} q_1^{\frac12 D_{v_{k-1}} } , \\
    \tilde{A}_{2,j} & = \prod_{k=1}^{j-1} q_1^{-\frac12 D_{v_{k-1}}} v_{j-1},
\end{split}
\end{align}
for $i,j =1,2,\cdots,N$.

Finally, we take the inverse $q$-Fourier transformation to retain the expressions in the $\bu$-space, yielding
\begin{subequations}
\begin{align}
    & \left( 1- \frac{\sqrt{m^+/a}}{Z} \right) \Pi^-(Z) = \bM^- {\mathbf{L}}_1 (Z) {\Pi}^+(Z), \\
    & \left( 1 - \frac{\kq q_1 ^{\frac 1 2}  \sqrt{\frac{a}{m^-q_1 ^N}}}{Z} \right)  q_1^{-D_Z} \Pi^+(Z) = \bM^+ {\mathbf{L}}_2(Z)  \Pi^-(Z),
\end{align}
\end{subequations}
where 
\begin{align} \label{eq:nngauge}
\begin{split}
    &\mathbf{L}_i (Z) = \mathbf{L}^+ _i - \frac{Z_i}{Z} \mathbf{L}^- _i,\qquad i=1,2,\\
    &Z_1 = \sqrt{\frac{m^+}{a}},\qquad Z_2 = \qe q_1 ^{\frac 1 2} \sqrt{\frac{a}{m^- q_1 ^N}},
\end{split}
\end{align}
is an $N\times N$ matrix with the entries given by $q_1$-difference operators in the monodromy parameters $(u_\o)_{\o=0}^{N-1}$.\\

We show that $\mathbf{L}_i (Z) \in \text{End}(\BC^N) \otimes \text{End}(\CalH)$ precisely provides the R-matrix of the quantum affine algebra $U_q (\widehat{\fgl}(N))$ intertwining evaluation modules. Let us first consider $\mathbf{L}_1 (Z)$. Its components are computed to be
\begin{align}
\begin{split}
 &\left(\mathbf{L}_1^+\right)_{ij} = \d_{i,j} q_1 ^{\frac{1}{2}D_{u_{i-1}}}  +  \th_{i>j} \left( q_1^{\frac{1}{2}D_{u_{i-1}}} - q_1^{-\frac12D_{u_{i-1}}}  \right) \prod_{k=j+1}^{i} q_1^{\frac12 D_{u_{k-1}} }  ,\\
 & \left(\mathbf{L}_1^-\right)_{ij} = \d_{i,j}  q_1 ^{- \frac{1}{2}D_{u_{i-1}}} - \th_{i<j} \left( q_1^{\frac{1}{2}D_{u_{i-1}}} - q_1^{-\frac12D_{u_{i-1}}}  \right) \prod_{k=i+1}^{j} q_1^{-\frac12 D_{u_{k-1}} }.
\end{split}
\end{align}
We show in section \ref{subsec:Nclus} that this is precisely an R-matrix of $U_q (\widehat{\fgl}(N))$ intertwining certain evaluation modules, after conjugating a simple prefactor.

Next, let us turn to $\mathbf{L}_2 (Z)$, for which the properties of the R-matrix is more straightforward. Its components are given by
\begin{align}
\begin{split}
   & \left(\mathbf{L}^+_2 \right)_{ij} = \sqrt{ \frac{m^+ _{i-1}}{m^- _{i-1} q_1} }  q_1 ^{-\frac{1}{2} D_{u_{i-1}}} \d_{i,j} +   B_{2,i} A_{2,j}\th_{i>j}, \\
   &\left(\mathbf{L}^-_2 \right)_{ij}  = \sqrt{ \frac{m^- _{i-1} q_1}{m^+ _{i-1}} } q_1 ^{\frac{1}{2} D_{u_{i-1}}} \d_{i,j} -  B_{2,i} A_{2,j} \th_{i<j},
\end{split}\qquad i,j=1,2,\cdots, N,
\end{align}
where we defined
\begin{align}
\begin{split}
    &B_{2,i} = u_{i-1} \prod_{k=1}^{i-1} \sqrt{ \frac{m^+ _{k-1}}{m^- _{k-1} q_1} } q_1^{-\frac12 D_{u_{k-1}} } ,\\
    &A_{2,i} = \prod_{k=1}^{i-1}  \sqrt{ \frac{m^- _{k-1} q_1}{m^+ _{k-1}} } q_1^{\frac12 D_{u_{k-1}}} \frac{1}{u_{i-1}} \left(  \sqrt{ \frac{m^+ _{i-1}}{m^- _{i-1} q_1} }  q_1^{-\frac{1}{2}D_{u_{i-1}}} -  \sqrt{ \frac{m^- _{i-1} q_1} {m^+ _{i-1}}}  q_1^{\frac12D_{u_{i-1}}} \right),
\end{split}
\end{align}
for $i=1,2,\cdots, N$.

Let us consider a map $\varrho_2 : U_q (\fgl(N)) \to \text{End}(\CalH)$ defined by (see appendix \ref{subsubsec:qea} for the presentation of the quantized enveloping algebra $U_q (\fgl(N))$)
\begin{align} \label{eq:glNmap}
    T^\pm \mapsto \mathbf{L}^\pm _2,
\end{align}
or, in terms of the Chevalley generators,
\begin{align}
\begin{split}
    &t_i \mapsto  \sqrt{ \frac{m^+ _{i-1}}{m^- _{i-1} q_1} } q_1 ^{-\frac{1}{2} D_{u_{i-1}}},\\
    & e_r \mapsto \frac{u_{i-1} u_i ^{-1} \left( \sqrt{\frac{m^+ _i}{m^- _i q_1}} q_1 ^{-\frac{1}{2} D_{u_{i}}}- \sqrt{\frac{m^- _i q_1}{m^+ _i}} q_1 ^{\frac{1}{2} D_{u_{i}}}  \right)}{q_1 ^{\frac 1 2} -q_1 ^{-\frac 1 2}} ,\\
    &f_r \mapsto \frac{u_{i} u_{i-1} ^{-1} \left(\sqrt{\frac{m^+ _{i-1}}{m^- _{i-1} q_1}} q_1 ^{-\frac{1}{2} D_{u_{i-1}}} - \sqrt{\frac{m^- _{i-1} q_1}{m^+ _{i-1}}} q_1 ^{\frac{1}{2} D_{u_{i-1}}}  \right)}{q_1 ^{\frac 1 2} -q_1 ^{-\frac 1 2}} ,
\end{split}
\begin{split}
    &i=1,2,\cdots, N, \\
    &r = 1,2,\cdots, N-1.
\end{split}
\end{align}
It is straightforward to check that this map is an algebra homomorphism, provided the following identification of the quantum parameter with the $\O$-background parameter,
\begin{align}
    q = q_1 ^{-\frac 1 2}.
\end{align}
It is crucial to note that there is a sign difference compared to the identification in the bispectral dual side (recall \eqref{eq:identhbar}).\footnote{This is not a contradiction, of course, but only means the quantum parameters in the two sides of the bispectral duality are different by a sign.}

Thus, the space $\CalH$ is equipped with a $U_q (\fgl(N))$-module structure. Note that it is irreducible (for generic mass parameters) and bi-infinite (i.e., not highest-weight nor lowest-weight). The quantum Casimirs are given by the mass parameters. For instance, the lowesst quantum Casimir is given by
\begin{align}
    C_q ^{(1)} = t_1 t_2 \cdots t_N \mapsto \sqrt{\frac{a}{m^- q_1^N}}.
\end{align}
The higher quantum Casimirs can also be obtained by computing the quantum Gel'fand invariants of $U_q (\fgl(N))$, which are given by $\text{Tr} \, D \left(T^- (T^+)^{-1} \right)^m$ with $m \in \BZ_{\geq 0}$ and $D = \text{diag} (q^{N-1}, q^{N-3}, \cdots , q^{-N+1} )$ \cite{Gould:1989mt,Jing:2024eha}, using the image $\mathbf{L} _2 ^\pm $ under the homomorphism $\varrho_2$ \eqref{eq:glNmap}. We leave the computation of these invariants to interested readers. See also \cite{Hopkins2006} for another formulation of the quantum Casimirs of $U_q (\fgl(N))$.\\

Then, we compose the map $\varrho_2$ with the evaluation map $\text{ev}_{Z_2} : U_q (\widehat{\fgl}(N)) \to U_q (\fgl(N))$ with the evaluation parameter given by the exponentiated complexified gauge coupling $Z_2 = \qe q_1 ^{\frac 1 2} \sqrt{\frac{a}{m^- q_1 ^N}} \in \BC^\times$, to view $\CalH$ as an evaluation module over the quantum affine algebra $U_q (\widehat{\fgl}(N))$. In this view, $\mathbf{L}_2 (Z)$, which is valued in $\text{End}(\BC^N) \otimes \text{End}(\CalH)$, is identified with the R-matrix of $U_q (\widehat{\fgl}(N))$ intertwining the two evaluation modules.

\section{Cluster realization of R-matrices} \label{sec:cluster}

In this section, we revisit the 5d $\EN=1$ gauge theoretical construction of R-matrices of the quantum affine algebra in the point of view of the quantum cluster algebra for the BPS quiver.

After reviewing the general notion of the quantum cluster algebra, we will apply it to the BPS quiver for the 5d $\EN=1$ $A_{M-1}$-quiver $U(N)$ gauge theory. We will construct the R-matrices of two quantum affine algebras $U_q (\widehat{\fgl}(M))$ and $U_q (\widehat{\fgl}(N))$ from the quantum Kasteleyn operator. These R-matrices precisely match with the ones we constructed previously as the constraints on the correlation functions of the defects.

\subsection{Quantum cluster algebra for BPS quiver}
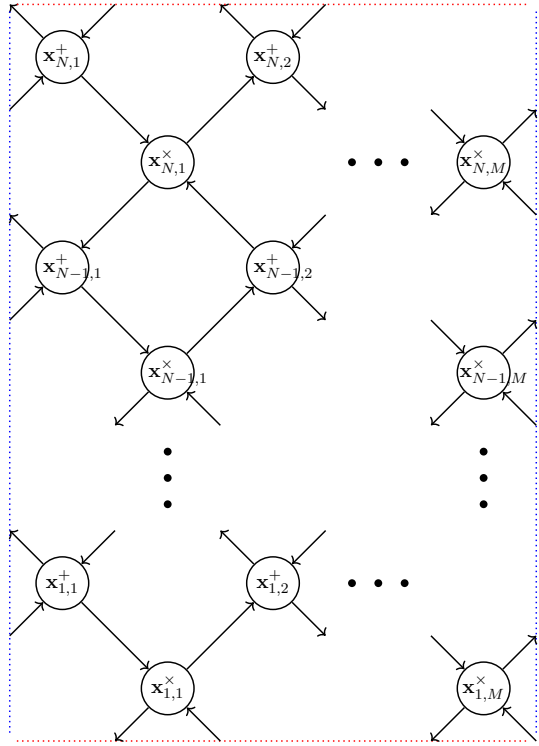
\begin{figure}[h!] \centering
\scalebox{0.7}{
\begin{tikzpicture}[square/.style={regular polygon,regular polygon sides=4}]
   
    \node[style={circle,draw},thick,minimum size=1cm] (a0) at (-2,-2) {};
    \node at (-2,-2) {$\bx^+ _{N,1}$};
    \node[style={circle,draw},thick,minimum size=1cm] (c0) at (2,-2) {};
    \node at (2,-2) {$\bx^+_{N,2}$}; 
 \node[style={circle,draw},thick,minimum size=1cm] (b0) at (0,-4) {};
 \node at (0,-4) {$\bx^\times _{N,1}$};
    \draw[->,thick] (c0) -- (1,-1);
    \draw[->,thick] (-1,-1) -- (a0);
    \draw[->,thick] (a0) -- (-3,-1);
    \draw[->,thick] (-3,-3) -- (a0);
    \draw[->,thick] (a0) -- (b0);
    \draw[->,thick] (b0) -- (c0);

   \node[style={circle,draw},thick,minimum size=1cm] (a1) at (-2,-6) {};
  \node at (-1.8,-6) {$\bx^+ _{N-1,1}$};
   
    \node[style={circle,draw},thick,minimum size=1cm] (c1) at (2,-6) {};
       \node at (2.2,-6) {$\bx^+ _{N-1,2}$};
       
    \node[style={circle,draw},thick,minimum size=1cm] (b1) at (0,-8) {};
    \node at (0.2,-8) {$\bx^\times _{N-1,1}$};
    
        \draw[->,thick] (b0) -- (a1);
         \draw[->,thick] (a1) -- (-3,-5);
        \draw[->,thick] (-3,-7) -- (a1); 
    \draw[->,thick] (a1) -- (b1);
    \draw[->,thick] (b1) -- (c1);
    \draw[->,thick] (c1) -- (b0);
        \draw[->,thick] (b1) -- (-1,-9);
    \draw[->,thick] (1,-9) -- (b1);

     \node at (0,-9.5) [circle,fill,inner sep=1.5pt]{};
        \node at (0,-10) [circle,fill,inner sep=1.5pt]{};
     \node at (0,-10.5) [circle,fill,inner sep=1.5pt]{};

          \node at (6,-9.5) [circle,fill,inner sep=1.5pt]{};
        \node at (6,-10) [circle,fill,inner sep=1.5pt]{};
     \node at (6,-10.5) [circle,fill,inner sep=1.5pt]{};

        \node at (3.5,-4) [circle,fill,inner sep=1.5pt]{};
          \node at (4,-4) [circle,fill,inner sep=1.5pt]{};
            \node at (4.5,-4) [circle,fill,inner sep=1.5pt]{};

        \node at (3.5,-12) [circle,fill,inner sep=1.5pt]{};
          \node at (4,-12) [circle,fill,inner sep=1.5pt]{};
         \node at (4.5,-12) [circle,fill,inner sep=1.5pt]{};

      \node[style={circle,draw},thick,minimum size=1cm] (a2) at (-2,-12) {};
      \node at (-2,-12) {$\bx^+ _{1,1}$};
        \node[style={circle,draw},thick,minimum size=1cm] (c2) at (2,-12) {};
        \node at (2,-12) {$\bx^+ _{1,2}$};

  \draw[->,thick] (c2) -- (1,-11);
    \draw[->,thick] (-1,-11) -- (a2);

    \node (t0) at (-3,-1) {};
    \node (t3) at (7,-1) {};
    \node (t4) at (-2,-1) {};
    \node (t5) at (2,-1) {};
     \node (t1) at (-1,-1) {};
     \node (t2) at (1,-1) {};

     \draw[dotted,thick,red] (t0) -- (t3);

     \node (bb1) at (-1,-15) {};
     \node (bb2) at (1,-15) {};
          \node (bb3) at (-2,-15) {};
     \node (bb4) at (2,-15) {};
          \node (bb5) at (-3,-15) {};
     \node (bb6) at (7,-15) {};
      \node[style={circle,draw},thick,minimum size=1cm] (b2) at (0,-14) {};
      \node at (0,-14) {$\bx_{1,1} ^\times$};
    \draw[->,thick] (a2) -- (b2);
     \draw[->,thick] (b2) -- (c2);

     \draw[dotted,thick,red] (bb5) -- (bb6);

     \draw[->,thick] (b2) -- (-1,-15);
     \draw[->,thick] (1,-15) -- (b2);
       \draw[dotted,thick,blue] (t0) -- (bb5);
        \draw[dotted,thick,blue] (t3) -- (bb6);
       
         \draw[->,thick] (a2) -- (-3,-11);
 \draw[->,thick] (-3,-13)--(a2);
 \draw[->,thick] (c2)--(3,-13);
  \draw[->,thick] (3,-11)--(c2);
 \draw[->,thick] (c1)--(3,-7);
  \draw[->,thick] (3,-5)--(c1);
 \draw[->,thick] (c0)--(3,-3);
  \draw[->,thick] (3,-1)--(c0);

  \node[style={circle,draw},thick,minimum size=1cm] (nm) at (6,-4) {};
  \node at (6,-4) {$\bx^\times _{N,M}$};
    \node[style={circle,draw},thick,minimum size=1cm] (nnm) at (6,-8) {};
 \node at (6.2,-8) {$\bx^\times _{N-1,M}$};    
    \node[style={circle,draw},thick,minimum size=1cm] (nnnm) at (6,-14) {};
    \node at (6,-14) {$\bx^\times _{1,M}$};

    \draw[->,thick] (nm) -- (5,-5);
    \draw[->,thick]  (5,-3)--(nm);
    \draw[->,thick] (nm) -- (7,-3);
    \draw[->,thick]  (7,-5)--(nm);

    \draw[->,thick]  (7,-9)--(nnm);
\draw[->,thick]  (nnm)--(7,-7);   
\draw[->,thick]  (5,-7)--(nnm);
\draw[->,thick]  (nnm)--(5,-9);

\draw[->,thick]  (7,-15)--(nnnm);
\draw[->,thick]  (5,-13)--(nnnm);
\draw[->,thick]  (nnnm)--(7,-13);
\draw[->,thick]  (nnnm)--(5,-15);

\end{tikzpicture}
}  \caption{The quiver ${\S_{M,N}}$. The horizontal dotted lines (red) on the top and the bottom are identified, as well as the vertical dotted lines (blue) on the left and right. In turn, the quiver is doubly periodic in the horizontal and the vertical directions. We adopt the notations used in \cite{Marshakov:2019vnz}.} \label{fig:quiverMN}
\end{figure}

The spectra of the BPS particles of the 5d $\EN=1$ gauge theory compactified on a circle are encoded in the $\EN=4$ supersymmetric quantum mechanics on their worldline. This supersymmetric quantum mechanics often admits a quiver description, called the 5d BPS quiver. When the 5d theory is realized by the geometric engineering on a toric Calabi-Yau threefold, its BPS quiver can be systematically obtained from the toric data, using the brane tiling method \cite{Feng:2000mi} (see also \cite{Closset:2019juk} for a study from the derived category of coherent sheaves on the resolution). In this work, we restrict our attention to the case of the 5d $\EN=1$ $A_{M-1}$-quiver $U(N)$ gauge theory. Its BPS quiver, which we denote by $\S_{M,N}$, is given as in Figure \ref{fig:quiverMN} \cite{Hanany:2005ss} (we adopt the notations used in \cite{Marshakov:2019vnz} for the classical analogue). 

It is crucial to note that the quiver $\S_{M,N}$ is invariant under the exchange $M \leftrightarrow N$. As we will see, it is the existence of this symmetry that gives rise to the bispectral duality between certain evaluation modules over the two quantum affine algebras $U_q (\widehat{\fgl}(M))$ and $U_q (\widehat{\fgl}(N))$. \\

To study the quantum cluster algebra for this 5d BPS quiver, let us briefly review the quantum cluster algebra which is defined upon a given \textit{cluster seed}.\footnote{In this work, we utilize the quantum cluster $\CalX$-algebra constructed in \cite{Fock:2003xxy}, instead of the quantum cluster $\CalA$-algebra firstly realized in \cite{BERENSTEIN2005405}. There is an isomorphism between the two quantum cluster algebras under a suitable condition on the seed.} A seed is a triple $(I,I_0,\varepsilon)$, where $I$ is a finite set; $I_0 \subset I$ is a subset; and $\varepsilon = (\varepsilon_{ij})_{i,j \in I}$ is a skew-symmetric $\frac{1}{2}\BZ$-valued matrix such that $\varepsilon_{ij} \in \BZ$ unless $i,j \in I_0$. To a given seed $\S$, we associate an algebraic torus $\CalX_\S = \left( \BC^\times \right)^{\vert I \vert}$, called the cluster $\CalX$-torus. Its coordinates $(X_i)_{i\in I}$ are called the cluster variables. The Poisson structure on $\CalX_\S$ defined by $\varepsilon$ can be quantized to yield the quantum torus algebra $\CalO_q (\CalX_\S)$. It is given as an algebra over $\BZ[q^{\pm\frac 1 2}]$, generated by the cluster variables $(X_i)_{i\in I}$ subject to the relations
\begin{align}
    X_i X_j = q^{2\varepsilon_{ji}} X_j X_i.
\end{align}

For a given pair of seeds $\S = (I,I_0,\varepsilon)$ and $\S' = (I',I_0',\varepsilon')$, fix $k \in I \setminus I_0$. An isomorphism $\m_k : I \to I'$ is called the \textit{seed mutation in direction $k$} if $\m_k (I_0) = I_0 '$ and 
\begin{align}
    \varepsilon_{\m_k (i), \m_k (j)} ' = \begin{cases}
        -\varepsilon_{ij},\qquad\qquad & \text{if $i=k$ or $j=k$} \\
        \varepsilon_{ij}  & \text{if $\varepsilon_{ik} \varepsilon_{kj} \leq 0$} \\
        \varepsilon_{ij} +\vert \varepsilon_{ik} \vert \varepsilon_{kj} & \text{if $\varepsilon_{ik} \varepsilon_{kj} >0$}.
    \end{cases}
\end{align}
For the seed mutation $\m_k$, we define the \textit{quantum cluster mutation} $\m_k^q : \CalO_q (\CalX_\S) \to \CalO_q (\CalX_{\S'})$ by
\begin{align}
    \m_k^q (X_i) = \begin{cases}
        X_k ^{-1} \qquad \qquad\qquad &\text{if $i=k$} , \\
        X_i \prod_{r=1} ^{\varepsilon_{ki}} (1+q^{2r-1 }X_k ^{-1} )^{-1} & \text{if $i \neq k $ and $\varepsilon_{ki} \geq 0$}, \\
        X_i \prod_{r=1} ^{-\varepsilon_{ki}} (1+q^{2r-1 }X_k  ) & \text{if $i \neq k $ and $\varepsilon_{ki} \leq 0$}.
    \end{cases}
\end{align}

The quantum cluster algebra associated with a seed $\S$ is defined as the subalgebra of the quantum torus algebra $\CalO_q (\CalX_\S)$ consisting of \textit{universally Laurent} elements, i.e., the ones that remain Laurent polynomials under all finite sequences of quantum cluster mutations. \\

The 5d BPS quiver $\S_{M,N}$ defines a cluster seed, which we denote by the same letter, for which $I_0 = \varnothing$. Namely, the finite set $I$ labels the $M N$ nodes of the quiver, while the $\BZ$-valued skew-symmetric matrix $(\varepsilon_{ij})_{i,j\in I}$ is determined by the number of arrows connecting the $i$-th node to the $j$-th node with orientation. Note also that this quiver $\S_{M,N}$ is \textit{balanced}, i.e., the numbers of incoming arrows and the outgoing arrows are the same for all the nodes. When it is apparent, we will omit the subscript $\S_{M,N}$ of $\CalX_{\S_{M,N}}$ to keep the notation concise.

Then, we label the cluster variables for the seed $\S_{M,N}$ by (see Figure \ref{fig:quiverMN})
\begin{align}
    \bx^\times _{i,a} ,\; \bx^+ _{i,a} ,\qquad \begin{split} &i\in \{1,2,\cdots, N\} \\ &a\in \{1,2,\cdots, M\} \end{split}\;\;,
\end{align}
which are subject to the relations
\begin{align} \label{eq:clustercomm}
\begin{split}
    &\bx^\times _{i,a} \bx^+ _{j,b} = q^{2\left(\d_{i,j}\d_{a,b+1} +\d_{i,j+1}\d_{a,b} - \d_{i,j+1} \d_{a,b-1} - \d_{i,j} \d_{a,b}  \right) }    \bx^+ _{j,b} \bx^\times _{i,a}, \\
    &[\bx^\times _{i,a},\bx^\times _{j,b}] = [\bx^+ _{i,a}, \bx^+ _{j,b}]=0,
\end{split}
\end{align}
to form the quantum torus algebra $\CalO_q (\CalX)$.


\subsection{R-matrices from quantum Kasteleyn operator}
 A bipartite graph $\G = (B,W,E)$ on a compact oriented surface $S$ is a triad consisting of a finite set of black nodes $B$, a finite set of white node $W$, and a finite set of edges $E$ connecting the black nodes to the white nodes on the surface $S$. We choose the orientation of the edges from the black to the white. The edges intersect only at the nodes. 

For a balanced quiver $\Sigma$, there is dual bipartite graph $\G_\S = (B,W,E)$, where the nodes of the quiver $\S$ correspond to the faces of the bipartite graph $\G_\S$. The edges of the bipartite graph $\G_\S$ are dual to the edges of the quiver $\S$, so that the orientation of the former is determined by the orientation of the latter. We color the nodes of the bipartite graph $\G_\S$ in the way that all the edge are oriented from a black node to a white node.

The group $H_1 (\G_{\S},\BZ)$ is generated by the faces of the graph $\G_\S$, i.e, the cluster variables for the quiver $\S_{M,N}$, augmented by the cycles in $H_1 (S,\BZ)$. They coordinatize the moduli space of line bundles with connections on the graph $\G_{\S}$. The intersection pairing defines a Poisson structure on this moduli space, and its quantization precisely yields the quantum torus algebra $\CalO_q (\CalX)$ for the quiver $\S_{M,N}$ augmented by additional quantum torus algebra generated by the basis elements in $H_1 (S,\BZ)$ \cite{goncharov2011dimers}. \\

\begin{figure}[h!] \centering
\scalebox{0.7}{
\begin{tikzpicture}[square/.style={regular polygon,regular polygon sides=4},decoration={markings, 
    mark= at position 0.5 with {\arrow[scale=1.3]{stealth}}}
 ]

     \node at (-2,-4) [circle,fill,inner sep=3pt] (b1) {};
     \node at (-1.5,-4) {${}_{N,1}$};
     \node at (-2,-8) [circle,fill,inner sep=3pt] (b2) {};
     \node at (-1.3,-8) {${}_{N-1,1}$} ;
    \node at (-2,-14) [circle,fill,inner sep=3pt] (b3) {};
    \node at (-1.5,-14) {${}_{1,1}$} ;
    
    \node at (0,-2) [style={circle,draw},thick,inner sep=3pt] (w1) {};
    \node at (.5,-2) {${}_{N,1}$} ;
     \node at (0,-6) [style={circle,draw},thick,inner sep=3pt] (w2) {};
     \node at (.7,-6) {${}_{N-1,1}$} ;
     \node at (0,-12) [style={circle,draw},thick,inner sep=3pt] (w3) {};
     \node at (.5,-12) {${}_{1,1}$} ;

    \node at (2,-4) [circle,fill,inner sep=3pt](b4){};
    \node at (2.5,-4) {${}_{N,2}$} ;
     \node at (2,-8) [circle,fill,inner sep=3pt](b5){};
     \node at (2.7,-8) {${}_{N-1,2}$} ;
    \node at (2,-14) [circle,fill,inner sep=3pt](b6){};
    \node at (2.5,-14) {${}_{1,2}$} ;

     \node at (6,-2) [style={circle,draw},thick,inner sep=3pt](w4){};
     \node at (6.5,-2) {${}_{N,M}$} ;
     \node at (6,-6) [style={circle,draw},thick,inner sep=3pt](w5){};
     \node at (6.7,-6) {${}_{N-1,M}$} ;
     \node at (6,-12) [style={circle,draw},thick,inner sep=3pt](w6){};
     \node at (6.5,-12) {${}_{1,M}$} ;

    \draw[postaction={decorate},thick] (b1)--(-3,-3);
     \draw[postaction={decorate},thick] (b1)--(-3,-5);
     \draw[postaction={decorate},thick] (b1)--(w1);
    \draw[postaction={decorate},thick] (b1)--(w2);
     \draw[postaction={decorate},thick] (b2)--(w2);
      \draw[postaction={decorate},thick] (b2)--(-1,-9);
       \draw[postaction={decorate},thick] (b2)--(-3,-7);
          \draw[postaction={decorate},thick] (b2)--(-3,-9);
       \draw[postaction={decorate},thick] (b3)--(w3);
          \draw[postaction={decorate},thick] (b3)--(-1,-15);
     \draw[postaction={decorate},thick] (b3)--(-3,-15);
        \draw[postaction={decorate},thick] (b3)--(-3,-13);
    \draw[postaction={decorate},thick] (-1,-11)--(w3);
      \draw[postaction={decorate},thick] (-1,-1)--(w1);

      \draw[postaction={decorate},thick] (b4)--(w1);
      \draw[postaction={decorate},thick] (b4)--(w2);
  \draw[postaction={decorate},thick] (b5)--(w2);
  \draw[postaction={decorate},thick] (b5)--(1,-9);
    \draw[postaction={decorate},thick] (1,-11)--(w3);
  \draw[postaction={decorate},thick] (b6)--(w3);
  \draw[postaction={decorate},thick] (b6)--(1,-15);

   \draw[postaction={decorate},thick] (b4)--(3,-3);
   \draw[postaction={decorate},thick] (1,-1)--(w1);
   \draw[postaction={decorate},thick] (b5)--(3,-7);
\draw[postaction={decorate},thick] (b4)--(3,-5);
\draw[postaction={decorate},thick] (b5)--(3,-9);
\draw[postaction={decorate},thick] (b6)--(3,-13);
\draw[postaction={decorate},thick] (b6)--(3,-15);

\draw[postaction={decorate},thick] (7,-1)--(w4);
\draw[postaction={decorate},thick] (5,-1)--(w4);
\draw[postaction={decorate},thick] (7,-3)--(w4);
\draw[postaction={decorate},thick] (5,-3)--(w4);

\draw[postaction={decorate},thick] (7,-5)--(w5);
\draw[postaction={decorate},thick] (7,-7)--(w5);
\draw[postaction={decorate},thick] (5,-5)--(w5);
\draw[postaction={decorate},thick] (5,-7)--(w5);

\draw[postaction={decorate},thick] (7,-11)--(w6);
\draw[postaction={decorate},thick] (7,-13)--(w6);
\draw[postaction={decorate},thick] (5,-11)--(w6);
\draw[postaction={decorate},thick] (5,-13)--(w6);

          \node at (0,-9.5) [circle,fill,inner sep=1.5pt]{};
        \node at (0,-10) [circle,fill,inner sep=1.5pt]{};
     \node at (0,-10.5) [circle,fill,inner sep=1.5pt]{}; 
          
          \node at (6,-9.5) [circle,fill,inner sep=1.5pt]{};
        \node at (6,-10) [circle,fill,inner sep=1.5pt]{};
     \node at (6,-10.5) [circle,fill,inner sep=1.5pt]{};

        \node at (3.5,-4) [circle,fill,inner sep=1.5pt]{};
          \node at (4,-4) [circle,fill,inner sep=1.5pt]{};
            \node at (4.5,-4) [circle,fill,inner sep=1.5pt]{};

        \node at (3.5,-12) [circle,fill,inner sep=1.5pt]{};
          \node at (4,-12) [circle,fill,inner sep=1.5pt]{};
         \node at (4.5,-12) [circle,fill,inner sep=1.5pt]{};

    \node (t0) at (-3,-1) {};
    \node (t3) at (7,-1) {};
    \node (t4) at (-2,-1) {};
    \node (t5) at (2,-1) {};
     \node (t1) at (-1,-1) {};
     \node (t2) at (1,-1) {};

     \draw[dotted,thick,red] (t0) -- (t3);

     \node (bb1) at (-1,-15) {};
         \node (bb2) at (1,-15) {};
          \node (bb3) at (-2,-15) {};
     \node (bb4) at (2,-15) {};
          \node (bb5) at (-3,-15) {};
     \node (bb6) at (7,-15) {};

     \draw[dotted,thick,red] (bb5) -- (bb6);

       \draw[dotted,thick,blue] (t0) -- (bb5);
        \draw[dotted,thick,blue] (t3) -- (bb6);

            \node at (-2,-2) {$\bx^+ _{N,1}$};
 \node at (2,-2) {$\bx^+_{N,2}$};
\node at (0,-4) {$\bx^\times _{N,1}$};

  \node at (-2,-6) {$\bx^+ _{N-1,1}$};
       \node at (2,-6) {$\bx^+ _{N-1,2}$};
    \node at (0,-8) {$\bx^\times _{N-1,1}$};
      \node at (-2,-12) {$\bx^+ _{1,1}$};
      \node at (2,-12) {$\bx^+ _{1,2}$};
     \node at (0,-14) {$\bx_{1,1} ^\times$};
  \node at (6,-4) {$\bx^\times _{N,M}$};
 \node at (6.2,-8) {$\bx^\times _{N-1,M}$};    
    \node at (6,-14) {$\bx^\times _{1,M}$};


    \node[left] at (-3,-2.7) {$\sigma_N\eta_{N,M}$} ;
    \node[left] at (-3,-5.3) {$\sigma_N\xi_{N,M}$} ;
    \node[left] at (-3,-6.7) {$\sigma_{N-1}\eta_{N-1,M}$} ;
    \node[left] at (-3,-9.3) {$\sigma_{N-1}\xi_{N-1,M}$} ;
    \node[left] at (-3,-13) {$\sigma_1\eta_{1,M}$} ;
    \node[left] at (-3,-15.2) {$\kappa_M\sigma_1\xi_{1,M}$} ;

    \node[left] at (-.8,-2.7) {$\xi_{N,1}$} ;
    \node[left] at (-.8,-5.2) {$\eta_{N,1}$} ;
    \node[left] at (-.8,-6.7) {$\xi_{N-1,1}$} ;
    \node[left] at (-.8,-9.2) {$\eta_{N-1,1}$} ;
    \node[left] at (-.8,-10.8) {$\eta_{2,1}$} ;
    \node[left] at (-.8,-12.8) {$\xi_{1,1}$} ; 
    \node at (-.8,-15.2) {$\kappa_1\eta_{1,1}$} ;

    \node[right] at (.8,-2.7) {$\eta_{N,1}$} ;
    \node[right] at (.8,-5.2) {$\xi_{N,1}$} ;
    \node[right] at (.8,-6.7) {$\eta_{N-1,1}$} ;
    \node[right] at (.8,-9.2) {$\xi_{N-1,1}$} ;
    \node[right] at (.8,-10.8) {$\xi_{2,1}$} ;
    \node[right] at (.8,-12.8) {$\eta_{1,1}$} ; 
    \node at (.8,-15.2) {$\kappa_1\xi_{1,1}$} ;

    \node[right] at (2.8,-2.7) {$\xi_{N,2}$} ;
    \node[right] at (2.8,-5.2) {$\eta_{N,2}$} ;
    \node[right] at (2.8,-6.7) {$\xi_{N-1,2}$} ;
    \node[right] at (2.8,-9.2) {$\eta_{N-1,2}$} ;
    \node[right] at (2.8,-12.8) {$\xi_{1,2}$} ; 
    \node[right] at (2.8,-15.2) {$\kappa_2\eta_{1,2}$} ;

    \node[left] at (5.4,-3.2) {$\xi_{N,M}$} ;
    \node[left] at (5.4,-4.8) {$\eta_{N,M}$} ;
    \node[left] at (5.4,-7.2) {$\xi_{N-1,M}$} ;
    \node[left] at (5.4,-10.8) {$\eta_{2,M}$} ;
    \node[left] at (5.4,-13.2) {$\xi_{1,M}$} ; 

\end{tikzpicture}
}
\caption{The bipartite graph $\G_{M,N}$ dual to the quiver $\S_{M,N}$. The horizontal dotted lines (red) on the top and the bottom are identified, as well as the vertical dotted lines (blue) on the left and right. In turn, the bipartite graph is drawn on a torus $T^2$.}\label{fig:bipartite}
\end{figure}
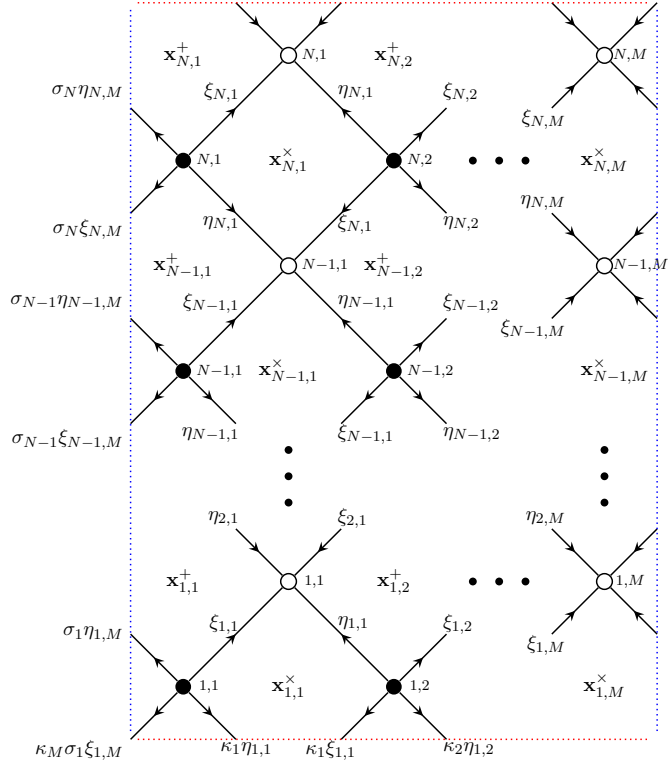

Now, let us turn to our example of the 5d BPS quiver $\S_{M,N}$. Its dual bipartite graph, called $\G_{M,N}$, is a graph on a torus $S=T^2$ drawn as in Figure \ref{fig:bipartite}.\\

In particular, we may fix the gauge of the line bundle on $\G_{M,N}$ so that the edge variables, i.e., the parallel transports of the connection along the edges, are assigned as in the Figure \ref{fig:bipartite}. Upon quantization, they satisfy the commutation relations given by
\begin{align}
    \begin{split}&\xi_{i,a} \eta_{j,b} = q^{ \d_{i,j} \d_{a,b}} \eta_{j,b} \xi_{i,a} \\ &[\s_i , \,\cdot\,] = [\k_a, \,\cdot\,] = 0 \end{split}\;, \begin{split} &i,j\in \{1,2,\cdots, N\} \\ &a,b\in \{1,2,\cdots, M\} \end{split}\;.
\end{align}
Then, the cluster variables are expressed by the corresponding faces in the bipartite graph,
\begin{align}
    \bx^\times _{i,a} = \xi_{i,a} \eta_{i,a}^{-1} \xi_{i,a} \eta_{i,a}^{-1}~,~
    \bx^+ _{i,a} = (\k_a\k_{a+1}^{-1})^{\d_{i,1}} (\s_i \s_{i-1}^{-1})^{\d_{a,N}} \eta_{i-1,a} \xi_{i-1,a+1}^{-1} \eta_{i,a+1} \xi_{i,a}^{-1},
\end{align}
It is straightforward to check that the commutation relation \eqref{eq:clustercomm} of the quantum torus algebra $\CalO_q (\CalX)$ is obeyed.\\

For a given bipartite graph $\G$, the quantum Kasteleyn operator $\fD_\G$ is defined by the $\vert W \vert \times \vert B \vert $ matrix (i.e., square matrix since $\vert W \vert = \vert B \vert$) whose entries are given by the edge variables connecting the black nodes to the white nodes. In our case of $\G_{M,N}$, the quantum Kasteleyn operator $ \fD_{M,N} \in \text{End}(\BC^N) \otimes \text{End}(\BC^M) \otimes \mathcal{O}_q(\mathcal{X}) \otimes \mathcal{O}_q (\BC^\times \times \BC^\times) $ is given by
\begin{align} \label{eq:qko}
\begin{split}
    &\mathfrak{D}_{M,N}= \sum_{i=1} ^N \sum_{a=1} ^M -\xi_{i,a} (E_{i,i} \otimes E_{a,a} + \k_a ^{\d_{i,N}} \s_i ^{\d_{a,1}} E_{i,i+1}\otimes E_{a-1,a}) \\
    & \qquad\qquad\qquad\quad +\eta_{i,a} (\k_a^{\d_{i,N}} E_{i,i+1} \otimes E_{a,a} +\s_{i} ^{\d_{a,1}} E_{i,i} \otimes E_{a-1,a} ),
\end{split}
\end{align}
where the matrices $(E_{i,j})_{i,j=1}^N$ and $(E_{a,b})_{a,b=1}^M$ are the standard basis elements of $\text{End}(\BC^N)$ and $\text{End}(\BC^M)$, respectively, defined with twisted periodicity,
\begin{align}
    E_{N,N+1} = Z E_{N,1} ~,~ E_{0,1} = X E_{M,1},
\end{align}
respectively. Here, $X$ and $Z$ are the generators of $H_1 (T^2,\BZ)$, giving rise to an additional quantum torus algebra $\mathcal{O}_q (\BC^\times \times \BC^\times )$ subject to the relation $XZ = q^2 ZX$ after quantization.

\subsubsection{R-matrices of quantum affine algebra of $\fgl(M)$}
We cast the quantum Kasteleyn operator \eqref{eq:qko} into the form of
\begin{align}
\begin{split}
    &\fD_{M,N} = \sum_{i=1}^N E_{i,i} \otimes A_i + E_{i,i+1} \otimes  C_i K^{\d_{N,i}},
\end{split}
\end{align}
where we defined the following $M \times M$ matrices valued in $\CalO_q (\CalX) \otimes \CalO_q (\BC^\times \times \BC^\times)$,
\begin{align}
\begin{split}
     & A_i(X) = \sum_{a=1}^M -\xi_{i,a} E_{a,a} + \eta_{i,a} \s_i ^{\d_{a,1}} E_{a-1,a}, \\
     &C_i(X) = \sum_{a=1}^M \eta_{i,a}E_{a,a} - \xi_{i,a} \s_i^{\d_{a,1}} E_{a-1,a}, \qquad i \in \{1,2,\cdots, N\} \\
     &K= \sum_{a=1}^M \k_a E_{a,a}
\end{split}
\end{align}
Let us consider the following $N$ vectors with $M$ components,
\begin{align}
    \psi_i = \sum_{a=1}^M \EQ_{i,a} e_a \in \text{End}(\BC^M) \otimes \CalO_q (\CalX) \otimes \CalO_q (\BC^\times \times \BC^\times) ,\qquad i \in \{1,2,\cdots, N\},
\end{align}
from which we form $\EQ = \sum_{i=1}^N e_i \otimes \psi_i$ that solves the quantum Kasteleyn operator, $\fD\EQ = 0$. The condition decomposes into $N$ constraints, 
\begin{align}
\begin{split}
        &  (ZK)^{\d_{i,N}} \psi_{[i+1]} = -C_i^{-1,L}(X) A_i(X) \psi_i =: L_i (X) \psi_i ,\qquad i \in \{1,2,\cdots, N\},
\end{split}
\end{align}
where we defined the left-inverse of $C_i$ by $C_i ^{-1,L} C_i = \text{id}$. Such a left-inverse may not always exist, but it does for $C_i$ in our case. It is given by
\begin{align}
    {C}_i^{-1,L} = \frac{1}{1-X\tilde\s_i} \sum_{a=0}^{M-1} (\bet_{i}^{-1} \bxi_i \boldsymbol\s_i \bU )^{a} \bet^{-1}_i,
\end{align}
where $\boldsymbol\eta_i = \text{diag}(\eta_{i,a})_{a=1}^M$, $\boldsymbol\xi_i = \text{diag}(\xi_{i,a})_{a=1}^M$, $\boldsymbol\s_i = \text{diag}(1,\dots,1,\s_{i})$, $\bU = \sum_{a=1}^M E_{a-1,a}$, and finally $\tilde\s_i = \s_i \prod_{a=1}^M \eta_{ia}^{-1} \xi_{ia}$. Note that $\s_i$ and $\tilde{\s}_i$ belong to the center of $\CalO_q (\CalX)$. In particular, the inverse in front is well-defined.

We thereby obtain an $M \times M$ matrix $L_i (X) = - C_i (X) ^{-1,L}  A_i (X)$ that transports the $i$-th vector to the $(i+1)$-th vector for each $i\in\{1,2,\cdots, N\}$. We claim that they are R-matrices of the quantum affine algebra of $\fgl(M)$, $U_q (\widehat{\fgl}(M))$. In this work, we will confirm this explicitly in the case of $M=2$.\\

When $M=2$, the $2 \times 2$ matrices $A_i $ and $C_i$ are given by
\begin{align}
    A_i = \begin{pmatrix}
        -\xi_{i,1} & \eta_{i,2} \\
        X\s_i\eta_{i,1} & -\xi_{i,2}
    \end{pmatrix}, \quad C_i = \begin{pmatrix}
        \eta_{i,1} & -\xi_{i,2} \\
        -X\s_i\xi_{i,1} & \eta_{i,2}
    \end{pmatrix}.
\end{align}
The left inverse of $C_i^{-1,L}$ of $C_i$ is computed to be
\begin{align}
 C_i^{-1,L} = (1 - X \sigma_iq^{-1}\eta_{i,1}^{-1}\xi_{j,1}\eta_{i,2}^{-1}\xi_{i,2})^{-1} \eta_{i,1}^{-1}\eta_{i,2}^{-1} \begin{pmatrix}
        \eta_{i,2} & \xi_{i,2} q^{-1} \\
        X \sigma_i \xi_{i,1} q^{-1} & \eta_{i,1}
    \end{pmatrix}.
\end{align}
The combination $\sigma_iq^{-1}\eta_{i,1}^{-1}\xi_{i,1}\eta_{i,2}^{-1}\xi_{i,2} = (m^+_{i-1})^{-1}$ is a zigzag loop in the bipartite graph, which belongs to the center of the quantum torus algebra $\CalO_q (\CalX)$. The inverse in front is therefore well-defined. Then, we compute 
\begin{align}
\begin{split}
  & (ZK)^{\d_{i,N}} \left( \frac{X}{m_{i-1}^+} -1\right) \psi_{[i+1]} \\
    & =  \begin{pmatrix}
        X \s_i \eta_{i,2}^{-1} \xi_{i,2} q^{-1} - \eta_{i,1}^{-1} \xi_{i,1}  &  (\eta_{i,1}^{-1}\eta_{i,2} - \eta_{i,1}^{-1} \eta_{i,2}^{-1} \xi_{i,2}^2 q^{-1} ) \\
        X \s_i( \eta_{i,2}^{-1} \eta_{i,1} - \eta_{i,1}^{-1} \eta_{i,2}^{-1} \xi_{i,1}^2 q^{-1} )& X \s_i \eta_{i,1}^{-1} \xi_{i,1} q^{-1} - \eta_{i,2}^{-1} \xi_{i,2}
    \end{pmatrix} \psi_i \\
    & = \frac{X}{m_{i-1}^+} L_i(X)  \psi_i,
\end{split}
\end{align}
where the $2 \times 2$ R-matrix is computed as
\begin{align}\label{eq:r2cl}
L_i(X) = L_i^+ - \frac{m_{i-1}^+}{X} L_i^-,\qquad i\in\{1,2,\cdots, N\},
\end{align}
with components explicitly given by
\begin{align}
\begin{split}
        & L_i^+ = \begin{pmatrix}
        \xi_{i1}^{-1}\eta_{i1} & 0 \\
        \xi_{i1}^{-1}\xi_{i2}^{-1}\eta_{i1}^2q - \xi_{i2}^{-1}\xi_{i1} & \xi_{i2}^{-1}\eta_{i2}
    \end{pmatrix}, \quad  L_i^- = \begin{pmatrix}
        \eta_{i1}^{-1}\xi_{i1} & \eta_{i1}^{-1}\eta_{i2}^{-1}\xi_{i2}^2q^{-1} - \eta_{i1}^{-1}\eta_{i2} \\ 0 & \eta_{i2}^{-1}\xi_{i2}
    \end{pmatrix}.
\end{split}
\end{align}
Now, it is straightforward to check that the map $ U_q (\widehat{\fgl}(2)) \to  \CalO_q (\CalX)$ defined by\footnote{We expect that the image is in fact universally Laurent, so that it is contained in the quantum cluster algebra for $\S_{M,N}$. We will present the detail in a separate work \cite{qclus}.}
\begin{align}
    T^+ (X) \mapsto L_i (X),\quad T^- (X) \mapsto -\frac{X}{m^+ _{i-1}} L_i (X),\quad  q \mapsto q
\end{align}
is an algebra homomorphism, factored through the evaluation map $\text{ev}_{m^+_{i-1}} : U_q (\widehat{\fgl}(2)) \to U_q ({\fgl}(2))$ (see appendix \ref{sec:qaa} for the presentation of $U_q(\widehat{\fgl}(2))$ and the definition of the evaluation map). It is crucial to note that the quantum parameters of the two sides, both of which are denoted by $q$, are exactly identified.
\\

We stress that these R-matrices $L_i (X)$ of the quantum affine algebra $U_q (\widehat{\fgl}(2))$ exactly recover with the ones obtained as the constraints on the correlation functions of the codimension-two and codimension-four observables in the 5d $\EN=1$ gauge theory in section \ref{subsec:rgl2}. Namely, we confirm an exact match between \eqref{eq:r2gauge} and \eqref{eq:r2cl}, provided with the dictionary,
\begin{align}\label{eq:dic}
\begin{split}
        & \eta_{i1}^{-1} \xi_{i1} = q_1^{-\frac12D_{u_{i-1}}} ~,~ \eta_{i2}^{-1} \xi_{i2} = \sqrt{\frac{m^-_{i-1}}{m^+_{i-1}}} q_1^{\frac12D_{u_{i-1}}+\frac12}~,~ \eta_{i2}^{-1} \eta_{i1} = \frac{1}{u_{i-1}} \left( q_1^{D_{u_{i-1}}} \frac{m_{i-1}^-}{m_{i-1}^+} - 1 \right)~,~ \\
    & \s_i = \frac{1}{\sqrt{m_{i-1}^+m_{i-1}^-}} ~,~ q = q_1 ^{\frac{1}{2}},
\end{split}
\end{align}
that maps the cluster variables to the $q_1$-difference operators in the monodromy parameters and the mass parameters.

\subsubsection{R-matrices of quantum affine algebra of $\fgl(N)$} \label{subsec:Nclus}

Due to the symmetry of the bipartite graph $\G_{M,N}$ (which is induced from the symmetry of the quiver $\S_{M,N}$) exchanging $M \leftrightarrow N$, we may alternatively express the quantum Kasteleyn operator \eqref{eq:qko} in the form of
\begin{align}
    \fD _{M,N}= \sum_{a=1}^M \tilde{A}_a \otimes E_{a,a} + \tilde{C}_a \tilde{K}^{\d_{a,1}} E_{a-1,a},
\end{align}
with the $N \times N$ matrices valued in $\CalO_q (\CalX) \otimes \CalO_q (\BC^\times \times \BC^\times)$ defined by
\begin{align}
\begin{split}
    &\tilde{A}_a(Z) = \sum_{i=1}^N \left( -\xi_{i,a} E_{i,i} + \eta_{i,a} \kappa_{a}^{\delta_{i,N}} E_{i,i+1} \right) 
    = -\tilde{\boldsymbol{\xi}}_a + \tilde{\boldsymbol{\eta}}_a \, \kappa_a \tilde{\boldsymbol{U}}, \\
    &\tilde{C}_a(Z) = \sum_{i=1}^N \left( \eta_{i,a} E_{i,i} - \xi_{i,a} \kappa_{a}^{\delta_{i,N}} E_{i,i+1} \right) 
    = \tilde{\boldsymbol{\eta}}_a - \tilde{\boldsymbol{\xi}}_a \, \kappa_a \tilde{\boldsymbol{U}}, \qquad a \in \{1,2,\dots,M\}, \\
    &\tilde{K} = \sum_{i=1}^N \sigma_i E_{i,i}m
\end{split}
\end{align}
where $\tilde\bet_a = \text{diag}(\eta_{i,a})_{i=1}^N$, $\tilde\bxi_a = \text{diag}(\xi_{i,a})_{i=1}^N$, $\bka_a = \text{diag}(1,\dots,1,\k_a)$, $\tilde\bU = \sum_{i=1}^N E_{i,i+1}$.
Let us re-arrange the solution $\EQ$ so that we have $N$ vectors with $M$ components,
\begin{align}
    \EQ = \sum_{a=1}^M {\psi}_a' \otimes e_a.
\end{align}
Let us further define $\P_a := \tilde{C}_a \psi_a'$. The condition $\fD\EQ = 0$ then translates into 
\begin{align}
\begin{split}
           & (X\tilde{K})^{\d_{a,M}} \Pi_{[a+1]} = - \tilde{A}_a(Z) \tilde{C}_a(Z)^{-1,L} \Pi_a =: \tilde{L}_a(Z) \Pi_a,\qquad a\in \{1,2,\cdots,M\}.
\end{split}
\end{align}
Here, 
\begin{align}
    \tilde{C}_a(Z)^{-1,L} = \frac{1}{1-Z\tilde\k_a} \sum_{j=0}^{N-1} (\tilde\bet_{a}^{-1} \tilde\bxi_a \bka_a \tilde\bU )^{j} \tilde\bet^{-1}_a 
\end{align}
is the left-inverse of $\tilde{C}_a (Z)$, satisfying $\tilde{C}_a ^{-1,L} \tilde{C}_a = \text{id}$. Note that $\tilde\k_a = \k_a \prod_{i=1}^{N} \eta_{i,a}^{-1} \xi_{i,a} $ belongs to the center of the quantum torus algebra $\CalO_q (\BC^\times \times \BC^\times)$, so that the inverse in front is well-defined.

Therefore, we construct an $N \times N$ matrix $\tilde{L}_a (Z) = - \tilde{A}_a (Z) \tilde{C}_a (Z)  ^{-1,L}$ that transports the $a$-th vector to the $(a+1)$-th vector for each $a \in \{1,2,\cdots, M\}$. Its components are explicitly computed to be
\begin{align} \label{eq:RclusterN}
\begin{split}
 \tilde{L}_a (Z) = \tilde{L}^+_a-\frac{1}{Z\tilde{\k}_a} \tilde{L}^- _a,\qquad a \in \{1,2,\cdots, M\},
\end{split}
\end{align}
with
\begin{align}
\begin{split}
        & \left(\tilde{L}_a^+\right)_{ij} = \d_{ij} \xi_{ia}\eta_{ia}^{-1} +\tilde{B}_{ia}\tilde{A}_{ja} \theta_{i>j}\\
    & \left(\tilde{L}_a^-\right)_{ij} = \d_{ij} \eta_{ia} \xi_{ia}^{-1} - \tilde{B}_{ia} \tilde{A}_{ja} \theta_{i<j}
\end{split}
\end{align}
Here, we defined the following elements in $\CalO_q (\CalX)$,
\begin{align}
\tilde{A}_{ia} = \prod_{l=1}^i (\xi_{ia}^{-1}\eta_{ia}) \eta_{ia}^{-1} ,\quad \tilde{B}_{ia} = \eta_{ia} (\xi_{ia}^{-1}\eta_{ia}-\eta_{ia}^{-1}\xi_{ia}) \prod_{l=1}^i (\eta_{la}^{-1}\xi_{la}),\quad i \in \{1,2,\cdots, N\},
\end{align}
for each $a\in \{1,2,\cdots, M\}$.\\

Consider the map $U_q (\widehat{\fgl}(N)) \to \CalO_q (\CalX)$ defined by
\begin{align}
    T^+ (Z) \mapsto \tilde{L}_a (Z),\quad T^- (Z) \mapsto - Z \tilde{\k}_a \tilde{L}_a (Z),\quad q\mapsto q^{-1}.
\end{align}
A straightforward computation shows that the map is an algebra homomorphism, factored through the evaluation map $\text{ev}_{\tilde{\k}_a ^{-1}} : U_q (\widehat{\fgl}(N)) \to U_q (\fgl(N)) $ (see appendix \ref{sec:qaa} for the presentation of $U_q (\widehat{\fgl}(N))$ and the definition of the evaluation map). In this sense, we identify $\tilde{L}_a$ as R-matrices intertwining evaluation modules for each $a\in \{1,2,\cdots, M\}$. It should be noted that the identification of the quantum parameters, both of which are denoted by $q$, involves a sign.\\

Let us restrict to the case of $M=2$. In this case, the two R-matrices $\tilde{L}_a (Z)$ recover the ones obtained from the fractional TQ equations, i.e., the constraints on the codimension-two and codimension-four
observables in section \ref{subsec:rmatN}. Indeed, we confirm the exact relation,
\begin{align}
\bL_a(Z) = \tilde\bet_1^{-1} \tilde{L}_a(Z) \tilde\bet_1,\qquad a=1,2,
\end{align}
between \eqref{eq:nngauge} and \eqref{eq:RclusterN} established by a conjugation of $\tilde\bet_1$, provided with the dictionary \eqref{eq:dic} mapping the cluster variables to the $q_1$-difference operators in the monodromy parameters and the mass parameters. 

Note that $\mathbf{L}_2 (Z) = \tilde\bet_1 ^{-1} \tilde{L}_2 (Z) \tilde\bet_1$ satisfies the same commutation relation as $\tilde{L}_2 (Z)$, since $\tilde{L}_2(Z)$ does not involve any $\tilde\bxi_1$ so that the conjugation does not affect the commutation relations among its entries. Thus, $\mathbf{L}_2(Z)$ is an R-matrix as we confirmed in section \ref{subsec:rmatN}. On the other hand, $\mathbf{L}_1 (Z) = \tilde\bet_1 ^{-1} \tilde{L}_1 (Z) \tilde\bet_1$ itself is \textit{not} an R-matrix since the non-trivial conjugation affects the commutation relations among its entries, compatible with what we found in section \ref{subsec:rmatN}. 

\section{Discussion} \label{sec:dis}
In this work, we have defined the canonical codimension-two defect in the 5d $\EN=1$ gauge theory by coupling a 3d $\EN=2$ sigma model. We defined the observables defined by insertion of the defect as the $Q$-observable or the $H$-observable, depending on whether the 3d $\EN=2$ theory is in the Coulomb phase or the Higgs phase. It was shown that both give rise to a $Q$-operator of XXZ spin chains, which are bispectral dual to each other. We derived the exact the analytic constraints on the correlation function of the defects, and rearranged them into the R-matrices of the quantum affine algebra involving bi-infinite evaluation modules. Using this construction, we showed that the monodromy codimension-two defect provides a basis of the space of states by its vevs, which simultaneously diagonalizes the $Q$-operators and the quantum Hamiltonians of the XXZ spin chain.

We believe our investigation can be further extended in several interesting directions.

\paragraph{Separation of variables}
 In our previous work \cite{Jeong:2024onv}, we established the quantum separation of variables for the $\fgl(N)$ Gaudin model and the $\fgl(2)$ XXX spin chain through the transition of the monodromy surface defect into a two-dimensional sigma model coupled to the 4d $\EN=2$ theory. This transition gives rise to an integral transformation for the common eigenfunction of the quantum Hamiltonians, where the integrand is written as a simple product of the ($\hbar$-)oper solutions, i.e., the vevs of the $Q$-observable or the $H$-observable in the 4d $\EN=2$ gauge theory.

We expect that the K-theoretic uplift of the procedure would lead to the quantum separation of variables for the $\fgl(2)$ XXZ spin chain and the $\fgl(N)$ XXZ spin chain studied in this work. Namely, the separation of variables transformation would be realized through the transition of the monodromy codimension-two defect into a 3d sigma model coupled to the 5d $\EN=1$ gauge theory. The integral transformation for the common eigenfunction of the quantum Hamiltonians would involve a simple product of the $q$-oper solutions, i.e., the vevs of the $Q$-observable or the $H$-observable in the 5d $\EN=1$ gauge theory. Note that in this 5d uplift, both $Q$-observable and $H$-observable give rise to $Q$-operators for XXZ spin chains, which are bispectral dual to each other.

In fact, when $q_2 \neq 1$ we expect to obtain a $q$-deformation of the KZ/BPZ correspondence \cite{Frenkel95,Stoyanovsky2000ARB,Ribault:2005wp}. The correspondence would relate the solutions to the quantum Knizhnik-Zamolodchikov (qKZ) equation to the degenerate correlation functions in the $q$-deformed $W$-algebra.

\paragraph{Quantum Knizhnik-Zamolodchikov equation and stable envelopes}

It was shown in \cite{Nekrasov:2021tik} that the vev of the monodromy surface defect in the 4d $\EN=2$ $U(N)$ gauge theory yields solutions to the Knizhnik-Zamolodchikov (KZ) equation for the affine Kac-Moody algebra $\widehat{\fgl}(N)$. Moreover, it was further shown in \cite{Jeong:2021bbh} that the solutions to the KZ equations with additional degenerate vertex operators can be obtained by inserting canonical surface defects intersecting the monodromy surface defect at the origin. In our 5d uplift setting, we expect a $q$-deformation of the statement to hold. Specifically, we propose that the vev of the monodromy codimension-two defect corresponds to a solution of the quantum Knizhnik-Zamolodchikov (qKZ) equation for the quantum affine algebra $U_q (\widehat{\fgl}(N))$ \cite{frenkel1992quantum}.\footnote{Extending the results of \cite{Jeong:2020uxz,Nekrasov:2020qcq}, the blowup formula for the intersecting codimension-two defect configuration is expected to yield a 5d $\EN=1$ gauge theoretical construction of the tau function for $q$-isomonodromy problems. See \cite{Jeong:2017mfh,Bonelli:2017gdk,Grassi:2019coc} for related works.}

When the associated modules are finite-dimensional evaluation modules, solutions to the qKZ equation can be constructed by inserting the images of the K-theoretic stable envelopes into the contour integral formula for the vortex partition function of a 3d $\EN=4$ theory \cite{Aganagic:2016jmx,Aganagic:2017smx}. The construction was extended to the highest-weight Verma modules in \cite{Tamagni:2023wan,Haouzi:2023doo} by adapting it to the context of 3d $\EN=2$ theory. We expect that our approach $-$ constructing qKZ solutions from the monodromy codimension-two defect $-$ will provide a further generalization of the K-theoretic stable envelopes to include bi-infinite evaluation modules. Such a development would offer new insights into the structure of the quantum $q$-Langlands correspondence in this broader setting.

\paragraph{$Q$-observable from sequence of mutation}
In this work, the $Q$-operators of the XXZ spin chain have been constructed by coupling a 3d $\EN=2$ theory to the 5d $\EN=1$ bulk theory. A similar construction for the relativistic Toda lattice was carried out in was carried out in \cite{Lee:2023wbf}. 

In \cite{schrader2018b}, the $Q$-operators for the open Toda lattice is realized as a sequence of mutations on the quantum cluster algebra associated with a specific cluster seed. This raises a natural question if such a cluster construction of $Q$-operators can be extended to other integrable models (closed Toda lattice, XXZ spin chain, etc.) Interestingly, in the context of 4d $\EN=1$ quiver gauge theories, cluster mutations are realized as Seiberg dualities. Understanding the relation between different construction of the $Q$-operators may lead to a deeper and more unified understanding of the interplay among 3d–5d supersymmetric gauge theories, 4d $\EN=1$ quiver gauge theories, and cluster algebra.

\appendix

\section{Dual $Q$-observable} \label{sec:dualQ}
For our main example of the 5d $\EN=1$ $U(N)$ gauge theory with $N$ fundamental and $N$ anti-fundamental hypermultiplets, there are two independent codimension-two $Q$-observables. In section \ref{subsubsec:Qobs}, we presented one of them $\bQ(X)$ as an equivariant K-theory class on the moduli space of instantons. Here, we give a construction of the other, denoted by $\tilde{\bQ}(X)$, and derive the TQ equation satisfied by its vacuum expectation value. 

\subsection{Observable expression} \label{subsec:dualq}
Consider the intersecting stacks of D3-branes on the orbifold $X = \BC^4 / \G_{34}$, where the action of $\G_{34} = \BZ_3$ is assigned by the weight $(0,0,+1,-1)$. The moduli space of D$(-1)$-branes on the stack of D3-branes gives rise to the moduli space of spiked instantons. The equivariant Chern characters of the framing bundles are written as
\begin{align}\label{eq:go}
\begin{split}
    \bN_{12} & = \sum_{\alpha=1}^N a_\alpha \CalR_0 + m_\alpha^- q_4^{-1} \CalR_1 + m_\alpha^+ q_3^{-1} \CalR_2 \\
    \bN_{13} & = X' q_1\CalR_1 \\
    \bN_{34} & = \varnothing \quad \text{or} \quad X \CalR_0
\end{split}
\end{align}
To get the observable $\tilde{\bQ}(X)$ only, we take $\bN_{34} = \varnothing$ first. 

We take the decoupling limit $\kq_1=\kq_2=0$ and denote $\kq=\kq_0$. The gauge origami partition function is obtained by picking up the weight zero part of under the action of $\G_{34}$ induced by its action on the spacetime. Note that $\bK_{13}$ can grow only along the $q_1$-direction, yielding
\begin{align}
    P_1\bK_{13} = X'(1 - q_1^{k_{13}}) \CalR_0.
\end{align}
The decoupling limit gives us a reduced 13-34 1-loop contribution to
\begin{align}
    -\hat{q}_{13} \bN_{34} \bN_{13}^* - \hat{q}_{34} \bN_{13} \bN_{34}^* .
\end{align}
The gauge origami partition function is given by 
\begin{align} \label{eq:gop}
\begin{split}
    \CalZ_\text{GO} = \sum_{\l_{12}} \sum_{k_{13}=0}^\infty \sum_{k_{34}=0}^1 &  \kq^{|\l_{12}|+k_{13}+k_{34}} \\
    & \hat{a} \left[ - \frac{P_3 \bS_{12}\bS_{12}^*}{P_{12}^*} - \frac{P_2(\bS_{13}\bS_{13}^*-\bN_{13}\bN_{13}^*)}{P_{13}} - \frac{P_1(\bS_{34}\bS_{34}^*-\bN_{34}\bN_{34}^*)}{P_{34}^*} \right. \\
    & - q_{12}\bS_{34} \bS_{12}^* + q_2 P_4 \frac{\bS_{13}\bS_{12}^*}{P_1^*} - q_{13} \bN_{34}\bN_{13}^* - q_{34}\bN_{13}\bN_{34}^* \\
    & \left. + P_2P_4 \bN_{13}\bK_{34}^* + P_2P_4 \bN_{34} \bK_{13}^* - P_1P_2P_3P_4 \bK_{34}\bK_{13}^* \right]^{\G_{34}}
\end{split}
\end{align}
The 12-13 cross term creates
\begin{align}
    \tilde{Q}_{k_{13}}(X')[\bl] := \frac{Q(X') [\bl]M(X'q_1^{k_{34}})}{Q(X'q_1^{k_{34}})[\bl] Q(X'q_2q_1^{k_{34}+1}) [\bl]}
\end{align}
with $M(X)=M(Xq_1^{-1})P(X)$. It satisfies
\begin{align}
    \frac{\tilde{Q}_{k_{13}}(X)[\bl]}{\tilde{Q}_{k_{13}+1}(Xq_1^{-1})[\bl]} = \EY(X)[\bl]. 
\end{align}
The so-obtained $\tilde{Q}$-observable, as an equivariant K-theory class, is written as
\begin{align}
    \tilde{Q}(X)[\bl] = \kq^{\frac{\log X}{\log q_1}} \sum_{d=0}^\infty \kq^d q_2^{-\frac{d}{2}} \frac{(q_1q_2;q_1)_d}{(q_1;q_1)_d} \tilde{Q}_d(X)[\bl].
\end{align}

\subsection{TQ equation} \label{subsec:dualtq}
To derive the TQ equation that the vev of $\tilde{\bQ}(X)$ satisfies, we consider the case $\bN_{34} = X \CalR_0$ in the gauge origami configuration \eqref{eq:go}. The gauge origami partition function \eqref{eq:gop} computes
\begin{align}
\begin{split}
    \kq^{\frac{\log Xq_2^{-1}}{\log q_1}} \sum_{d=0}^\infty \kq^d q_2^{-\frac{d}{2}} \frac{(q_1q_2;q_1)_d}{(q_1;q_1)_d} & \left[ \left( \sqrt{\frac{X'}{X}} - \sqrt{\frac{X}{X'}} \right) q_2^{\frac12} \frac{1-\frac{X'q_1^d}{Xq_2}}{1-\frac{X'q_1^d}{X}} Y(Xq_1q_2) \tilde{Q}_d(X') \right. \\
    & + \kq \left. \left( \sqrt{\frac{X'}{Xq_2}} - \sqrt{\frac{Xq_2}{X'}} \right) q_2^{-\frac12} \frac{1-\frac{X'q_2q_1^{d+1}}{X}}{1-\frac{X'q_1^{d+1}}{X}} \frac{P(X)}{Y(X)} \tilde{Q}_d(X') \right]
\end{split}
\end{align}

Let us take $X=X'$. Then,
\begin{align}
\begin{split}
    & \tilde{T}(X=X',X')\tilde{Q}(X') \\
    & = - \kq^{\frac{\log X'}{\log q_1}} q_2^{\frac12} \left( 1 - q_2^{-1} \right) Y(X'q_1q_2) \tilde{Q}_0(X') \\
    & \quad + \kq^{\frac{\log X'}{\log q_1}} \sum_{d=0}^\infty \kq^{d} q_2^{-\frac{d}{2}} \frac{(q_1q_2;q_1)_d}{(q_1;q_1)_d} \kq \left( q_2^{-\frac12} - q_2^{\frac12} \right) q_2^{-\frac12} \frac{1-q_2q_1^{d+1}}{1-q_1^{d+1}} \frac{P(X')}{Y(X')} \tilde{Q}_{d}(X') \\
    & = - \left( q_2^{\frac12} - q_2^{-\frac12} \right) \kq^{\frac{\log X'}{\log q_1}} Y(Xq_1q_2) \frac{M(X')}{Q(X'q_1q_2)} \\
    & \quad - \left( q_2^{\frac12} - q_2^{-\frac12} \right) P(X') \kq^{\frac{\log X'q_1^{-1}}{\log q_1}+1} \sum_{d=0}^\infty \kq^{d+1} q_2^{-\frac{d+1}{2}} \frac{(q_1q_2;q_1)_{d+1}}{(q_1;q_1)_{d+1}}   \tilde{Q}_{d+1}(X'q_1^{-1}) \\
    & = - \left( q_2^{\frac12} - q_2^{-\frac12} \right) \kq^{\frac{\log X'}{\log q_1}}  \frac{M(X')}{Q(X'q_2)} \\
    & \quad  - \left( q_2^{\frac12} - q_2^{-\frac12} \right) \kq P(X') \left[ \tilde{Q}(X'q_1^{-1}) - \kq^{\frac{\log X'q_1^{-1}}{\log q_1}}\tilde{Q}_0(X'q_1^{-1}) \right] \\
    & = - \left( q_2^{\frac12} - q_2^{-\frac12} \right) \kq P(X') \tilde{Q}(X'q_1^{-1}) .
\end{split}
\end{align}
Also, if we take instead $Xq_2=X'$ we get 
\begin{align}
\begin{split}
    & \tilde{T}(X=X'q_2^{-1};X') \tilde{Q}(X') \\
    & = \kq^{\frac{\log X'}{\log q_1}} \sum_{d=1}^\infty \kq^d q_2^{-\frac{d}{2}} \frac{(q_1q_2;q_1)_d}{(q_1;q_1)_d} \left( q_2^{\frac12} - q_2^{-\frac12} \right) q_2^{\frac12} \frac{1-q_1^{d}}{1-q_2q_1^d} Y(X'q_1) \tilde{Q}_d(X') \\
    & = \left( q_2^{\frac12} - q_2^{-\frac12} \right) \kq \kq^{\frac{\log X'}{\log q_1}} \sum_{d=0}^\infty \kq^d q_2^{-\frac{d}{2}}  \frac{(q_1q_2;q_1)_d}{(q_1;q_1)_d} \tilde{Q}_d(X'q_1) \\
    & = \left( q_2^{\frac12} - q_2^{-\frac12} \right) \tilde{Q}(Xq_1) .
\end{split}
\end{align}
By subtracting the two equations from each other, we get the TQ equation, 
\begin{align}
\begin{split}
    \left\langle \tilde\ET(X) \tilde{Q}(X) \right\rangle & = \langle \tilde{Q}(Xq_1) \rangle + \kq P(X) \langle \tilde{Q}(Xq_1^{-1}) \rangle  ,
\end{split}
\end{align}
satisfied by the vev of the dual $Q$-observable and its correlation function with the observable $\tilde{\ET}(X)$, which we defined by
\begin{align}
     \langle \tilde\ET(X') \tilde{Q}(X') \rangle = \left \langle \frac{ \tilde{T}(X=X'q_2^{-1};X') \tilde{Q}(X')-\tilde{T}(X=X';X') \tilde{Q}(X')}{q_2^{\frac12}-q_2^{-\frac12}} \right \rangle.
\end{align}

In the $q_2 \to 1$ limit with limit shape $\bLm$: 
\begin{align}
\begin{split}
    \tilde{\ET}(X)[\bLm] \tilde{Q}(X)[\bLm]
    & = \left[ Y(Xq_1)[\bLm] + \kq \frac{P(X)}{Y(X)[\bLm]} \right] \tilde{Q}(X) [\bLm] = \ET(X)[\bLm] \tilde{Q}(X)[\bLm]
\end{split}
\end{align}
where $\ET(X)$ is for the original $Q(X)$. We conclude $\tilde\ET(X)[\bLm] = \ET(X)[\bLm]$.

\subsection{Fractional TQ equation}
Now, we consider the intersecting stacks of D3-branes on the orbifold $X = \BC ^4 / (\G_{24}\times \G_{34})$, where the actions of $\G_{24} = \BZ_N$ and $\G_{34}= \BZ_3$ are assigned by the weights $(0,+1,0,-1)$ and $(0,0,+1,-1)$, respectively.

Correspondingly, we modify the equivariant Chern character of the framing bundle by
\begin{align}
    \hat\bN_{13} = \sum_{\o=0}^{N-1} X'_\o q_1 \hat{q}_2^\o \CalR_0 \otimes \fR_\o,
\end{align}
where $\fR_\o$ is the one-dimensional representation of $\BZ_N$ of weight $\o \in [N]$.

We denote $\bX'=\{X_0',\dots,X'_{N-1}\}$. 
The instanton in 13 space is $N$ single column Young diagrams $\bd=(d_0,\dots,d_{N-1})$: 
\begin{align}
    \hat\bK_{13} = \sum_{\o=0}^{N-1} X'_\o \hat{q}_2^\o \sum_{j=1}^{d_\o} q_1^j ~ \CalR_0 \otimes \fR_\o 
\end{align}
The 13-13 term is given by
\begin{align}
\begin{split}
    \fZ_\bd(\bX') & = \hat{a} \left[ \hat{P}_2 \hat{P}_4 \hat\bN_{13} \hat\bK_{13}^* - \hat{P}_1\hat{P} _2 \hat{P}_3 \hat\bK_{13} \hat\bK_{13}^* \right]^{\BZ_3 \times \BZ_N} \\
    & = \hat{a} \left[ \hat{P}_2 (\hat\bN_{13}-P_1 \hat\bK_{13} ) \hat\bK_{13}^* \right] \\
    & = \hat{a} \left[ \sum_{\o=0}^{N-1} \sum_{j=1}^{d_\o} q_1^j - \sum_{\o=0}^{N-2} \sum_{j=1}^{d_{\o+1}} \frac{X'_\o}{X'_{\o+1}} q_1^{d_\o+1-j} - q_2 \sum_{j=1}^{d_0} \frac{X'_{N-1}}{X'_0} q_1^{d_{N-1}+1-j} \right] \\
    & = \prod_{\o=0}^{N-1} \frac{\prod\limits_{j=1}^{d_{[\o+1]}} \left( \sqrt{\frac{X'_\o}{X'_{[\o+1]}}} q_1^{\frac{d_\o+1-j}{2}} q_2^{\frac{\delta_{\o,N-1}}{2}} - \sqrt{\frac{X'_{[\o+1]}}{X'_\o}} q_1^{-\frac{d_\o+1-j}{2}} q_2^{-\frac{\delta_{\o,N-1}}{2}} \right) }{ \prod\limits_{j=1}^{d_\o} \left(q_1^{\frac{j}{2}} - q_1^{-\frac{j}{2}} \right) }
\end{split}
\end{align}
The 12-13 contribution is
\begin{align}
\begin{split}
    & \hat{a} \left[ \hat{q}_2 \hat{P}_4 \frac{\hat\bS_{13}\hat\bS_{12}^*}{P_1^*} \right]^{\BZ_3\times \BZ_N} \\
    & = \hat{a} \left[ \sum_{\o=0}^{N-1} X'_\o \left( q_1^{d_\o+1} q_2^{\d_{\o,N-1}} \frac{\bS_{12,[\o+1]}^*}{P_1^*} +  (q_1^{d_\o}-1) \frac{(\bS_{12,\o}-M_\o^-)^*}{P_1^*} - q_1^{d_\o} \frac{(M_\o^+)^*}{P_1^*} \right) \right] \\
    & = \prod_{\o=0}^{N-1} \frac{Q_\o(X'_\o) M_\o(X_\o'q_1^{d_\o})}{Q_\o(X'_\o q_1^{d_{\o}})Q_{[\o+1]}(X'_\o q_1^{d_\o+1}q_2^{\d_{\o,N-1}}) } \\
    & := \tilde\EQ_\bd(\bX').
\end{split}
\end{align}
It satisfies
\begin{align}
\begin{split}
    \tilde\EQ_\bd(\bX') Y_\o(X'_\o q_1) = \tilde\EQ_{\bd-e_\o}(\bX'q_1^{e_\o})~,~ \tilde\EQ_\bd(\bX') \frac{1}{Y_\o(X'_\o)} = \tilde\EQ_{\bd+e_\o}(\bX' q_1^{-e_\o})~,
\end{split}
\end{align}
with the notation
\begin{align}
\begin{split}
    \bd+e_\o & = \{d_0,\dots,d_{\o-1},d_\o+1,d_{\o+1},\dots,d_{N-1}\}~, \\
    \bX'q_1^{e_\o} & = \{ X'_0,\dots,X'_{\o-1},X'_{\o}q_1,X'_{\o+1},\dots,X'_{N-1} \} ~.
\end{split}
\end{align}
We define the dual fractional $Q$-observable as in the absence of 34-brane by
\begin{align}
    \tilde\EQ(\bX') = \sum_\bd \prod_{\o=0}^{N-1} \kq_\o^{\frac{\log X'_\o}{\log q_1} + d_\o}  \fZ_\bd(\bX') \tilde\EQ_\bd(\bX')
\end{align}
Last but not least is the 13-34 term, when $\bK_{34}=0$: 
\begin{align}
\begin{split}
    & \hat{a} \left[ -X'_\o X^{-1} + X(X'_\o)^{-1} q_1^{-d_\o} - X(X'_{[\o+1]})^{-1} q_1^{-d_{[\o+1]}}q_2^{\d_{\o,N-1}} \right] \\
    & = \left( \sqrt{\frac{X'_\o}{X}} - \sqrt{\frac{X}{X'_\o}} \right)  \frac{ \sqrt{\frac{Xq_2^{\d_{\o,N-1}}}{X'_{[\o+1]}q_1^{d_{[\o+1]}}}} - \sqrt{\frac{X'_{[\o+1]}q_1^{d_{[\o+1]}}}{Xq_2^{\d_{\o,N-1}}}} }{ \sqrt{\frac{X}{X'_\o q_1^{d_\o}}} -\sqrt{\frac{X'_\o q_1^{d_\o}}{X}}} 
\end{split}
\end{align}
when $\bK_{34}=\bN_{34}$: 
\begin{align}
\begin{split}
    & \hat{a} \left[ - X'_{[\o+1]}X^{-1}q_2^{-\d_{\o,N-1}} + X(X'_\o)^{-1} q_1^{-d_\o-1} - X(X'_{[\o-1]})^{-1} q_1^{-d_{[\o-1]}-1}q_2^{-\d_{\o,0}} \right] \\
    & = \left( \sqrt{\frac{X'_{\o+1}}{Xq_2^{\d_{\o,N-1}}}} - \sqrt{\frac{Xq_2^{\d_{\o,N-1}}}{X'_{[\o+1}}} \right)  \frac{ \sqrt{\frac{X}{X'_{[\o-1]}q_1^{d_{[\o-1]}+1}q_2^{\d_{\o,0}}}} - \sqrt{\frac{X'_{[\o-1]}q_1^{d_{[\o-1]}+1}q_2^{\d_{\o,0}}}{X}} }{ \sqrt{\frac{X}{X'_\o q_1^{d_\o+1}}} -\sqrt{\frac{X'_\o q_1^{d_\o+1}}{X}}}
\end{split}
\end{align}
The gauge origami partition function gives the $qq$-character
\begin{align}
\begin{split}
    \CalZ_\text{GO}
     = & \sum_{\hat\l_{12}} \sum_{\bd} \prod_{\o=0}^{N-1} \kq_\o^{k_\o^{12}+d_\o} \hat\CalZ_{12}(\ba,\bm) \fZ_\bd(\bX') \\
     & \left[ \left( \sqrt{\frac{X'_\o}{X}} - \sqrt{\frac{X}{X'_\o}} \right)  \frac{ \sqrt{\frac{Xq_2^{\d_{\o,N-1}}}{X'_{[\o+1]}q_1^{d_{[\o+1]}}}} - \sqrt{\frac{X'_{[\o+1]}q_1^{d_{[\o+1]}}}{Xq_2^{\d_{\o,N-1}}}} }{ \sqrt{\frac{X}{X'_\o q_1^{d_\o}}} -\sqrt{\frac{X'_\o q_1^{d_\o}}{X}}} \times Y_{[\o+1]}(Xq_1q_2^{\d_{\o,N-1}})\right. \\
     & \left. + \kq_\o \left( \sqrt{\frac{X'_{[\o+1]}}{Xq_2^{\d_{\o,N-1}}}} - \sqrt{\frac{Xq_2^{\d_{\o,N-1}}}{X'_{[\o+1]}}} \right)  \frac{ \sqrt{\frac{X}{X'_{[\o-1]}q_1^{d_{[\o-1]}+1}q_2^{\d_{\o,0}}}} - \sqrt{\frac{X'_{[\o-1]}q_1^{d_{[\o-1]}+1}q_2^{\d_{\o,0}}}{X}} }{ \sqrt{\frac{X}{X'_\o q_1^{d_\o+1}}} -\sqrt{\frac{X'_\o q_1^{d_\o+1}}{X}}}\frac{P_\o(X)}{Y_\o(X)}  \right] \tilde\EQ_\bd(\bX').
\end{split}
\end{align}
Denote 
\begin{align}
     \prod_{\o=0}^{N-1} \kq_\o^{\frac{\log X'_\o}{\log q_1}} \times \CalZ_\text{GO} = \langle \tilde{T}_\o(X;\bX') \tilde\EQ(\bX') \rangle 
\end{align}
\begin{align}
\begin{split}
    \tilde{T}_\o(X;\bX') \tilde\EQ(\bX')
    & = \sum_{\bd} \prod_{\o=0}^{N-1} \kq_\o^{\frac{\log X'_\o}{\log q_1}+d_\o} \fZ_\bd (\bX') \\
    & \left[ \left( \sqrt{\frac{X'_\o}{X}} - \sqrt{\frac{X}{X'_\o}} \right)  \frac{ \sqrt{\frac{Xq_2^{\d_{\o,N-1}}}{X'_{[\o+1]}q_1^{d_{[\o+1]}}}} - \sqrt{\frac{X'_{[\o+1]}q_1^{d_{[\o+1]}}}{Xq_2^{\d_{\o,N-1}}}} }{ \sqrt{\frac{X}{X'_\o q_1^{d_\o}}} -\sqrt{\frac{X'_\o q_1^{d_\o}}{X}}} \times Y_{[\o+1]}(Xq_1q_2^{\d_{\o,N-1}})\right. \\
     & \left. + \kq_\o \left( \sqrt{\frac{X'_{\o+1}}{Xq_2^{\d_{\o,N-1}}}} - \sqrt{\frac{Xq_2^{\d_{\o,N-1}}}{X'_{[\o+1}}} \right)  \frac{ \sqrt{\frac{X}{X'_{[\o-1]}q_1^{d_{[\o-1]}+1}q_2^{\d_{\o,0}}}} - \sqrt{\frac{X'_{[\o-1]}q_1^{d_{[\o-1]}+1}q_2^{\d_{\o,0}}}{X}} }{ \sqrt{\frac{X}{X'_\o q_1^{d_\o+1}}} -\sqrt{\frac{X'_\o q_1^{d_\o+1}}{X}}} \frac{P_\o(X)}{Y_\o(X)}  \right] \tilde\EQ_\bd(\bX')
\end{split}
\end{align}
First we take $X=X'_{[\o+1]}q_2^{-\d_{\o,N-1}}$:
\begin{align}
\begin{split}
    & \tilde{T}_\o(X=X'_{[\o+1]}q_2^{-\d_{\o,N-1}};\bX') \tilde\EQ(\bX') \\
    & = \left( \sqrt{\frac{X_\o' q_2^{\d_{\o,N-1}}}{X'_{[\o+1]}}} - \sqrt{\frac{X'_{[\o+1]}}{X'_\o q_2^{\d_{\o,N-1}}}}  \right) \tilde\EQ(\bX'q_1^{e_{[\o+1]}})
\end{split}
\end{align}
Second we take $X=X'_\o$:
\begin{align}
\begin{split}
    & \tilde{T}_\o(X=X'_\o;\bX') \tilde\EQ(\bX') \\ 
    & = \left( \sqrt{\frac{X'_{[\o+1]}}{X'_\o q_2^{\d_{\o,N-1}}}} - \sqrt{\frac{X_\o' q_2^{\d_{\o,N-1}}}{X'_{[\o+1]}}} \right) \kq_\o P_\o(X_\o') \tilde\EQ(\bX'q_1^{-e_\o})
\end{split}
\end{align}
By subtracting the two equations, we finally arrive at the fractional TQ equation,
\begin{align}
    \langle \tilde\ET_\o(\bX) \tilde\EQ(\bX') \rangle = \langle \tilde\EQ(\bX'q_1^{e_{[\o+1]}}) \rangle + \kq_\o P_\o(X'_\o) \langle \EQ(\bX'q_1^{-e_{\o}}) \rangle ,
\end{align}
satisfied by the correlation functions of the dual $Q$-observable and the codimension-four observables. Here, we defined
\begin{align}
    \tilde\ET_\o(\bX') = \frac{\tilde{T}_\o(X=X'_{[\o+1]}q_2^{-\d_{\o,N-1}};\bX')-\tilde{T}_\o(X=X'_\o;\bX')}{\sqrt{\frac{X_\o' q_2^{\d_{\o,N-1}}}{X'_{[\o+1]}}} - \sqrt{\frac{X'_{[\o+1]}}{X'_\o q_2^{\d_{\o,N-1}}}}}.
\end{align}

To calculate $\tilde{\ET}_\o(\bX')$, we expand $T_\o(X;\bX')$ in $X>\!\!>1$ and $X<\!\!<1$ to obtain
\begin{align}
    \tilde{T}_\o(X,\bX') = \tilde\spadesuit_\o X + \tilde\diamondsuit_\o + \frac{\tilde\clubsuit_\o}{X}
\end{align}
with
\begin{subequations}
    \begin{align}
    -\tilde\spadesuit_\o = & ~ (X'_{[\o+1]})^{-\frac12} a_{[\o+1]}^{-\frac12} q_1^{\frac{1}{2}(k_\o-k_{\o+1}+1) } q_1^{\frac12(d_\o-d_{[\o+1]})}q_2^{\d_{\o,N-1}} \\
    & + \hat\kq_\o (X'_{[\o+1]})^{-\frac12} (X'_\o)^{\frac12} (X'_{[\o-1]})^{-\frac12} a_\o^{\frac12}(m_\o^+)^{-\frac12}(m_\o^-)^{-\frac12} q_2^{\frac{1}{2}(\d_{\o,N-1}-\d_{\o,0})} q_1^{-\frac{1}{2}(k_{\o-1}-k_\o)+\frac{1}{2}(d_\o-d_{[\o-1]})} \nonumber \\
    -\tilde\clubsuit_\o = & ~ (X'_{[\o'+1]})^{\frac12} a_{[\o+1]}^{\frac12} q_1^{-\frac{1}{2}(k_\o-k_{\o+1}+1) } q_1^{-\frac12(d_\o-d_{[\o+1]})}q_2^{-\d_{\o,N-1}} \\
    & + \hat\kq_\o (X'_{[\o+1]})^{\frac12} (X'_\o)^{-\frac12} (X'_{[\o-1]})^{\frac12} a_\o^{-\frac12}(m_\o^+)^{\frac12}(m_\o^-)^{\frac12} q_2^{-\frac{1}{2}(\d_{\o,N-1}-\d_{\o,0})} q_1^{\frac{1}{2}(k_{\o-1}-k_\o)-\frac{1}{2}(d_\o-d_{[\o-1]})}.  \nonumber
\end{align}
\end{subequations}
Thus, by definition of $\tilde{\ET}_\o (\bX')$ we obtain
\begin{align}
\begin{split}
    \tilde\ET_\o(\bX') 
    & = \frac{\tilde{T}_\o(X=X'_{[\o+1]}q_2^{-\d_{\o,N-1}};\bX')-\tilde{T}_\o(X=X'_\o;\bX')}{\sqrt{\frac{X_\o' q_2^{\d_{\o,N-1}}}{X'_{[\o+1]}}} - \sqrt{\frac{X'_{[\o+1]}}{X'_\o q_2^{\d_{\o,N-1}}}}} \\
    & = \frac{\tilde\spadesuit_\o (X'_\o-X'_{[\o+1]}q_2^{-\d_{\o,N-1}}) + \tilde\clubsuit_\o \left( \frac{1}{X'_\o} - \frac{1}{X'_{[\o+1]}q_2^{-\d_{\o,N-1}}}  \right)   }{\sqrt{\frac{X_\o' q_2^{\d_{\o,N-1}}}{X'_{[\o+1]}}} - \sqrt{\frac{X'_{[\o+1]}}{X'_\o q_2^{\d_{\o,N-1}}}}} \\
    & = - \tilde\spadesuit_\o (X_\o'X'_{[\o+1]}q_2^{-\d_{\o,N-1}})^{\frac12} - \tilde\clubsuit_\o (X_\o'X'_{[\o+1]}q_2^{-\d_{\o,N-1}})^{-\frac12} \\
    & = (X'_{\o})^{\frac12} a_{[\o+1]}^{-\frac12} q_1^{\frac{1}{2}(k_\o-k_{\o+1}+1) } q_1^{\frac12(d_\o-d_{[\o+1]})} q_2^{\frac12\d_{\o,N-1}} \\
    & \quad + \hat\kq_\o (X'_\o) (X'_{[\o-1]})^{-\frac12} a_\o^{\frac12}(m_\o^+)^{-\frac12}(m_\o^-)^{-\frac12} q_2^{-\frac{1}{2} \d_{\o,0}} q_1^{-\frac{1}{2}(k_{\o-1}-k_\o)+\frac{1}{2}(d_\o-d_{[\o-1]})} \\
    & \quad + (X'_{\o})^{-\frac12} a_{[\o+1]}^{\frac12} q_1^{-\frac{1}{2}(k_\o-k_{\o+1}+1) } q_1^{-\frac12(d_\o-d_{[\o+1]})}q_2^{-\frac12\d_{\o,N-1}} \\
    & \quad + \hat\kq_\o (X'_\o)^{-1} (X'_{[\o-1]})^{\frac12} a_\o^{-\frac12}(m_\o^+)^{\frac12}(m_\o^-)^{\frac12} q_2^{\frac{1}{2}\d_{\o,0}} q_1^{\frac{1}{2}(k_{\o-1}-k_\o)-\frac{1}{2}(d_\o-d_{[\o-1]})} . \nonumber 
\end{split}
\end{align}
Denote
\begin{align}
    \Psi = u_{0}^{\frac{\log q_2}{\log q_1}} \prod_{\o=0}^{N-1} u_\o^{\frac{\log \frac{m_\o^+}{a_\o}}{\log q_1}} \times \hat\CalZ_{12}
\end{align}
such that we can rewrite $\tilde\ET_\o(\bX')$ as 
\begin{align}
\begin{split}
    \tilde\ET_\o(\bX') = & ~(X'_{[\o+1]})^{\frac{1}{2}} (m_\o^+)^{-\frac12} q_1^{\frac12D_{u_{[\o+1]}}+\frac12} - (X'_{[\o+1]})^{-\frac{1}{2}} (m_\o^+)^{\frac12} q_1^{-\frac12D_{u_{[\o+1]}}-\frac12} \\
    & + \kq_\o \left[ (X'_{\o})^{\frac{1}{2}} (m_\o^-)^{-\frac12} q_1^{-\frac12D_{u_{\o}}} - (X'_{\o})^{-\frac{1}{2}} (m_\o^-)^{\frac12} q_1^{-\frac12D_{u_{\o}}} \right] \\
    = & ~\ET_\o(\bX')
\end{split}
\end{align}
acting on $\langle \tilde\EQ(\bX') \rangle \Psi$.

\section{Quantum affine algebra of $\fgl(n)$ and R-matrices} \label{sec:qaa}

\subsection{Quantum affine algebra of $\fgl(n)$}
We give a brief review of the definition of the quantum affine algebra of $\fgl(n)$, $U_q (\widehat{\fgl}(n))$, as well as its coproduct.

\subsubsection{Definition}
Define the R-matrix $R(z,w) \in \text{End}(\BC^n) \otimes \text{End}(\BC^n)$ by
\begin{align}
\begin{split}
    R(z,w) &= (z-w) \sum_{a\neq b } E_{aa} \otimes E_{bb} + (q^{-1} z - q w) \sum_{a=1 } ^n E_{aa} \otimes E_{aa}  \\
    &\quad +(q^{-1} - q) z \sum_{a>b} E_{ab} \otimes E_{ba} + (q^{-1 } - q)w \sum_{a<b} E_{ab} \otimes E_{ba} \\
    & = z R^- - w R^+,
\end{split}
\end{align}
where 
\begin{align}\begin{split}
    &R^+ =  q\sum_{a=1} ^n E_{aa } \otimes E_{aa} + \sum_{a\neq b} E_{aa} \otimes E_{bb} +(q-q^{-1}) \sum_{a<b} E_{ab} \otimes E_{ba} \\
    &R^- =q^{-1} \sum_{a=1} ^n E_{aa } \otimes E_{aa} + \sum_{a\neq b} E_{aa} \otimes E_{bb} -(q-q^{-1}) \sum_{a>b} E_{ab} \otimes E_{ba} \\
    & R^+ - R^- = (q-q^{-1}) \sum_{a=1} ^n E_{ab} \otimes E_{ba} = (q-q^{-1} ) P.
\end{split}
\end{align}
Here, $E_{ab} \in \text{End}(\BC^n)$ is the standard basis element of $\text{End}(\BC^n)$ and $P$ is the exchange operator $P(v \otimes w ) = w \otimes v$. Also note that
\begin{align} \label{eq:Rspecial}
    R(z,zq^{-2}) = z(1-q^{-2}) \left( \mathds{1}- P^q \right) = 2 z (1-q^{-2}) A^{(2)}.
\end{align}

Let us also define
\begin{align}
\begin{split}
    \bar{R}(z) &:= \frac{R(z,1)}{q^{-1 } z -q} \\
    & = \sum_{a=1} ^n E_{aa} \otimes E_{aa} + \frac{1-z}{q- q^{-1} z} \sum_{a\neq b} E_{aa} \otimes E_{bb} \\
    &\quad + \frac{(q-q^{-1}) z}{q-q^{-1} z} \sum_{a>b} E_{ab} \otimes E_{ba} + \frac{q-q^{-1}}{q- q^{-1} z} \sum_{a<b} E_{ab} \otimes E_{ba}.
\end{split}
\end{align}
and
\begin{align}
    R(z) := f(z) \bar{R}(z),
\end{align}
where
\begin{align}
    f(z) = \frac{(z;q^{2n})_\infty (z q^{2n} ; q^{2n})_\infty }{(z q^2 ; q^{2n})_\infty (z q^{2n-2}; q^{2n})_\infty} ,\qquad f(z q^{2n})= f(z) \frac{(1- z q^2)(1- z q^{2n-2})}{(1-z )(1- z q^{2n})}.
\end{align}
It satisfies the trigonometric Yang-Baxter equation,
\begin{align}
    R_{12} (z) R_{13} (zw) R_{23}(w) = R_{23} (w) R_{13} (zw) R_{12} (z).
\end{align}
The R-matrix also satisfies the crossing symmetry relations,
\begin{align}
    \left( R_{12} (z) ^{-1}  \right)^{t_2} D_2 R_{12} (z q^{2n} )^{t_2} = D_2, \qquad R_{12} (x q^{2n}) ^{t_1} D_1 \left( R_{12} (x)^{-1} \right)^{t_1} = D_1,
\end{align}
where $D$ is a diagonal matrix given by
\begin{align}
    D = \text{diag}(q^{n-1} ,q^{n-3},\cdots, q^{-n+1}).
\end{align}

The quantum affine algebra $U_q (\hfgl(n))$ is generated by elements
\begin{align}
\begin{split}
    &T^+ _{ab} [s] ,\qquad T^- _{ab} [s] ,\qquad \text{with }\;\; 1\leq a,b \leq n,\quad s=0,1,2,\cdots 
\end{split}
\end{align}
and an invertible central element $q^c$. The defining relations are
\begin{align}
\begin{split}
    &T^+ _{ab} [0] = T^- _{ba } [0] = 0 ,\qquad \text{for }\quad 1\leq a<b \leq n, \\
    &T^+ _{aa} [0] T^- _{aa} [0] = T^- _{aa} [0] T^+ _{aa} [0] = 1 ,\qquad \text{for} \quad a=1,2,\cdots, n,
\end{split}
\end{align}
and
\begin{align} \label{eq:RTT}
\begin{split}
    &R (z/w) T^\pm _1 (z)  T^\pm _2 (w) = T^\pm _2 (w) T^\pm _1 (z) R(z/w) \\
    &R(zq^{-c} /w) T^- _1 (z)  T^+ _2 (w) = T^+ _2 (w) T^- _1 (z) R(z q^c/w).
\end{split}
\end{align}
For the last two relations, we defined the generating matrices $T^\pm (z) = \sum_{a,b=1}^n E_{ab} \otimes T^\pm _{ab} (z) \in \text{End}(\BC^n) \otimes U_q (\hfgl(n)) [[z^ \pm]]$ where
\begin{align}
    T^+ _{ab} (z) = \sum_{s=0} ^\infty T^+ _{ab} [s] z^{-s} ,\qquad T^- _{ab} (z) = \sum_{s=0} ^\infty T^- _{ab} [s] z^{s}.
\end{align}

\subsubsection{Coproduct and quantum determinant} \label{subsubsec:cop}
The quantum affine algebra $U_q (\hfgl(n))$ is a Hopf algebra. It is endowed with a coproduct $\D : U_q (\hfgl(n)) \to U_q (\hfgl(n)) \otimes U_q (\hfgl(n)) $ defined by
\begin{align}
    \D(T^\pm _{ab} ( z )) = \sum_{c=1 }^n T_{ac} ^\pm (z q^{\pm \frac{c_2}{2}}) \otimes T_{cb} ^\pm (z q^{\mp \frac{c_1}{2}}),
\end{align}
where $q^{c_1} = q^{c \otimes 1} $ and $q^{c_2} = q^{1\otimes c}$. This is equivalent to
\begin{align}
    (\text{id} \otimes \D) T^\pm (z) = T^\pm _{1} (z q^{\pm \frac{c_2}{2}}) T^\pm _{2} (z q^{\mp \frac{c_1}{2}}) \in \text{End}(\BC^n)\otimes U_q (\hfgl(n)) \otimes U_q (\hfgl(n)),
\end{align}
where the subscripts indicate where the generating matrices are valued in. For instance, $T^\pm _1 (z) = \sum_{a,b=1}^n E_{ab} \otimes T^\pm _{ab} (z) \otimes \text{id}$.

The quantum determinant\footnote{The \textit{quantum determinant} (denoted by $\text{qdet})$ should not be confused with the \textit{$q$-determinant} (denoted by $\det_q$) for $q$-Manin matrices. We will shortly construct a particular $q$-Manin matrix whose $q$-determinant produces the quantum determinant of the generating matrices $T^\pm (z)$.} of the generating matrices $\text{qdet} \, T^\pm (z)$ are series in $z^{-1}$ and $z$, respectively, defined by
\begin{align} \label{eq:qdet}
\begin{split}
    \text{qdet} \, T^\pm (z) &=  \sum_{\s \in S_n} (-q)^{-l(\s)} T^\pm _{\s(1) 1} (z) T^\pm _{\s(2) 2} (z q^{-2})  \cdots T^\pm _{\s(n) n} (z q^{-2n+2}) \\
    &= \sum_{\s \in S_n} (-q)^{l(\s)} T^\pm _{n \s(n) } (zq^{-2n+2}) \cdots T^\pm _{2\s(2) } (z q^{-2}) T^\pm _{1\s(1) } (z),
\end{split}
\end{align}
where $S_n$ is the permutation group of $\{1,2,\cdots,n\}$ and $l(\s)$ is the length of the permutation $\s \in S_n$. Note that
\begin{align} \label{eq:factor}
    \D(\text{qdet} \, T^\pm (z)) = \text{qdet} \, T^\pm (z) \otimes \text{qdet} \, T^\pm (z) . 
\end{align}
The coefficients of $\text{qdet} \, T^\pm (z)$ lie in the center of the quantum affine algebra $U_q (\hfgl(n))$ at the critical level $c=-n$ \cite{Frappat:2015osj}.

\subsection{R-matrix and monodromy matrix}

\subsubsection{Quantized enveloping algebra of $\fgl(n)$} \label{subsubsec:qea}
The quantized enveloping algebra of $\fgl(n)$, denoted by $U_q (\fgl(n))$, is an associative algebra generated by $t_i$, $t_i ^{-1}$, $i=1,2, \cdots n$, and $e_i$, $f_i$, $i=1,2,\cdots n-1$, with the defining relations
\begin{align}
\begin{split}
    &t_i t_j = t_j t_i ,\qquad t_i t_i ^{-1} = t_i ^{-1} t_i = 1, \\
    &t_i e_j t_i ^{-1} = q^{\d_{i,j} - \d_{i,j+1}} e_j, \qquad t_i f_j t_i ^{-1} = q^{-\d_{i,j} + \d_{i,j+1}} f_j, \\
    &[e_i, f_j] = \frac{t_i t_{i+1} ^{-1} - t_{i+1} t_i ^{-1}}{q- q^{-1}} \d_{i,j} ,\\
    & [e_i ,e_j ] = [f_i,f_j] = 0 , \qquad \text{if } |i-j| >1, \\
    & [e_i, [e_{i\pm 1} , e_i]_q]_q = [f_i,[f_{i\pm 1} , f_i]_q]_q = 0, 
\end{split}
\end{align}
where we have used the $q$-commutator, $[a,b]_q = a b - q b a$.

There is an R-matrix presentation of $U_q (\fgl(n))$. Consider the R-matrix given by
\begin{align}
    R^+ = q \sum_{i=1} ^n E_{ii} \otimes E_{ii} + \sum_{i\neq j} E_{ii} \otimes E_{jj} + (q- q^{-1}) \sum_{i<j} E_{ij} \otimes E_{ji},
\end{align}
valued in $\text{End}(\BC^n) \otimes \text{End}(\BC^n)$. The $q$-deformed enveloping algebra $U_q (\fgl(n))$ is generated by the elements $t_{ij} ^+$ and $t_{ij} ^-$ with $1 \leq i,j \leq n$, subject to the relations
\begin{align}
\begin{split}
    &t_{ij } ^+ = t_{ji} ^- = 0 ,\qquad 1\leq i <j \leq n ,\\
    &t_{ii}^+ t_{ii} ^{-} = t_{ii} ^{-} t_{ii} ^+ = 1,\quad 1\leq i\leq n,\\
    &R^+ T_1 ^+ T_2 ^+ = T_2 ^+  T_1 ^+ R^+ , \qquad R^+ T^- _1 T^- _2 = T_2 ^- T_1 ^- R^+, \qquad R^+ T^- _1 T^+ _2 = T_2 ^+ T_1 ^- R^+,
\end{split}
\end{align}
where we defined
\begin{align}
    T ^+= \sum_{i,j=1} ^n E_{ij} \otimes t_{ij} ^+  , \qquad     T^- = \sum_{i,j=1} ^n E_{ij} \otimes t_{ij} ^-,
\end{align}
valued in $ \text{End}(\BC^n) \otimes U_q (\fgl(n))$. The last three relations are valued in $ \text{End}(\BC^n) \otimes \text{End}(\BC^n) \otimes U_q (\fgl(n)) $, where the subscripts of $T^\pm$ indicate which of the two $\text{End}(\BC^n)$ it is valued in. 

The last three relations can also be explicitly written as
\begin{align}
\begin{split}
    &q^{\d_{i,j}} t_{ia} ^\pm t_{jb} ^\pm - q^{\d_{a,b}} t_{jb} ^\pm t_{ia} ^\pm = (q-q^{-1}) (\th_{b<a} -\th_{i<j}) t_{ja} ^\pm t_{ib} ^\pm, \\
    & q^{\d_{i,j}} {t}^- _{ia} t^+ _{jb} -q^{\d_{a,b}} t ^+ _{jb} {t}^-_{ia} = (q-q^{-1}) (\th_{b<a} t^+_{ja} {t}^-_{ib} -\th_{i<j} t^-_{ja} t^+_{ib}).
\end{split}
\end{align}

An isomorphism between the two presentations is given by
\begin{align}
    t_i \mapsto t_{ii} ^+ ,\quad t_i ^{-1} \mapsto t_{ii} ^- ,\quad e_i \mapsto -\frac{t_{i,i+1} ^- t_{ii} ^+}{q-q^{-1}} ,\quad f_i \mapsto \frac{t_{ii} ^- t_{i+1,i} ^+}{q-q^{-1}}.
\end{align}

\subsubsection{Evaluation homomorphism}
The quantized enveloping algebra can be identified as a Hopf subalgebra of the quantum affine algebra, by the natural embedding $\iota:U_q ({\fgl}(n)) \hookrightarrow U_q (\widehat{\fgl}(n))$ defined by
\begin{align}
    t^+ _{ij} \mapsto T^+ _{ij} [0],\qquad t^- _{ij} \mapsto T^- _{ij} [0].
\end{align}

Given any evaluation parameter $a \in\BC^\times$, the evaluation homomorphism $\text{ev}_a : U_q (\hfgl(n)) \to U_q (\fgl(n))$ is defined by \cite{molev2004}
\begin{align}
    T^+ (z) \mapsto T^+ - \frac{a}{z} T^-  ,\quad T^- (z) \mapsto T^- - \frac{z}{a} T^+ ,\quad q^c \mapsto 1.
\end{align}
In particular, $T^\pm (z)$ are not independent of each other in the image of the evaluation homomorphism: $\text{ev}_a (T^+ (z)) = - \frac{a}{z} \text{ev}_a (T^- (z))$. Note that $\text{ev}_a \circ \iota = \text{id}$ for any $a \in \BC^\times$.

Given $N$ modules $ \r_{\CalH_\o} : U_q (\fgl(n)) \to \text{End} (\CalH_\o)  $, $\o=0,1,\cdots, N-1$, and $N$ evaluation parameters $\left(a_\o \right)_{\o=0} ^{N-1} \in \left(\BC^\times\right)^{\times N}$, we construct a module $\CalH(\ba)$ over $U_q (\widehat{\fgl}(n))$ by
\begin{align}
    (\r_{\CalH_{N-1}} \otimes \cdots \otimes\r_{\CalH_0} ) \circ (\text{ev}_{a_{N-1}} \otimes \cdots \otimes \text{ev}_{a_0} )\circ \Delta^{N-1} : U_q (\widehat{\fgl}(n)) \to \text{End}\left( \CalH\right),\quad \CalH = \otimes _{\o=0 } ^{N-1} \CalH_\o.
\end{align}
The generating matrix $T^+ (z)$ is represented on $\CalH$ as
\begin{align}
    T^+ (z) \vert_{\CalH} =T^+ (z) \vert_{\CalH_{N-1}} \cdots T^+ (z) \vert_{\CalH_0} \in \text{End}(\BC^n) \otimes \text{End}(\CalH),
\end{align}
where the product is taken as matrices in $\text{End}(\BC^n)$.

\bibliographystyle{utphys}
\bibliography{reference}

\end{document}